\documentclass[twocolumn]{aastex63} 
\usepackage{placeins}
\usepackage[T1]{fontenc}
\usepackage[scaled]{helvet}
 
\usepackage[version=4]{mhchem} 

\shorttitle{Super-Earths and Earth-like Exoplanets}
\shortauthors{Lichtenberg \& Miguel}
\graphicspath{{./}{figures/}}

\begin{document}

\title{Super-Earths and Earth-like Exoplanets}

\correspondingauthor{Tim Lichtenberg}
\email{tim.lichtenberg@rug.nl}

\author[0000-0002-3286-7683]{Tim Lichtenberg}
\affiliation{Kapteyn Astronomical Institute, University of Groningen, P.O. Box 800, 9700 AV Groningen, The Netherlands}
\author[0000-0002-0747-8862]{Yamila Miguel}
\affiliation{Leiden Observatory, University of Leiden, Niels Bohrweg 2, 2333 CA Leiden, The Netherlands}
\affiliation{SRON Netherlands Institute for Space Research, Niels Bohrweg 4, 2333 CA Leiden, The Netherlands}

\begin{abstract}
In the last few years astronomical surveys have expanded the reach of planetary science into the realm of small and dense extrasolar worlds. These share a number of characteristics with the terrestrial and icy planetary objects of the Solar System, but keep stretching previous understanding of the known limits of planetary thermodynamics, material properties, and climate regimes. Improved compositional and thermal constraints on exoplanets below $\sim$2 Earth radii suggest efficient accretion of atmosphere-forming volatile elements in a fraction of planetary systems, pointing to rapid formation, planet-scale melting, and chemical equilibration between the core, mantle, and atmosphere of rocky and volatile-rich exoplanets. Meaningful interpretation of novel observational data from these worlds necessitates cross-disciplinary expansion of known material properties under extreme thermodynamic, non-solar conditions, and accounting for dynamic feedbacks between interior and atmospheric processes. Exploration of the atmosphere and surface composition of individual, short-period super-Earths in the next few years will enable key inferences on magma ocean dynamics, the redox state of rocky planetary mantles, and mixing between volatile and refractory phases in planetary regimes that are absent from the present-day Solar System, and reminiscent of the conditions of the prebiotic Earth. The atmospheric characterization of climate diversity and the statistical search for biosignatures on terrestrial exoplanets on temperate orbits will require space-based direct imaging surveys, capable of resolving emission features of major and trace gases in both shortwave and mid-infrared wavelengths.
\linebreak
\end{abstract}



\section*{Key points} \label{sec:keypoints}

\begin{itemize}
    \item Exoplanet observations are on the brink of characterizing the atmospheres and surface composition of low-mass exoplanets orbiting M-dwarf stars.
    \item The majority of transiting exoplanets up to the super-Earth size regime receive irradiations beyond the runaway greenhouse threshold, suggesting liquid rock mantles for an $\gtrsim$Earth-like volatile endowment.
    \item Volatile-stripped rocky exoplanets may host localized lava ponds on their daysides and silicate-rich atmospheres that reflect the magma composition.
    \item Characterization of exoplanets with dominantly molten mantles (magma oceans) can probe planetary thermodynamic regimes that are inaccessible in the present-day Solar System, providing a testbed for the transition from primary to secondary atmospheres.
    \item Super-Earths with lower densities than Earth may have a greater bulk fraction of atmospheric volatiles and global magma oceans, which store the majority of atmophile elements.
    \item In highly molten planetary regimes, atmospheric volatiles chemically equilibrate rapidly between core, mantle and atmosphere: derived bulk volatile fractions must account for interior phase state.
    \item Interaction between atmospheric and magma-dissolved volatiles creates emergent co-evolution of the interior and atmosphere of rocky and volatile-rich exoplanets.
    \item The distribution and composition of secondary atmospheres on super-Earths can potentially discriminate interior redox states, revealing timescales and efficacy of chemical segregation between core, mantle, and atmosphere.
    \item The surface mineralogy and spatial extent of dayside magma oceans of atmosphere-less rocky exoplanets offers insight into planetary geodynamics and tectonic history.
    \item Deriving robust physical and chemical constraints from astronomical observations will require experimental extension of known material properties under extreme thermodynamic and non-solar conditions.
\end{itemize}

\section{Introduction} \label{sec:introduction}

The observed variety and anticipated number of extrasolar planets in the galaxy breaches the limits of human imagination. Over the past $\sim$30 years, the rapid increase in detected exoplanets opened a whole new field of science that transcends canonical disciplinary boundaries. While data on exoplanets can -- within the foreseeable time -- only be acquired by remote observation; understanding their meaning requires interpretation in light of physical and chemical concepts that are derived from measurements and experiments on Earth and across the Solar System. Deciphering signals from extrasolar worlds to understand the diversity of planets afar therefore interconnects astronomy, physics, earth and atmospheric sciences, chemistry, biology, and more. At the time of writing, more than 5500 exoplanets are known, spanning planetary classes that are not represented in the Solar System, like super-Earths and sub-Neptunes, planets that are intermediate in size between the terrestrial planets and the ice giants. In this review, we focus on exoplanets with bulk densities that approach values approximately consistent with the terrestrial planets and moons of the Solar System, even though we will highlight that the boundaries between planetary classes at high entropy/energy become increasingly diffuse. We limit our discussion to exoplanets up to $\sim$2 Earth radii and $\sim$10 Earth masses -- exoplanets in this regime are typically referred to as sub-Neptunes or super-Earths. Sub-Neptunes fill the mass-radius-density space toward Neptune, while we use the term super-Earth to refer to planets that approach Earth-like densities. However, while qualitative separation in radius-density space has become more evident in the past few years, the physical and chemical distinction between these two classes is ambiguous as of yet, but promises to reveal fundamental processes of planetary formation and evolution that closely interweave with the very concept of being \emph{like Earth}.

Information on individual exoplanets is scarce -- remote signals can never be as detailed as in-situ measurements -- but the great strength of exoplanet science lies in breadth: exoplanetary systems span the full range of evolutionary snapshots since the formation of the galaxy, and reveal the wide variety of elemental compositions and system architectures that are physically viable. This opens the possibility of filling crucial gaps in the Solar System record, which is patchy due to billions of years of planetary and dynamical evolution, overwriting much of the geochemical and geologic evidence of the past. Rocky exoplanets, such as ultrashort-period super-Earths with orbits of less than one day, open observational windows into global geodynamic and climate regimes that are inaccessible in the present Solar System: seemingly exotic planetary regimes, such as steam runaway greenhouse climates, magma oceans, heat-pipe tectonics, atmospheric blow-off, or snowball cycles, are likely important periods of planetary evolution in the distant past that shaped the long-term climate and surface of the Earth and terrestrial planets. In particular, favoured by the inherent bias introduced by the dominant planet search techniques, the transit and radial velocity methods, the exoplanet population provides access to planets in high entropy states -- a regime crucially important for the interpretation of the earliest evolution of the terrestrial planets right after formation, and the most difficult to study experimentally. Therefore, while, at first glance, exoplanet science seems to be disconnected from the study of Earth and other planets of the Solar System, characterization of qualitatively different evolutionary trajectories promises to reveal global physical and chemical processes that shaped our own world in the distant past.

Characterising planetary regimes that are potential analogs for the prebiotic Earth will crucially aid studies of the chemical origins of life and the distinction between potentially habitable and lifeless worlds. Finding unambiguous signs of possible extrasolar life -- remote biosignatures -- can thus only succeed by building a solid foundation of the abiotic physical and chemical processes of rocky planets -- from their birth in the circumstellar disk to their long-term geodynamic and climatic evolution. Giving an overview of some of the major related developments of this endeavour in the past years is the major goal of this review. At the time of writing, the James Webb Space Telescope (JWST) has recently started to take its first measurements of the atmospheres of Earth-sized exoplanets: a feat that was out of reach just a few years ago. Planned ground- and space-based surveys in the 2020s and 2030s will increase the number and information on known exoplanets in the Earth to super-Earth size-regime manifold, qualitatively change our currently theory-dominated concepts, and pave the way toward the telescopic characterization of terrestrial exoplanets on temperate orbits in the 2040s. This review is therefore necessarily a snapshot of the research that underpins the current thinking and observational tests of theoretical concepts that will be performed in the next decade.

We start the review with a recap of currently accessible observational evidence on rocky exoplanets (Section \ref{sec:observations}), with a focus on the compositional variability exhibited in mass-radius and radius-density space. We then chart the expansion of planet formation theory in light of new data from exoplanets and disks, with a brief connection on its relation to the Solar System in Section \ref{sec:formation}. In Section \ref{sec:atmospheres1} we outline the transition from planetary accretion toward the potential formation of secondary atmospheres on rocky exoplanets, with an emphasis on the geophysical and geochemical scenarios that ought to be tested with novel experimental and observational evidence. This is followed by an overview of the main classes and observational techniques to analyse planetary atmospheres in Section \ref{sec:atmospheres2}. The connection to the deep interiors of rocky exoplanets, their mantles and cores, is described in Section \ref{sec:interiors}, both from an experimental and theoretical point of view. We present a brief outlook on the developments we expect in the upcoming years with data from new astronomical surveys and experimental facilities in Section \ref{sec:outlook}, and summarise in Section \ref{sec:summary}.

\section{Observational evidence} \label{sec:observations}

\begin{figure*}[tbh]
 	\centering
        \includegraphics[width=0.99\textwidth]{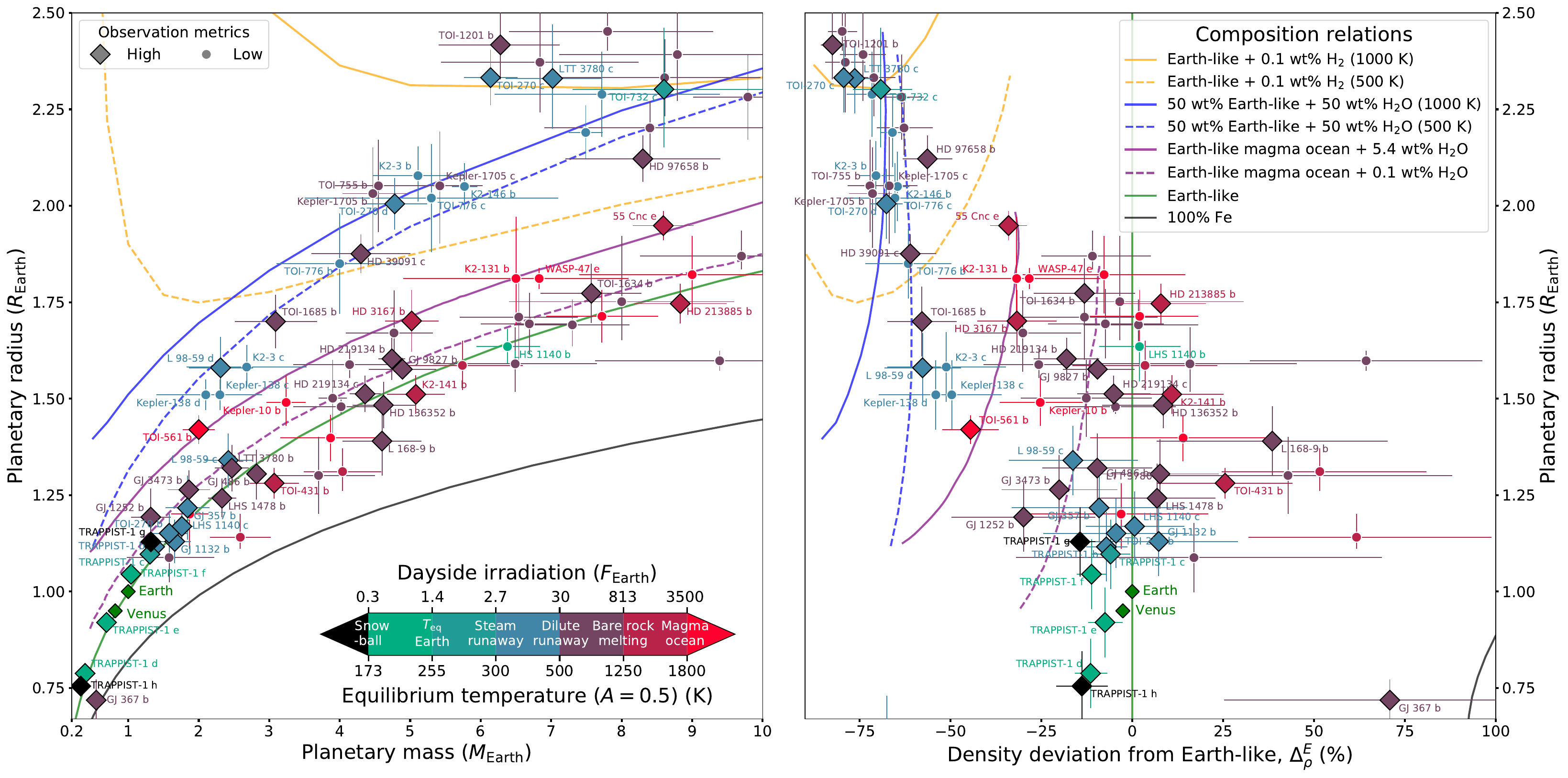}
 	\caption{\textsf{\textbf{(Left) Measured exoplanet radii as a function of mass relative to idealized composition curves. (Right) Radii versus density deviation relative to Earth-like.} Colored symbols represent the best-characterized exoplanets with mass uncertainties below 50\% and radius uncertainties below 30\%. Data points from \citet{2020A&A...634A..43O}, \citet{2022MNRAS.511.4551L}, \citet{2022Sci...377.1211L}, \citet{Diamond-Lowe2022}, and \citet{2023NatAs...7..206P}. Mass-radius relations from \citet{2019PNAS..116.9723Z} and \citet{2021ApJ...922L...4D}. Diamond symbols with attached names indicate planets with high transmission (TSM) or emission (ESM) metrics, which makes these planets favourable targets for observational characterization: TSM $>$ 90 if $R_{\rm{p}} \geq 1.5$ $R_{\rm{Earth}}$, otherwise TSM $>$ 10, or ESM $>$ 7.5 \citep{2018PASP..130k4401K}. The translation of dayside irradiation to equilibrium temperature assumes efficient global heat redistribution and a constant bond albedo of 0.5. Color deviations and trend interpretation are discussed in the main text.}}
    \label{fig:M-R}
\end{figure*}

Detecting exoplanets is a major technical challenge because the intrinsic and reflected light emitted from any planet is orders of magnitude smaller than that from its host star. This extreme contrast together with the vast distances between the Solar System and extrasolar planetary systems so far preclude observations of rocky exoplanets via direct imaging. All observational evidence on rocky exoplanets we possess at this point is through indirect measurements of the exoplanets' host star light, via transit, radial velocity, or microlensing techniques. Because these techniques, however, are limited in obtaining information on temperate exoplanets, a shift toward direct-imaging surveys will take place on a multi-decadal timeline. A general overview of observational techniques for exoplanet detection and general demographics is beyond the scope of this review, and we refer the reader to more specialised summaries, with a focus on giant planet demographics \citep{2021ARA&A..59..291Z,2021exbi.book....2G}, transit \citep{2021exbi.book....3P} and radial velocity \citep{2021exbi.book....4F} surveys, direct imaging \citep{2021exbi.book....5H,2023ASPC..534..799C}, transit spectroscopy \citep{2021exbi.book....7D}, high-resolution spectroscopy \citep{2021exbi.book....8B}, space missions \citep{2021exbi.book....9S}, and gravitational microlensing \citep{2012ARA&A..50..411G,2018haex.bookE.123U}. 

Transit techniques measure the dip in received stellar light when an exoplanet passes through the line of sight between the telescope and host star (transit or primary eclipse) or vanishes fully behind the star during its orbit (secondary eclipse). The variation in brightness of the star-planet system over the course of one entire orbit (phase curve) is also useful as it shows the different phases of the planet. From these observations a host of information can be deduced, in particular the ratio between the radius of the planet and that of the star at different wavelengths (transit), but also information about the planet's flux divided by that of the star (secondary eclipse), which could also lead to a determination of the atmospheric composition, and finally redistribution of heat and circulation of the atmosphere (phase curves, Section \ref{sec:3d}). Crucially important in the characterisation of rocky exoplanets is the precision of mass determinations with radial velocity measurements. When a planet orbits a star, it exerts a gravitational force on the star, causing it to move in a small orbit around the system's center of mass. The Doppler shift in the spectral lines of the stellar light can thus be measured with a spectrograph to determine the minimum mass of the planetary companion. Because the amplitude of the radial velocity variation depends on the mass of the planet as well as the distance between the planet and the star, small planets require extreme precision radial velocity measurements. Just as an example, the effect of the Earth orbiting around the Sun causes a velocity amplitude of 9 cm s$^{-1}$, which is, at the time of writing, out of reach for even the most powerful spectrographs \citep[CARMENES, ESPRESSO, and MAROON-X,][]{2014SPIE.9147E..1FQ,2021A&A...645A..96P,2022SPIE12184E..1GS,SuarezMascareno2023}. These measurements are not only challenging from an engineering perspective, but are in particular limited by our understanding of stellar activity \citep[e.g.,][]{2018A&A...614A..76J,2019MNRAS.489.2555D,2023A&A...670A..24S}. 

The earliest characterizations of rocky exoplanets were enabled by spaced-based transit photometry supported by ground-based radial velocity measurements of CoRoT-7b \citep{2009A&A...506..287L}, 55 Cancri e \citep{2011A&A...533A.114D,2011ApJ...737L..18W}, and Kepler-10b \citep{2011ApJ...729...27B}, yielding insight into the radius and mass of these planets and therefore their bulk density. Mass estimates from radial velocity or transit timing variations, and size estimates from transit photometry are the only available information on the vast majority of potentially rocky exoplanets to date, with a few notable exceptions, which we discuss below. 

Figure \ref{fig:M-R} shows the known population of low-mass exoplanets at the time of writing with reasonably-well constrained mass and radius measurements below 2.5 Earth radii ($R_\mathrm{Earth}$) and below 10 Earth masses ($M_\mathrm{Earth}$), compared to theoretical models of planetary structure, overall approximately 100 planets. Symbols in the figure indicate individual exoplanets, color-coded by their dayside (substellar) irradiation. The vast majority of these exoplanets orbit M stars, which are much smaller and cooler than G-type (Sun-like) stars, because these stars are much more numerous in the galaxy and their small radii relative to Sun-like stars increases the transit depth and signal-to-noise ratio of exoplanet transits \citep{2007AsBio...7...85S,2019A&A...624A..49W} (Section \ref{sec:futureObs}). Because the transit probability and depth decreases with increasing stellar mass and wider planetary orbit, we do not yet know of a transiting Earth-sized exoplanet on a temperate orbit around a Sun-like star. While in principle the combined yield from the on-going TESS mission \citep[Transiting Exoplanet Survey Satellite,][]{2015JATIS...1a4003R} and the upcoming PLATO \citep[PLAnetary Transits and Oscillations of stars,][]{2014ExA....38..249R,2016AN....337..961R} survey may change this, their survey strategies may not cover enough consecutive transits on any single target to unambiguously detect these planets. Mass-radius data points in Figure \ref{fig:M-R} are color-coded by the planets' received stellar irradiation, which is translated into equilibrium temperature on the color scale, assuming a constant albedo of 0.5 (Earth's albedo $\approx$ 0.3, Venus' albedo $\approx$ 0.77) and efficient heat redistribution from the day- to nightside. This represents an approximate lower limit to the actual surface temperature on these worlds, because it assumes that the planets are not tidally locked by the gravitational tides from their parent star or that an atmosphere is present that redistributes dayside irradiation toward the nightside via global circulation, and that additionally there is no substantial greenhouse effect operating. Colored lines represent isolines of a fixed composition and structure. For example, the Earth-like line (solid green) keeps the bulk composition and structure (core-mantle ratio) constant when scaling up the planet in mass, but accounts for isostatic compression of materials at high pressure (Section \ref{sec:interiors}). Here, we summarise the main observational findings in this regime to date. In the following sections, we will outline ranges of physical and chemical processes that contribute to shaping the planetary distribution.

On a population level, Figure \ref{fig:M-R} illustrates that planets cluster around compositional isolines. Most planets are consistent with an Earth-like, non-melted bulk composition and density (green solid line) or hydrogen-rich composition (yellow lines). For $\gtrsim$ 5 $M_\mathrm{Earth}$, and between the Earth-like (green) and water-rich (blue) composition lines, there is a dearth of planets. This occurrence gap, commonly referred to as the 'radius valley', was predicted by theory \citep{2013ApJ...775..105O,2013ApJ...776....2L,2014ApJ...795...65J,Rogers2015} and confirmed empirically using data from the NASA Kepler Mission \citep{2010Sci...327..977B} combined with increased precision on stellar radii from high-resolution Keck data \citep{2017AJ....154..109F,Martinez2019}, precise stellar distances from Gaia \citep{GaiaCollaboration2018,Fulton&Petigura2018}, and asteroseismology \citep{2018MNRAS.479.4786V}. The location of the radius valley shifts to smaller planet radii with decreasing stellar mass \citep{Cloutier2020}. At about 4--6 $M_\mathrm{Earth}$ and $\approx$2 $R_\mathrm{Earth}$, we find a cluster of planets, and one at $\gtrsim$6 $M_\mathrm{Earth}$ and $\gtrsim$ 2.2 $R_\mathrm{Earth}$, both with $\approx 75 \%$ lower density than Earth ($\Delta_\rho^E$). An important feature of the separation between this larger population, consistent with either a water- (blue) or hydrogen-rich (yellow) composition, is that these planets are generally less irradiated than the smaller and denser population around the Earth-like composition lines (green and purple). The latter planets are, with a few exceptions, all highly irradiated to $\gtrsim$ 30 Earth's instellation at present day ($F_\mathrm{Earth}$). This translates into equilibrium temperatures above 500 K (dilute runaway, purple), and for a few above 1250 K (bare rock melting, dark red), and even 1800 K (magma ocean, bright red). The latter two temperatures mean that the mantles of these planets are at least partially molten, irrespective of whether they host an atmosphere or not. As we discuss in Section \ref{sec:atmospheres1}, this change in phase state has an important effect on the interpretation of the low-mass exoplanet census because molten planetary interiors are highly efficient in storing atmospheric volatiles. Planets above 300 K (steam runaway, blue) and above 500 K (dilute runaway, purple) will undergo a runaway greenhouse climate, in which water catastrophically evaporates in a feedback loop (Section \ref{sec:atmospheres1}), if they host even minimal quantities of water. There are only six exoplanets (potentially) below the runaway greenhouse transition (black, and light and dark green): LHS 1140 b, and TRAPPIST-1 c, d, e, f, and g. We will discuss the significance of these climate categories in sections \ref{sec:atmospheres1} and \ref{sec:atmospheres2}, and the implications for the interior structure and dynamics in Section \ref{sec:interiors}.

Most demographic characteristics in the current exoplanet population are statistically significant only at larger sizes \citep{2015ARA&A..53..409W,2021ARA&A..59..291Z}, such as the giant planet and sub-Neptune size-regimes, but a few trends provide tentative guidance for the super-Earth to Earth-sized regime. While exact probabilistic estimates differ, occurrence rates from the Kepler and TESS space mission provide convincing evidence that small, potentially rocky exoplanets are numerous in the galaxy \citep{2015ApJ...807...45D,2019ApJ...883L..15P,2021AJ....161...36B, 2023AJ....165..265M}. Small exoplanets form around stars with a wide range of metallicities (elements heavier than He) \citep{2012Natur.486..375B,2014Natur.509..593B}, so far \citep[cf.][]{2023AAS...24143003B} showing no signs of a positive trend with host star metallicity or mass like giants \citep{2005ApJ...622.1102F,2021ApJS..255...14F}. In contrast, Kepler \citep{2012ApJS..201...15H,2015ApJ...798..112M} and the CARMENES radial velocity survey \citep{2014SPIE.9147E..1FQ,2021A&A...653A.114S} both suggest that the occurence rate of super-Earths decreases with increasing stellar mass \citep{Mulders2018,2021ApJ...920L...1M}. Combined analyses of stellar metallicities and planetary bulk composition tentatively indicates that density and metallicity are positively correlated, suggesting that rocky planets around higher-metallicity stars host larger iron cores \citep{2021Sci...374..330A}. In general, and in strong contrast to the classically-ordered structure of the Solar System, planetary size seems highly correlated within most exoplanetary systems \citep{2017ApJ...849L..33M,2018AJ....155...48W,2023ASPC..534..863W}, even though different architectures, from size-ordered to stochastic, exist in a (smaller) fraction of planetary systems \citep{2023A&A...670A..68M}. In addition, the presence of giant planets may impact the existence and composition of Earths and super-Earths in the same system \citep{2018AJ....156...92Z,2019ApJ...874...81F,Schlecker2021,Rosenthal2022}.

Multiple planets are found between the Earth-like composition and a pure iron composition (black line in Figure \ref{fig:M-R}), reminiscent of Mercury's internal structure. For example, combined data from TESS \citep{2015JATIS...1a4003R} and ground-based radial velocity indicate that the sub-Earth GJ 367 b \citep{2021Sci...374.1271L} features a very high density, suggesting the fraction of its metal core to be $\approx$90\%. Other notable examples include K2-229 b \citep{2018NatAs...2..393S}, HD 137496 b \citep{2022A&A...657A..68A}, K2-106 b \citep{2017AJ....153...82A, 2023AJ....165...97R} and Kepler-107 c \citep{2019NatAs...3..416B}. The latter system is of interest because with a nearly identical radius to its sister-planet Kepler-107 b, the density of Kepler-107 c is twice as high, suggesting that stochastic events, such as giant impacts \citep{2018SSRv..214...34S,2021arXiv210302045C}, may shape the bulk density of individual exoplanets (Section \ref{sec:formation}). While still subject to debate, further detections of such super-Mercury exoplanets 
\citep{2018NatAs...2..393S,2019NatAs...3..416B,2021Sci...374.1271L,2021Sci...374..330A,2023AJ....165...97R,2023AJ....165...47E} confirm that these planets are frequent, suggesting the formation pathway to form high-density exoplanets must be rather common \citep{2022A&A...662A..19J,2023A&A...673A..17M,2024MNRAS.529.2577D}. On the other side of the green Earth-like composition line, multiple notable exceptions fill the disconnect between Earth-like densities and hydrogen-rich planets: 55 Cnc e, Kepler-10 b, HD 3167 b, Kepler-138 d+c, L 98-59 d, K2-3 c, TOI-561 b, and TOI-1685 b are all inconsistent with a pure rocky (Earth-like) composition, but potentially too small to be explained with H-He gas accretion from the protoplanetary disk \citep{2022MNRAS.511.4551L,2022Sci...377.1211L,2023NatAs...7..206P,2023AJ....165...88B,2023A&A...679A..92P}. In a comparable argument to Kepler-107, the K2-3 system points to physical processes bifurcating the evolution of individual planets within a single system. In the K2-3 system architecture, the sub-Neptune world K2-3 b orbits closest to the host star, interior to the smaller planets, K2-3 c and d \citep{Diamond-Lowe2022}. The outer worlds K2-3 c and d have radii that place them on the super-Earth side of the radius valley, though the mass of K2-3 d is not well-constrained with existing radial velocity data. This may suggest that K2-3 c+d are inflated, potentially undergoing active mass loss in a runaway greenhouse state, hosting a molten, volatile-rich magma ocean below \citep{2021ApJ...922L...4D,2024ApJ...962L...8S}. A similar inference can be drawn for the unusually low-density super-Earth GJ 1018 b (TOI-244 b), which despite orbiting a very bright M-dwarf provides evidence for substantial incorporation of atmospheric volatiles \citep{2023A&A...675A..52C}.

In addition to analyses via the two primary observables radius and mass, astronomical studies are beginning to investigate the atmospheric compositions of small exoplanets. Before 2024,  there has not been a definitive detection of an atmosphere on a rocky exoplanet, but the beginning of JWST science operations are poised to change this. To a large extent within the achieved observational uncertainties of the pre-JWST era, the featureless transmission spectra of GJ 1132 b \citep{Diamond-Lowe2018,2021AJ....161..284M,2022AJ....164...59L,2023ApJ...959L...9M}, LHS 1140 b \citep[][]{2020AJ....160...27D,2021AJ....161...44E,2024arXiv240313265D}, LHS 3844 b \citep{2020AJ....160..188D,2023ApJ...952L...4I}, LTT 1445Ab \citep{Diamond-Lowe2023}, TRAPPIST-1 b--h \citep{2016Natur.537...69D,2018NatAs...2..214D,Wakeford2019,Garcia2022,Gressier2022}, L 98 59 b--d \citep{2022AJ....164..225D, 2022AJ....164..203Z, 2023RAA....23b5011Z, 2023arXiv230110866B}, LHS 475 b \citep{2023NatAs...7.1317L}, GJ 486 b \citep{Ridden-Harper2023,2023ApJ...948L..11M}, GJ 341 b \citep{2024AJ....167...90K}, and TOI-836b \citep{2024arXiv240400093A} have been measured to a precision high enough to rule out cloud-free, hydrogen-dominated atmospheres on these planets. JWST provides additional capabilities to put constraints on higher mean-molecular weight atmospheres, which, at the time of writing, has led to upper limits on CO$_2$ for TRAPPIST-1 b+c \citep{Greene2023,Zieba2023,2024ApJ...960...44T} and GJ 367 b \citep{2024ApJ...961L..44Z}. Notably, thermal emission from 55 Cnc e, observed with JWST, rules out the possibility of the planet being a lava world cloaked by a thin atmosphere of vaporized rock. Instead, they suggest the presence of a genuine volatile atmosphere, likely abundant in \ce{CO2} or CO \citep{Hu2024}, in equilibrium with an underlying magma ocean (Section \ref{sec:atmospheres1}). For larger sub-Neptunes, which have larger transmission features due to the increased scale heights of their H-rich atmospheres, JWST recently enabled the first transmission spectra, namely for K2-18 b \citep{2023ApJ...956L..13M} and TOI-270 d \citep{2024arXiv240303325B,2024A&A...683L...2H}. Full phase curves can provide additional information regarding atmospheric composition and circulation. So far full phase curves have been measured for four rocky exoplanets: 55 Cnc e \citep{2016Natur.532..207D,2023A&A...669A..64D,Mercier2022}, LHS 3844 b \citep{2019Natur.573...87K}, K2-141 b \citep{2022A&A...664A..79Z}, and GJ 367 b \citep{2024ApJ...961L..44Z}, which we will discuss in more detail in Section \ref{sec:atmospheres2}.

High-resolution spectroscopy from the ground has proven to be a successful technique for characterising giant planets, but so far most searches of molecular species on rocky exoplanets remained limited to ruling out H-dominated atmospheres for 55 Cnc e \citep{2020AJ....160..101J} and GJ 486 b \citep{2023AJ....165..170R}, non-detection of Fe-dominated species \citep{2023AJ....166..155R}, or inconclusive results \citep{2014ApJ...797L..21D,2016A&A...593A.129R,2017AJ....153..268E}. However, with the advent of the next generation of 30-meter telescopes \citep{2013ApJ...764..182S, 2019ApJ...871L...7S, 2024MNRAS.528.3509V}, this technique will offer a promising avenue for characterizing the atmospheres of a few of the closest rocky exoplanet atmospheres from the Sun (Section \ref{sec:outlook}). 

It is worth noting that most of the rocky exoplanets just discussed orbit close to small, cool stars called M dwarfs. The small planet-to-star radius ratio provided by M dwarfs make them the most favorable hosts for detecting rocky worlds and potentially investigating their atmospheres; and generally, the closer a planet is to its star, the larger its atmospheric signal. However, M dwarfs are more active than their Sun-like counterparts, and it is an open question as to whether low-mass, predominantly rocky exoplanets in tight orbits around M dwarfs can retain gaseous volatile envelopes at all. Analyses of the phase curve data of LHS 3844 b \citep{2019Natur.573...87K} and K2-141 b \citep{2022A&A...664A..79Z}, and eclipse data of GJ 1252 b \citep{2022ApJ...937L..17C}, TRAPPIST-1 b+c \citep{Greene2023,Zieba2023,2024ApJ...960...44T}, and GJ 367 b \citep{2024ApJ...961L..44Z} provide evidence for the absence of thick ($\gtrsim$10 bar) atmospheres around these worlds. This is typically interpreted to suggest the complete absence of a volatile envelope on these planets \citep[e.g.,][]{2022ARA&A..60..159W,2024arXiv240312617D}. Alternatively, initially incorporated volatiles may be locked up into the interior of these planets \citep{2021ApJ...914L...4L,2021ApJ...922L...4D,2024arXiv240116394L}. Though atmospheric detection in itself is currently a main focus of the field, the absence of a volatile envelope may enable attempts to assess the surface composition \citep{2012ApJ...752....7H}, geodynamics, and phase state of rocky exoplanets \citep{2011ApJ...735...72G,2021ApJ...908L..48M,2023A&A...678A..29M,2023arXiv230813614B}. JWST promises to make great strides for detecting and characterizing rocky planet atmospheres and even surfaces with transmission, emission, and phase curve observations at precisions and wavelengths that had been impossible to reach with prior intrumentation.

Among noteworthy planetary systems, the very low-mass M star TRAPPIST-1 deserves special attention \citep{2017Natur.542..456G,2021PSJ.....2....1A,2023arXiv231015895T}. This system hosts seven nearly Earth-sized planets, across the range of instellations where water in a terrestrial planet atmosphere would change its phase from gas to liquid to solid, providing a unique chance to explore the range of climates on small exoplanets orbiting a low-mass star in a distance from the Earth that makes this technically viable \citep{2020SSRv..216..100T,2020SSRv..216...98G,Greene2023,Zieba2023}. Given the faintness of M8 stars, we have so far been unable to survey a large sample of M8 stars for planets, making TRAPPIST-1 even more unique in the sample of known rocky world systems. The planet-to-star radius ratio of the TRAPPIST-1 planets is highly favorable for atmospheric investigation, meaning that for the forseeable future, these worlds are among the best characterizable Earth-sized exoplanets. The closest stellar system to the Solar System hosts a non-transiting planet, Proxima Centauri b \citep{2016Natur.536..437A}, consistent with an Earth-like mass, whose potential formation, climate and geophysical conditions have been theorized in great detail \citep{2016A&A...596A.111R,2016A&A...596A.112T,2021A&A...651A.103N}. Characterization of such close-by, non-transiting planetary systems will require either direct imaging or the detection of orbital phase variations in reflected starlight or thermal emission to explore their potential climates \citep{2015A&A...576A..59S,2016ApJ...832L..12K,2017A&A...599A..16L}.

\section{Formation} \label{sec:formation}

Large-scale surveys of extrasolar planets have revealed that Earth- to super-Earth-sized planets are among the most abundant planets in the solar neighborhood \citep{Howard2010, Mulders2018, Hsu2019, Kunimoto2020}. This is reproduced by planet formation models, which predict that small and low-mass planets should be dominating the exoplanet census \citep{Ida2005, Payne2007, Ida2008, Mordasini2009, Ida2020, Miguel2011disk,2021A&A...650A.152I,Emsenhuber2021b}. Small, rocky planets can form even in small disks around low-mass stars \citep{Miguel2020,Liu2020,Burn2021}. In particular, and after the discovery of the TRAPPIST-1 system \citep{2017Natur.542..456G}, major theoretical effort has gone into studying the formation of close-in rocky planets orbiting small, cool M stars \citep{Raymond2007a,Ogihara2009, Ronco2014, Alibert2017, Coleman2019, Schoonenberg2019, Miguel2020, Dash2020, Liu2020, Ormel2017, Burn2021, Zawadzki2021, Sanchez2022, Clement2022, Pan2022, Ogihara2022,2022ApJ...938L...3L,Schlecker2022}. Forming larger super-Earths around small stars is a challenge for planet formation theory, specifically to reproduce the occurrence and density contrast between the sub-Neptune and super-Earth classes, even though recently several models start to converge on compositional differences driven by disk migration and simultaneous envelope accretion \citep{2020A&A...643L...1V,2022ApJ...939L..19I,2024NatAs.tmp...33B}. In the astrophysical planet formation community (in contrast to, e.g., the geochemical/cosmochemical community) a large emphasis in the past has been on the dynamics of the planet formation process, i.e., how mass accretes onto protoplanets in a global view of planet formation. We refer to \citet{2020plas.book..287R}, \citet{2023ASPC..534..717D}, \citet{2023ASPC..534.1031K}, \citet{2023ASPC..534..501M}, and \citet{2023ASPC..534..863W} for different views on this issue, and focus here on highlighting a number of key developments that interface with our wider discussion on the internal and atmospheric evolution of rocky exoplanets.

\subsection{Planet formation theory}

Different theories have been discussed to explain the formation of rocky planets, which can be roughly divided in models classically inherited from the canonical theory of Solar System formation, in which planets accrete through mutual collisions of large $\gtrsim$km-sized planetesimals \citep{Kokubo1998,2020plas.book..287R,2023NatAs.tmp...10B} or models assuming that rocky planets can form through the direct accretion of mm- to cm-sized coagulated dust particles \citep[pebbles,][]{2017ASSL..445..197O,2017AREPS..45..359J}. The main difference between the two theories comes from the role of the gas disk drag force, which can play an important dynamical role. Pebbles that are marginally coupled with the gas experience a drag force when entering the Hill sphere of a growing protoplanet, which increases their effective cross-section for accretion. Planetesimals that are large enough to decouple from the gas do not feel this drag force and thus collisional accretion for planetesimal-sized objects is mainly ballistic \citep{2023ASPC..534..717D}. In both scenarios, the first step toward planetary growth is the formation of seed planetesimals in the disk via local gravitational collapse of dense dust clouds \citep{2016SSRv..205...41B}, which are condensed via disk-wide dust coagulation and drift processes \citep{2023ASPC..534..717D}. Hydrodynamical simulations, dynamical arguments, and crater counts on Solar System objects indicate that these bodies form with a typical size scale of $\sim$100 km \citep{2017ApJ...847L..12S,2017Sci...357.1026D,2018Icar..304...14T,2019NatAs...3..808N,2019Sci...363..955S,2020ApJ...901...54K}. After this initial rapid collapse, planetesimal growth proceeds via mutual collisions, because the gravitational influence of these birth planetesimals is insufficient to attract pebbles directly from the disk gas \citep{2016A&A...586A..66V,2019A&A...624A.114L}. Planetesimal growth in this phase (runaway growth) is rapid, depending mainly on the local planetesimal surface density and disk characteristics \citep{Kokubo2000,Kokubo2002,2010ApJ...714L..21K}.

\begin{figure}[tbh] 
 	\centering
 	\includegraphics[width=0.45\textwidth]{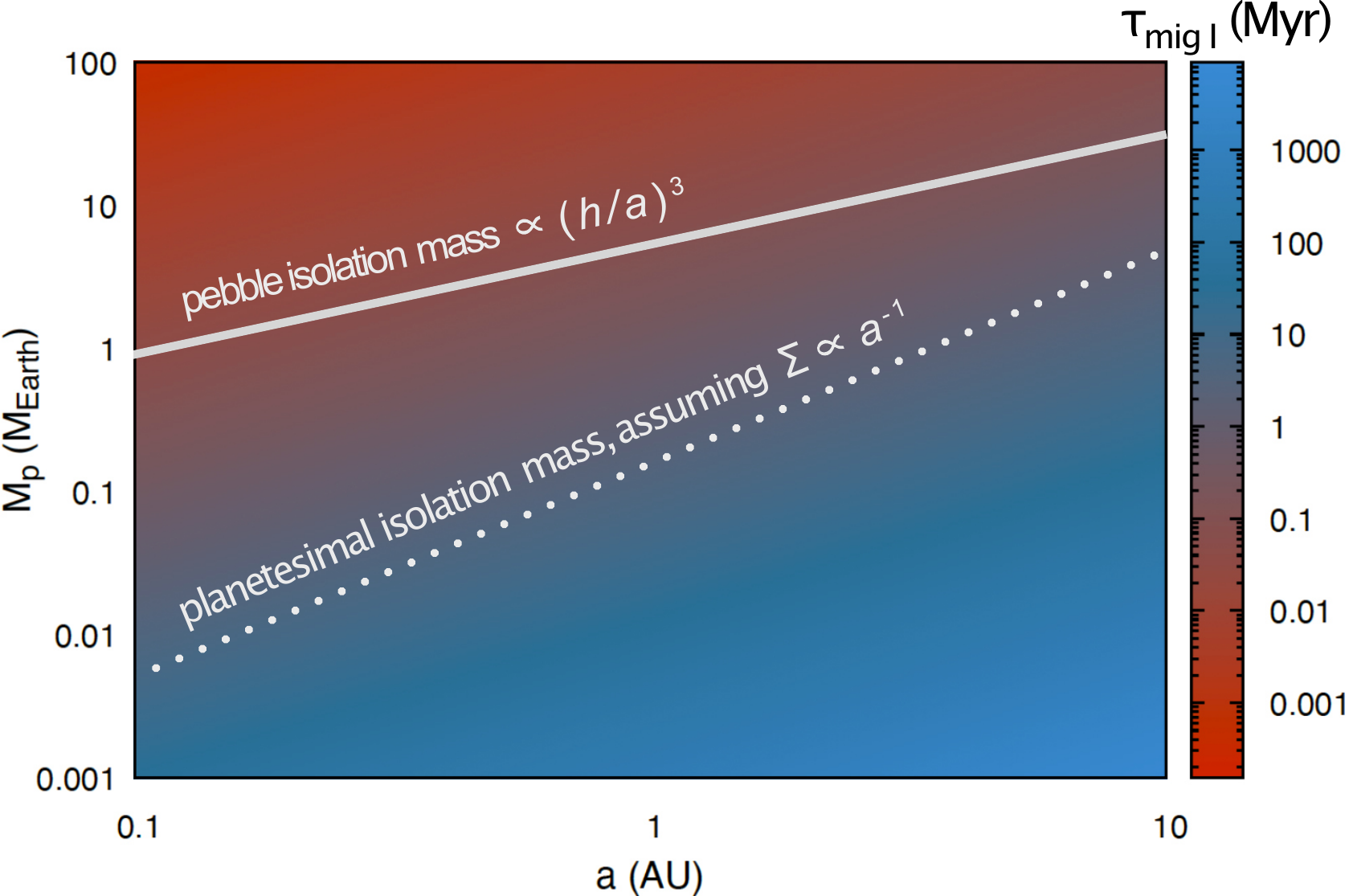}
 	\caption{\textsf{\textbf{Characteristic timescale for type I planetary migration (color scale) for different planet masses and semimajor axes.} In the calculations we adopt a planet orbiting a solar-type star and assume a typical disk mass of 0.01M$_{\odot}$. $\Sigma$ is the disk surface density, $a$ is the orbital distance, and $h$ is the disk scale height. Lines indicate the maximum mass that can be reached due to planetesimal accretion, also called isolation mass (dotted line), as dependent on the disk density profile, and due to pebble accretion (solid line), as dependent on the disk scale height. Pebble and planetesimal isolation curves from \citet{2023ASPC..534..863W}.}}
    \label{fig:migration}
\end{figure}

In collisional models (planetesimal accretion), this phase of accretion finishes when the biggest planetary embryos would reach the oligarchic growth phase, where a small set of larger planetary embryos, which are more or less equally spaced, would dominate the excited planetesimal population. This process typically operates on timescales that range between 10$^5$ to 10$^6$ years \citep{Wetherill1993, Weidenschilling1997, Kokubo1998, Kokubo2000, Kokubo2002}. After the disk gas is depleted, this is followed by a late-stage accretion phase dominated by giant embryo-planetesimal and embryo-embryo impacts that can last to about 10$^8$ years, until the final architecture of the system is reached \citep{Raymond2014}. In pebble-based models \citep{Levison2015, Johansen2021}, following the initial runaway growth phase, mass accretion is assumed to be dominated by pebbles \citep{Lambrechts2019,2019A&A...624A.109B,2021A&A...650A.132S}.  Once the disk is dissipated, this is followed by a phase of giant impacts that is similar to the classical planet formation scenario, with similar time-cales for their formation of the order of  10$^8$ years \citep{Levison2015, 2021A&A...650A.152I}, but potentially fewer left-over protoplanets and thus fewer giant impact events. On a disk level, pebble accretion is highly sensitive to ambient turbulence and the local pebble flux \citep{2018A&A...615A.138L,2018A&A...615A.178O,2023ASPC..534..863W}. With a constant flux, pebble accretion is thought to terminate when the gravitational influence of the planet starts to generate a wake-induced pressure bump outside its own orbit \citep{Morbidelli2012a,Lambrechts2014}. Recent work has additionally highlighted the role of accreting dust grains on envelope opacity, which may decelerate or inhibit growth at a specific mass scale \citep{2020ApJ...900...96A,2020ApJ...898..108R,2021A&A...647A.175O}. Rapid formation through pebble accretion may have substantial implications for the thermodynamics and chemical segregation of rocky planets \citep{2018SSRv..214..101N}, distinct to collisional accretion, which has recently been started to be explored in more detail \citep{2019PEPI..29406294O,2023E&PSL.62218418O,2022E&PSL.58717537O,Johansen2022b,2023A&A...676L...8V}.

Pebble and planetesimal accretion differ in defining the end of the accretion process as is shown in Figure \ref{fig:migration}. The runaway growth phase of planetesimal accretion stops when the planetary embryo has no more planetesimals to accrete in its area of gravitational influence or feeding zone ('planetesimal isolation mass'). On the other hand, pebble accretion continues until the planet induces a pressure bump outside its orbit ('pebble isolation mass') or when the outer disk has run out of pebbles. When terminated by the gas pressure bump, there is a change in the local pressure gradient of the gas and the upcoming pebbles get trapped in the pressure bump, stopping or at least decelerating the accretion process \citep{Lambrechts2014, Bitsch2018}. In the first case the final mass of the planet depends on the local planetesimal density and semi-major axis, while in the latter scenario it depends on global disk parameters such as the turbulent viscosity and disk aspect ratio. In Figure \ref{fig:migration}, this is scaled as proportional to the disk scale height as shown by \citep{Ormel2017}: pebble accretion typically allows for the formation of more massive embryos than pure planetesimal accretion, but is also more sensitive to migration within the gas disk.  

As the planetary embryos grow, so does their gravitational interaction with the massive gaseous disk, which causes a drift of the planet's orbit or migration \citep{2023ASPC..534..685P}. Once the planet has approximately Mars' mass \citep{Kley2012}, an imbalance in the corotation and Lindblad torques causes the planet to migrate. In this regime (type I migration), the embryos migrate with a speed that scales linearly with their mass and can be as fast as the accretion timescale itself. Figure \ref{fig:migration} shows the typical type I migration timescale for planets of different masses and semi-major axes for a star like the Sun, a disk of 0.01 M$_{\odot}$, and a disk density profile proportional to the inverse of the orbital distance to the host star, a similar profile as adopted for the calculation of the isolation mass. The figure shows that planets with masses smaller than Mars (approximately 0.1 Earth mass) have very long timescales of the order of Gyr, which decreases substantially for larger bodies. We also note that planets that orbit closer to their host star (smaller semi-major axis) migrate faster, with timescales of the order of 1000 years for the most massive planets. When the planets are massive enough to open up a gap in their orbit, torques on the disk are strong and the disk response becomes non-linear \citep[type II migration, ][]{Lin1986, Hasegawa2013, Kanagawa2018}. While this regime affects mostly giant planets, it can also affect smaller planets, especially if they form in disks with low viscosity. 

\subsection{The challenge of sub-Neptune formation}\label{sec:superEarths}

Sub-Neptunes are compatible with voluminous gaseous atmospheres mainly composed of gases such as hydrogen and helium \citep[][planets between the two yellow lines in Figure \ref{fig:M-R}]{Lopez2012, Rogers2015}, and they potentially share a common formation pathway with higher-density super-Earths and terrestrial planets. The H-dominated atmospheres of sub-Neptunes might be the remainder of an accreted gaseous atmosphere from the circumstellar disk gas, which fits nicely in the context of the core-accretion scenario for planet formation that we have discussed in this section. According to the theory, after the planets accreted a significant amount of solids, they start accumulating gas at a rate that is determined by the atmosphere’s cooling rate, until the gas accreted is approximately equal to the mass of solids accreted. After that, a runaway gas accretion begins that ends with the formation of a giant planet if there is enough gas left in the disk \citep{Mizuno1980, Wuchterl1993, Pollack1996, Ikoma2000, Rafikov2006, Lee2014, Piso2014, Lee2015}. Therefore, the total amount of primordial gas in a planet depends on the availability of sufficient solid material to enable runaway growth before much of the gas disk is dissipated \citep{Schlecker2021b}. Otherwise, the planet will remain a big 'core' with low-mass gaseous envelope similar to the observed population of low-density super-Earths or sub-Neptunes \citep{Bodenheimer2014}.

Because many low-density super-Earths are orbiting their stars at close-in orbits \citet{2015ApJ...807...45D}, many studies have focused on the formation of these planets assuming in-situ formation \citep{2012ApJ...753...66I, Bodenheimer2014, Inamdar2015, 2016ApJ...817...90L, 2016ApJ...825...29G}. Nevertheless, these studies show that gas accretion inside the snow line during the disk’s lifetime is unlikely. On the other hand, even when the planets form outside the snow line, the conditions to form these planets with such small gaseous envelopes require very specific conditions \citep{Terquem2007, Rogers2011, Liu2015, Ogihara2015, 2016ApJ...817...90L, Venturini2017}. However, most of the observed super-Earths and sub-Neptunes are old planets, and evolution played a big role in shaping the mass and radius observed today \citep{2021A&A...650A.152I,2022ApJ...939L..19I,2023ApJ...947L..19R}. It is expected that many of the planets we observed today were born with larger gaseous envelopes and are the result of extensive mass loss during their lifetimes \citep{2012ApJ...753...66I, Lopez2012, Owen2012, 2016ApJ...825...29G, Owen2016, Misener2021, Modirrousta-Galian2023}.  

\subsection{Planetesimal compositional evolution during accretion}\label{sec:planetesimal_evolution}

The basic compositional architecture of the Solar System -- dry and small planets on the inside, and large and wet/less dense planets on the outside -- has given rise to a wide range of dynamic arguments for explaining this structure. For example, specific emphasis has been traditionally given to the water snowline as a spatial divider between drier planetary materials inside, and wetter materials outside \citep{1981PThPS..70...35H}. This system-focused viewing angle has been exported directly onto exoplanetary systems \citep{Bean2021}. The basic compositional principle of the snowline explains a wealth of dynamical and architectural evidence, however, exoplanet occurrence rates have made clear that this picture must be much more dynamic than estimated before. For example, the above mentioned density-size dichotomy between super-Earths and sub-Neptunes can be explained either by clustering of planets across two major compositional classes: dry, rocky planets and wet, icy planets \citep{2019PNAS..116.9723Z,2020A&A...643L...1V,2022Sci...377.1211L,2022ApJ...939L..19I,2023AJ....165..167C}, or alternatively through differences in hydrogen content, as outlined before.

From a geophysical perspective this is interesting, because the left-over asteroidal and meteoritic populations of the Solar System clearly indicate a substantial compositional evolution during planet formation. Planetesimals in the Solar System did not accrete with a chemical composition inherited from the disk and stayed like this. Rather, substantial evidence in all known bodies points to a high degree of chemical overprinting (sometimes termed 'reset' in the astronomical community) by geophysical evolution \citep{2018SSRv..214...36A,2023ASPC..534..907L}. This included melting of water ice, aqueous alteration inside of planetesimals, degassing of volatile elements, and core-mantle separation, in general substantial chemical segregation driven by thermal evolution. Sources of thermal energy in early-formed planetesimals (cf. Section \ref{sec:atmospheres1}) are dominated by (i) the decay of short-lived radioactive isotopes such as $^{26}$Al and (ii) the potential energy release from accretion itself \citep{2012AREPS..40..113E,2021ChEG...81l5735C}. The specifics of these processes and Solar System-related timescales are out of the scope of this review, but we want to point out that these processes are either widespread or universal across and between planetary systems. In particular the prevalence of short-lived radioactive isotopes is subject to continuous debate in the star formation community, as it is generated by time-sensitive processes such as supernovae, Wolf-Rayet winds, and AGB stars in star-forming environments \citep{2018PrPNP.102....1L,2022EPJP..137.1071R,2022PrPNP.12703983D}. 

Specifically important for the exoplanet debate is that the Solar System's compositional dry-wet dichotomy can be explained mainly by $^{26}$Al-driven heating of planetesimals \citep{1993Sci...259..653G,2021Sci...371..365L}, without relying on the disk gradient induced solely by a static snowline. Recent experimental evidence hardens the case that $^{26}$Al is a significant source of volatile degassing (devolatilization) in planet-forming planetesimals \citep{2017GeCoA.211..115S,2023GeCoA.340..141P,2023Natur.615..854N,2024NatAs.tmp...13G}, and has been suggested to operate on a system-wide level, introducing a qualitative compositional deviation between $^{26}$Al-rich and $^{26}$Al-poor exoplanet systems \citep{2019NatAs...3..307L}. Furthermore, evidence from polluted white dwarfs indicates planetesimal compositions across evolved planetary systems \citep{2019Sci...366..356D,2023NatAs...7...39B} that are quantitatively consistent with estimates of short-lived radioisotope distributions across planetary systems from studies of star-forming regions \citep{2014MNRAS.437..946P,2023MNRAS.521.4838P,2016MNRAS.462.3979L,2020A&A...644L...1R,2021NatAs...5.1009F} and planetesimal degassing models with varying short-lived radioisotope content \citep{2024MNRAS.528.6619E}. Its influence on the primordial composition of planets has not yet been synchronized with stochastic and longer-term effects on a population level, such as impact stripping \citep{2016ApJ...817L..13I,2024PSJ.....5...28L}, atmospheric escape \citep{2020SSRv..216..129O}, or the runaway greenhouse effect \citep{2024PSJ.....5....3S}, but with ever-increasing resolution on lower-mass exoplanets, we anticipate increasing quantitative tests of this mechanism to generate compositional differences across planetary systems, when trying to interpret the redox states of atmosphere-stripped short-period exoplanets, detailed compositional abundances of white dwarf pollutants \citep{2022MNRAS.515..395C,2020MNRAS.492.2683B}, and the atmospheric diversity of low-mass exoplanets.

\section{Chemical Differentiation of Rocky Planets} \label{sec:atmospheres1}

\subsection{Atmosphere Formation}

The formation of short- and long-lived planetary atmospheres is tightly coupled to the main energy sources on and inside rocky exoplanets during accretion, leading to chemical (re-)equilibration or segregation between the structural reservoirs of (metal) core, mantle, and gas envelope \citep{2023ASPC..534..907L}. The primary heat source during accretion is the release of gravitational potential energy and the decay of short-lived ($\sim$Myr) radioactive isotopes \citep[][and references therein]{2021ChEG...81l5735C}. Radiative cooling of atmosphere-less planets would be governed by blackbody radiation, which is rapid, leading to important questions surrounding the phase state (liquid, solid, or gas phase) of the mantle of rocky planets during formation \citep{1990orea.book...69M}. However, multiple recent independent threads of investigation, from early Solar System geochronology \citep{2011Natur.473..489D,2014Sci...344.1150K,2021NatGe..14..369G}, the refractory mass budget in disks \citep{2023ASPC..534..863W,2023ASPC..534..717D}, to the abundance of close-in super-Earths and sub-Neptunes \citep{2020SSRv..216..129O,2023ASPC..534..863W}, provide mutually consistent evidence that the formation of rocky protoplanets of Mars-sized embryos and larger is comparable to or shorter than the timescale of the dispersal of protoplanetary disks \citep{2020ARA&A..58..483A}, and that atmosphere-forming volatile elements are incorporated on a similar timescale during the main phase of planetary accretion from volatile ices \citep[through pebbles and planetesimals,][]{2022Natur.611..245B,2023GeCoA.340..141P} and disk gas \citep{2019Natur.565...78W,2022Sci...377..320P,2023Natur.616..306Y}. Rapid formation increases the maximal temperatures reached because it decreases the cooling timescale between distinct accretion events and increases the peak heating efficiency of short-lived radionuclides \citep{2016Icar..274..350L,2022JGRE..12707020S}. Radiatively-active atmospheric molecules like \ce{H2O} and \ce{H2} strongly increase the cooling timescale of protoplanets through the infrared opacity of the primary gas envelope of protoplanets. Hence, rocky planets are largely molten \citep{2018SSRv..214...76I,2018ApJ...854...21C} or even vaporized \citep{2017Natur.549..511H,2022Sci...377.1529F} during accretion. This realization has important consequences for all stages of planetary evolution, as it facilitates global redistribution of chemical elements between core, mantle, and atmosphere, and charts the thermodynamic pathways of long-term climate and interior geophysics. 

\paragraph{Formation and loss of primary atmospheres}
One of the currently most debated processes potentially shaping the low-mass exoplanet census is the direct acquisition and escape of primordial H-He gas from the protoplanetary disk. We provide here an overview of some important developments in this direction, and refer to \citet{2020JGRA..12527639G} and \citet{2020SSRv..216..129O} for specialized reviews on the physics of atmospheric escape, and \citet{2023IAUS..370...56D} for a review on the observational perspective. During the earliest stages of exoplanet formation, the amount and incorporation of H-He gas from the disk onto protoplanets is sensitively coupled to the heating and cooling of the planetary structure, because for gas capture onto the planet, the kinetic energy of the gas molecules must be overcome by gravitational binding energy. Therefore, at a specific time and protoplanet mass, solid and gas capture rate are expected to be anti-correlated \citep{1979E&PSL..43...22H,2016ApJ...825...29G}, suggesting a threshold mass of $\approx$1 $M_\mathrm{Earth}$ for the capture of a $\gtrsim 1$ wt\% He-He gas envelope. However, recent numerical simulations of gas accretion onto super-Earths embedded in the disk suggest cyclic flows into and out of the proto-envelope, which complicates this simplified picture \citep{2015MNRAS.447.3512O,2015MNRAS.446.1026O,2017A&A...606A.146L,2021A&A...646L..11M,2022A&A...661A.142M}. Once the protoplanetary disk gas is dispersed, protoplanets start to lose their primary envelopes through stellar irradiation \citep{2018A&ARv..26....2L}, residual internal energy from formation \citep{2012ApJ...753...66I,2018MNRAS.476..759G,2019MNRAS.487...24G}, and mutual giant impacts \citep{2015ApJ...812..164L,2019MNRAS.485.4454B,2020ApJ...901L..31K,2020MNRAS.496.1166D,2022ApJ...937...39C,2024PSJ.....5...28L}. Analogous but reversed to nebular capture, atmospheric escape is facilitated when the kinetic energy of the envelope gas molecules overcomes the planetary gravitational potential. 

Sensitively important for short-period exoplanets \citep{2019AREPS..47...67O}, this escape is facilitated by increased temperatures in the exosphere of the planet, which is strongly heated by stellar X-ray and extreme ultraviolet radiation (XUV) because typical atmospheric molecules are highly opaque in XUV wavelengths \citep{1982RvGSP..20..280Z}. The heating of the uppermost atmospheric levels can power very efficient hydrodynamic escape during the early phases of planetary evolution, which had been predicted to lead to a divergence in primary H-He gas envelopes in the sub-Neptune regime \citep{2012ApJ...753...66I,2013ApJ...776....2L,2013ApJ...775..105O,2014ApJ...795...65J}, in close consistency with the statistical detection of the radius valley in Kepler planets \citep{2017AJ....154..109F,2018MNRAS.479.4786V}. This regime of hydrodynamic escape is just one out of many different escape mechanisms, but is regarded as the most important one shaping the early evolution of planetary atmospheres, particularly affecting short-period exoplanets and thus, shaping the exoplanet population seen today. However, theoretical analyses of this issue typically only consider bulk outflow from the gaseous envelope, which is a limiting assumption for constraining the residual, secondary atmospheres of rocky exoplanets across the entire range of masses, compositions, and irradiations, because molecules of different masses can diffusively separate in the outflow, which tends to decrease the amount of heavier species to escape. In the case of a vigorous and well-mixed outflow, however, heavier species, such as \ce{H2O}, can be dragged along with the hydrogen gas \citep{1987Icar...69..532H,1990Icar...84..502Z,2018Icar..307..327O,2018AJ....155..195W}. Importantly, by interaction with photolytic destruction of molecules in the upper atmospheres, this can lead to the buildup of residual atmospheres enriched in heavier species, such as oxygen \citep{2014ApJ...785L..20W,2015AsBio..15..119L,2016ApJ...829...63S} or noble gases \citep{2015ApJ...807....8H,2019GeCoA.244...56Z,2020ApJ...896...48M,2023NatAs...7...57M}. Differential escape could, in principle, enable the identification of isotope fractionations on even low-mass exoplanets \citep{2019AJ....158...26L,2019A&A...622A.139M,2024arXiv240210690C}. The desiccation of rocky exoplanets is regarded to be particularly pervasive for M star planets, which, as outlined above, are the most promising targets for transit detection missions. M star planets undergo a very luminous pre-main sequence phase that lasts up to several hundred million years \citep{1998A&A...337..403B,2015A&A...577A..42B,2000A&AS..141..371G}, which strongly enhances atmospheric escape in these systems \citep{2014ApJ...797L..25R,2015NatGe...8..177T,2017MNRAS.464.3728B}. Therefore, a major question of current exoplanet science is whether small exoplanets around M stars can host atmospheres. Estimated loss efficiencies vary, but typically range between a few Earth oceans to up to a few tens of Earth oceans for the most highly irradiated exoplanets \citep{2015AsBio..15..119L,2018AJ....155..195W,2018A&A...619A...1B,2021AsBio..21.1325B}. These upper range estimates are 1--2 orders of magnitude below the uncertainties for the delivery to and incorporation of atmospheric volatiles into M star planets \citep{2019PNAS..116.9723Z,2022ApJ...938L...3L,2021ApJ...922L...4D,2024NatAs.tmp...33B}, which is why the total content of atmospheric volatiles in M star exoplanets is a major uncertainty of present planetary evolution models.

Many exoplanets have confirmed observations of atmospheric escape occurring in their atmospheres \citep{2023IAUS..370...56D}. Traditional observations of atmospheric escape are made through transmission spectroscopy in ultraviolet wavelengths, particularly using the Lyman-$\alpha$ line at 1215.67 \AA, which probes the escape of hydrogen. Some examples of early detections of atmospheric escape from giant exoplanets are \citet[][]{2003Natur.422..143V} and \citet{2008A&A...483..933E}. Observations using Balmer and Paschen series in the H lines have also been made in some Hot Jupiters \citep[e.g.,][]{2020A&A...638A..87W, 2022A&A...666L...1S}. These observations take advantage of the fact that a planet with escaping gases would have a tail that can be probed through a decrease in the stellar flux in the blue wing during the planet's transit. Because these observations require measurements in the UV, they can only be made using the Hubble Space Telescope, the only current telescope with instruments able to observe in these wavelengths at high enough resolution. While most observations of atmospheric escape using H lines have been made in Hot Jupiters, observations have also been made in Neptune-like planets \citep[e.g.,][]{2015Natur.522..459E, 2018A&A...620A.147B}, and even a few in sub-Neptunes \citep{2020A&A...634L...4D, 2022AJ....163...68Z}. For the range of radii that we are focusing on in this review, only non-detections of H escape have been reported in the literature for a number of exoplanets. These include the TRAPPIST-1 system \citep{2017A&A...599L...3B}, the Kepler-444 system \citep{2017A&A...602A.106B}, HD 97658 b \citep{2017A&A...597A..26B}, 55 Cnc e \citep{2018A&A...615A.117B}, and GJ 1132 b \citep{2019AJ....158...50W}. Other, heavier species also escape from exoplanetary atmospheres and can be detected with similar methods. In particular, detections of the escape of metals have been made in a few Neptune-like planets \citep{2021ApJ...907L..36G, 2022NatAs...6..141B,2024arXiv240305614V} but have not been reported in smaller, rocky planets. One disadvantage of observing atmospheric escape in Ly-$\alpha$ is that the signals detected can be strongly affected by interstellar absorption and emission from the stars, and can only be carried out using the Hubble Space Telescope. However, helium escape can also be detected using the absorption line of a meta-stable state of helium at 10830 \AA. A key advantage of this method is that measurements can be done from the ground using high-resolution spectroscopy \citep{2000ApJ...537..916S, 2018ApJ...855L..11O}. Some of the sub-Neptune-like planets in the range explored in this review with He escape detected include TOI-1430 b and TOI-1683 b \citep{2023AJ....165...62Z}, and no He escape has been reported in 55 Cnc e \citep{2021AJ....161..181Z}, HD 97658 b \citep{2020AJ....160..258K}, the TRAPPIST-1 system \citep{2021AJ....162...82K}, GJ 9827 b+d \citep{2021AJ....161..136C}, and the HD 63433 system \citep{2022AJ....163...68Z}. 

Sensitively important for the analogy between early Earth and low-mass exoplanets, which we aim to to draw in this review, is a recent debate on whether the early Earth or its precedeing protoplanets experienced a period of XUV-driven bulk hydrodynamic escape \citep[cf.][]{2018A&ARv..26....2L,2019GeCoA.244...56Z}. Whether XUV-driven escape was important for the Hadean Earth quantitatively changes the outlook for the timescales of atmosphere-interior equilibration and the analogies we can draw from M-dwarf and short-period exoplanets.

\paragraph{Secondary atmospheres}
Secondary atmospheres have typically been thought of as being volatile envelopes replenished through long-term magmatism (volcanic emission) from the interior of rocky planets once the planet has shed off its primary, disk-derived envelope \citep[for reviews see, e.g.,][]{2014AREPS..42..151C,2020plas.book....3Z,2021SSRv..217...22G}. The clear distinction between primary and secondary atmospheres had been supported by the prevailing view that the atmospheric volatiles of the terrestrial planets were delivered after the main stage of planetary accretion and that the protoplanets only acquired a very thin -- if any -- primary envelope from disk gas \citep[see the discussions in][]{2009Natur.461.1227A,2013GeCoA.105..146H,2015GMS...212...71M}. In such a scenario, where mass growth and volatile acquisition are neatly divided in time, the distinction into two primary types of atmospheres yields a useful framework with which these can be studied. However, the realisation that planets grow rapidly on the timescale of the disk itself (Section \ref{sec:formation}), and that late accretion stages in the Solar System were predominantly dry and volatile-poor \citep{2016AmMin.101..540H,2017Natur.541..521D,2017Natur.541..525F} has also changed the general outlook for how the composition of secondary atmospheres on rocky exoplanets is established \citep{2023ASPC..534..907L,2022ARA&A..60..159W}: throughout accretion and during loss of the primary envelope the planetary interior (core+mantle) and atmosphere are in rapid chemical and physical exchange with each other, which means their compositional evolution can only be understood as a coupled, emergent phenomenon. Therefore, while mixing timescales during accretion can  be influenced by detailed physics \citep[e.g.,][]{2018E&PSL.487..117N,2020E&PSL.53015885L}, the general trend during formation is rapid mixing and chemical exchange between core, mantle, and atmosphere -- a process that is currently ignored when interpreting exoplanet compositions, and not included in typical astrophysical planet formation models.

Before we get to the two-way interaction between interior and atmosphere, we need to establish a more general classification describing the types of atmospheres that can be distinguished. This can be established through quantification of the planetary redox state \citep[e.g.,][]{2021SSRv..217...22G,2022ARA&A..60..159W,2023ASPC..534..907L}, which can be thought of as a measure of the local availability of atoms that donate or accept electrons, which shapes the type of chemical compounds that form in chemical reservoirs, such as the mantle or atmosphere. In other words, redox state can be used to quantitatively express the local or global chemical equilibrium state of planets. When discussing the redox state in the following, we mainly focus on the potentially observable redox state of a planet through transmission or emission spectroscopy, which is typically the interface between atmosphere and interior -- most likely the upper mantle or magma ocean surface, which governs the exchange of volatile elements between planetary interior and atmosphere. From an atmospheric point of view, hydrogen-rich atmospheres are 'reduced', which means they have an overabundance of valence electrons in chemical compounds, while oxygen-rich are 'oxidized'--poor of valence electrons because oxygen accepts electrons in chemical compounds. An example for a reduced atmosphere is Titan, while the modern Earth is an example of an oxidized atmosphere. In known exoplanet systems this distinction is not yet clear--naively one may expect the atmospheres of sub-Neptunes to be highly reduced, which would be the case if they indeed consist of a rocky core plus a disk-derived gaseous envelope. Recent compositional information from the best-studied sub-Neptunes, such as K2-18 b \citep{2019ApJ...887L..14B,2019NatAs...3.1086T,2023ApJ...956L..13M} and TOI-270 d \citep{2024arXiv240303325B,2024A&A...683L...2H} can be interpreted either in a water world ('Hycean')  \citep{2021ApJ...914...38Y,2021ApJ...921L...8H,2021ApJ...922L..27T} or magma ocean scenario \citep{2024ApJ...962L...8S,2024arXiv240214072R}. This uncertainty questions the primary formation mechanism of mid-sized planets, including super-Earths, and highlights the potential interactions between the atmosphere and deeper planetary interior. Before we discuss this in more detail, however, we need to have a framework within which to discuss the distinctions and relations between chemical classes of atmospheres.

\begin{figure*}[tbh]
 	\centering
        \includegraphics[width=0.99\textwidth]{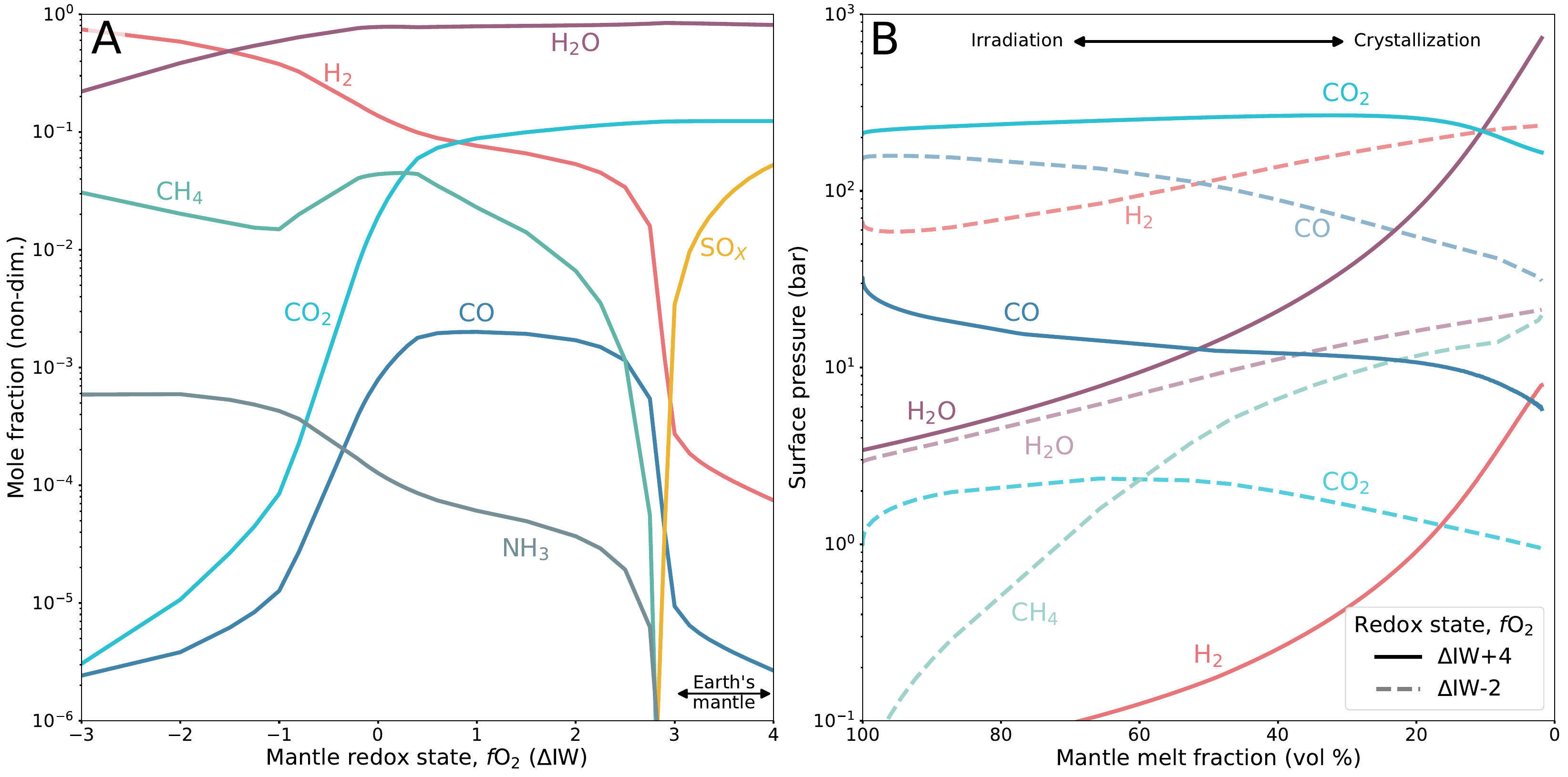}
 	\caption{\textsf{\textbf{Simulation of outgassing of atmospheric volatiles from a solid planetary mantle with varying redox state (A), and from a crystallizing mantle with varying melt fraction and redox state (B).} Generally, degassing is highly sensitive to mantle composition (expressed by the redox state through oxygen fugacity, $f$O$_2$). Reduced planets tend to degas H-rich species, such as \ce{H2}, \ce{CH4}, or \ce{NH3}, while oxidised planets degas O-rich species, such as \ce{H2O}, \ce{CO2}, or SO$_\mathrm{X}$-compounds. Numerical data from \citet{2022JGRE..12707123L} and \citet{2022PSJ.....3...93B}. Mantle volatile contents in (A) are set to match the \ce{H2O} and \ce{CO2} contents of the mantle after solidification in \citet{2008E&PSL.271..181E}, while S and N contents are scaled to mid-ocean ridge basalts, the degassing temperature is set to 800 K. In (B) the total water abundance in the mantle plus atmosphere is 3 Earth oceans and the H/C ratio is unity.}}
    \label{fig:volatileoutgassing}
\end{figure*}

Figure \ref{fig:volatileoutgassing} demonstrates the transition of the abundance of major H-C-N-S atmospheric compounds from reduced to oxidized atmospheres as a function of oxygen fugacity ($f$\ce{O2}), which -- for simplicity -- can be thought of as partial pressure of oxygen in a chemical assemblage if oxygen were an ideal gas. Oxygen fugacity is the scale on which redox state is typically expressed because oxygen is the dominant element in terrestrial rocks. A comprehensive recent summary of the importance of redox state in the context of Solar System science (with a focus on Venus) is given in \citet{2023SSRv..219...51S}. The oxygen fugacity is shown as deviation relative to the iron-wüstite buffer ($\Delta$IW = 0), which is the equilibrium state between iron (Fe) and wüstite (FeO) at a fixed pressure and temperature (\ce{Fe + O  <=> FeO}). Figure \ref{fig:volatileoutgassing}A illustrates the case of a solid rocky planet with a surface temperature of 800 K that is magmatically generated \citep[e.g., through volcanic degassing,][]{2022JGRE..12707123L}, and where the degassed volatiles are in chemical equilibrium at the surface. From left to right the atmospheric composition transitions from highly reduced to mildly reduced to oxidized. Highly reduced atmospheres ($f$O$_2\lesssim$ $\Delta$IW-0.5) are dominated by \ce{H2}, \ce{H2O}, and \ce{CH4}, with traces of \ce{NH3}; mildly reducing atmospheres ($\Delta$IW-0.5 $\lesssim$ $f$O$_2\lesssim$ $\Delta$IW-3) show increasing abundances of \ce{CO2} and \ce{CO} (O-rich compounds) and declining \ce{H2}, \ce{CH4}, and \ce{NH3} (H-rich compounds); while oxidized atmospheres ($f$O$_2\gtrsim$ $\Delta$IW+3) are devoid of \ce{CH4} and \ce{NH3}, with only minor amounts of \ce{H2} and \ce{CO}, but \ce{CO2} and \ce{SO}$_\mathrm{X}$ form dominant secondary species in the atmosphere. The exact values of these figures vary between different estimates and are dependent on initial conditions and assumptions, such as the degassing temperature and pressure, but broad consensus on the qualitative chemical classes of outgassing atmospheres have been reached in the past few years \citep{2014E&PSL.403..307G,2011ApJ...729....6S,2017ApJ...843..120S,2020NatSR..1010907O,2021PEPI..32006788G,2021SSRv..217...22G,2023A&A...675A.122B}. Within this classification the Earth falls into the oxidized regime: magmatic degassing on Earth is dominated by \ce{H2O}, \ce{CO2}, and \ce{SO2}. These general classes of atmospheres could in principle be distinguished from one another, as the primary volatiles should produce different transmission and emission signals, that can fall potentially within the range of current and next-generation astronomical facilities (Section \ref{sec:atmospheres2}).

However, these general compositional classes change substantially if the chemical equilibration between mantle and atmosphere is enhanced, such as in the case of fully or partially molten mantles, as demonstrated in Figure \ref{fig:volatileoutgassing}B. In this figure the surface partial pressure of individual molecules in the atmosphere is plotted against the global mantle melt fraction, meaning how much volume of the mantle is liquid magma, rather than a solid. Importantly, in these simulations, the entire magma reservoir interacts chemically with the atmosphere, which means that the chemical potential for atmospheric volatiles to partition (mix) into the liquid becomes important. In the figure there are two redox states plotted each for melt fractions of the mantle between 100 vol\% to 0 vol\% from left to right. For $f$O$_2 = \Delta$IW+4, the atmosphere of the planet is dominated by \ce{H2O} and \ce{CO2}, with minor traces of \ce{H2} and \ce{CO}. For high melt fractions, however, the \ce{H2O} and \ce{H2} abundances decrease by about two orders of magnitude, which makes \ce{CO2} and \ce{CO} the primary atmospheric constituents at very high melt fraction, in the magma ocean regime. At lower redox state, $f$O$_2 = \Delta$IW-2, the nominally dominating atmospheric constituents are \ce{H2} and \ce{CO}, while the behaviour of \ce{H2O}, \ce{CH4}, and \ce{CO2} is sensitively dependent on melt fraction. We will discuss the partition behaviour of individual compounds in Section \ref{sec:magmaocean}, which explains the reason for the order-of-magnitude transitions in Figure \ref{fig:volatileoutgassing}B. For the atmospheric composition in the case of exoplanets, however, it is important to realise that atmospheres are not isolated systems: the major atmospheric compounds are strong functions of both composition (as here expressed by $f$\ce{O2}) and planetary phase state (melt fraction), which influence each other. This inter-dependency leads to non-linear evolution of planetary atmospheres: the amount of atmospheric greenhouse gases sets the heat loss of the planet to space, the interior phase state (and geodynamic regime, Section \ref{sec:interiors}) controls the amount of volatiles in the atmosphere.

\begin{figure*}[tbh]
 	\centering
        \includegraphics[width=0.99\textwidth]{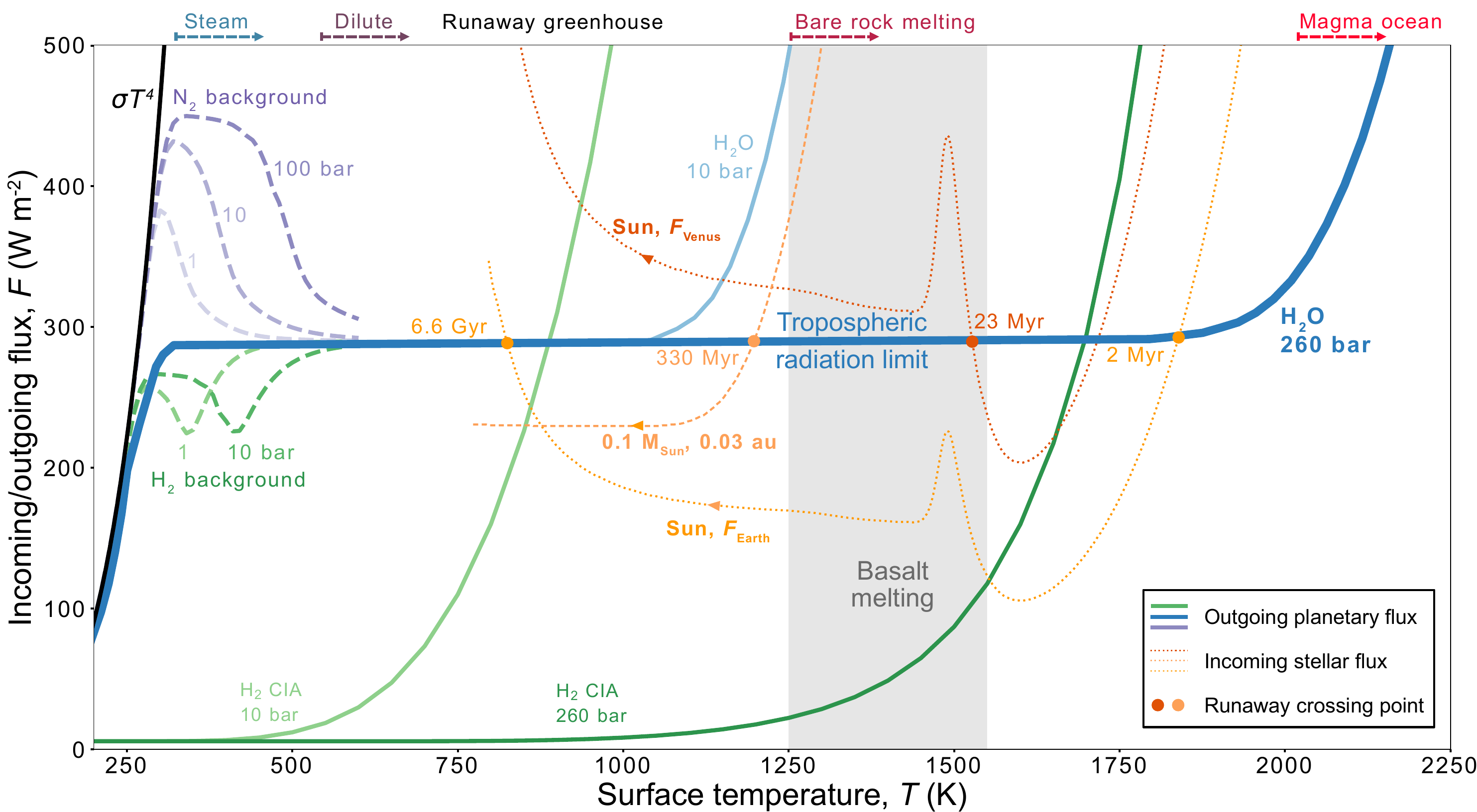}
 	\caption{\textsf{\textbf{Competition of outgoing planetary and incoming stellar radiation for various types of atmospheres as a function of surface temperature.} Green, blue, and purple lines indicate the maximum outgoing longwave flux of cloudless atmospheres. The solid black line (top left) illustrates blackbody radiation. The solid blue lines indicate a pure steam atmosphere with 260 bar (equivalent $\sim$ 1 vaporised Earth ocean) or 10 bar surface pressure \citep{2021ApJ...919..130B}; the solid green lines show the effect of collision-induced absorption in pure H$_2$ atmospheres with 260 bar or 10 bar \citep{2021JGRE..12606711L}, respectively. The dashed purple and green curves assume an H$_2$O-satured atmosphere with 1, 10, 100 bar N$_2$ or 1, 10 bar H$_2$ non-absorbing background gas \citep{2019ApJ...881..120K}. The yellow and orange dotted lines indicate the temporal evolution of the incoming stellar flux for a planet around a Sun-like star on an orbit like Venus (orange, top) or Earth (yellow, bottom); and a planet around an M dwarf star with $0.1$ Solar masses at 0.03 au. Stellar evolution data from \citet{1998A&A...337..403B,2015A&A...577A..42B,2000A&AS..141..371G}. In this idealised view, planets cool when their irradiation (yellow) is below the outgoing planetary flux (blue, green, purple), and heat up when the situation is reversed. For a fixed atmospheric composition and pressure, the equilibrium point is where the incoming and outgoing flux balance each other. The onset of melting for basaltic rocks of various water content at 1 bar is illustrated with a grey-shaded background. The boundaries for various atmospheric classes indicated at the top of the figure illustrate the climate categories set in Figure \ref{fig:M-R}.}}
    \label{fig:radiationlimits}
\end{figure*}

Figure \ref{fig:radiationlimits} illustrates the flux balance of rocky planets with different primary atmospheric volatiles by comparing the in- and outgoing energy flux from planet and star, respectively, as a function of surface temperature. The thick blue line in the figure shows the classical thermal radiation limit for a pure steam atmosphere case \citep{1929QJRMS..55Q..73.,1988JAtS...45.3081A,1988Icar...74..472K,1992JAtS...49.2256N}. This asymptotic limit underpins the 'habitable zone' concept, defining the inner orbital transition between hot runaway greenhouse climates and planets where liquid water is possible at the surface \citep{1993Icar..101..108K,2013ApJ...765..131K}. Why the emphasis on pure steam? Because water vapour is a strong infrared (IR) absorber and the only atmospheric molecule that is both stable in its liquid and vapour form in the terrestrial environment. In outgassing simulations and experiments of bulk silicate Earth materials \citep{2014ApJ...784...27L} or carbonaceous meteorites \citep{2021NatAs...5..575T} water is the dominant molecule \citep[even though newer experiments suggest a muted role for \ce{H2O}; see Fig. \ref{fig:volatileoutgassing} and][]{2023E&PSL.60117894S}. Historically, the steam atmosphere scenario also has gained attention because of the previously discussed thinking that secondary atmospheres are derived from secondary volatiles that are delivered after the main phase of planetary accretion. On a physical basis, the thermal limit (runaway greenhouse threshold = inner edge of the liquid water habitable zone) appears when the atmosphere becomes sufficiently hot and deep that the tropospheric temperature structure aligns with the water vapour saturation curve \citep[for an in-depth discussion, see, e.g.,][]{2013NatGe...6..661G,2013Natur.504..268L}. In this regime, changes to the total abundance of water do not affect the total outgoing radiation anymore, but only the near-surface temperature. For example, a 260 bar pure water atmosphere (condensed + vapour; 260 bar is approximately equivalent to one Earth ocean, i.e., the total liquid water at the surface of the Earth) reaches the thermal limit at about 300 K. At this point the amount of water in the atmosphere is enough for the radiation to be dominated by the opacity of water vapour in the troposphere. Increasing temperature then does not change this limit, which is approximately 280 W m$^2$ \citep{2013NatGe...6..661G,2019ApJ...875...31K,2021JGRE..12606711L}. Cooling in this regime is dominated by the two water vapour opacity windows in the short- to mid-infrared, between 3.5--4.5 and 8--20 $\mu$m, respectively. Only at temperatures of $\gtrsim$1900 K does the blue thick line in Figure \ref{fig:radiationlimits} increase again. This is when the surface becomes sufficiently hot such that the atmospheric temperature structure is not dominated by the latent heat of water condensation and evaporation, and instead aligns with a dry/supercritical adiabat. Different amounts of total water vapour in the atmosphere change this inflection point, for example a 10 bar atmosphere would lead to a change in thermal radiation at about 1200 K. Beyond this threshold, steam atmospheres start to strongly radiate in visible and ultraviolet wavelengths \citep{2021ApJ...919..130B}, which makes a combination of IR and UV monitoring a suitable combination to potentially discern the cooling paths and volatile inventory of steam runaway greenhouse climates. 

While water is abundant in the galaxy \citep{2021PhR...893....1O} and water vapour is by far the most efficient greenhouse gas present in terrestrial planetary atmospheres, different climate scenarios impact these predictions. In Figure \ref{fig:radiationlimits} we plot additional lines, that show the influence of background gases or alternating primary gases in the atmosphere. The purple lines between surface temperatures of 250 to 500 K illustrate the change in behaviour if \ce{N2} is a dominant background gas in the atmosphere. \ce{N2} itself is not radiatively active, and thus changes the outgoing radiation by its compositional effect on the temperature structure \citep{2019ApJ...881..120K,2020MNRAS.494..259R,2022A&A...658A..40C}. A similar but inverted scenario appears when \ce{H2} is a dominant background gas but does not act as an absorber itself (green dashed lines). In both cases, the onset of the thermal limit is shifted to higher surface temperatures, which can (in the \ce{N2} case) lead to colder climates being stable up to $\approx$ 500 K. This delayed onset of the thermal limit defines the 'dilute runaway' climate category in Figure \ref{fig:M-R}.

If instead of water \ce{H2} is the dominant atmospheric absorber (e.g., in chemically reduced super-Earths or sub-Neptunes), these thermal limits will be different. The green solid lines illustrate the idealised case of a pure \ce{H2} atmosphere with either 10 or 260 bar. \ce{H2} at high pressures is also an efficient greenhouse gase because collisions between individual molecules lead to excited molecular states that strongly interact with infrared photons. These \ce{H2} cases are particularly relevant for young super-Earths. As many super-Earths may undergo an initial phase of primary envelope loss (being transformed from sub-Neptunes to super-Earths by atmospheric loss), their initial atmospheric abundances of \ce{H2} will set the cooling of the planet. As the green solid lines demonstrate, radiative cooling in thick \ce{H2} atmospheres is strongly diminished. However, because of measurement uncertainties, the total \ce{H2} in sub-Neptunes is poorly constrained, which motivates proposals to measure the envelope thickness and mantle phase state via atmospheric composition in sub-Neptunes \citep{2021ApJ...914...38Y,2021ApJ...921L...8H,2021ApJ...922L..27T,2024ApJ...962L...8S}.

The discussion above was related to the outgoing radiation. The global energy balance of the planet is established, however, through the net balance between incoming stellar irradiation and outgoing planetary radiation, which is dependent on stellar type, planetary orbit, and time after star formation. These parameters are introduced with the dotted orange and yellow lines in Figure \ref{fig:radiationlimits}, which presents a stripped-down, simplistic view of planetary energy balance, from which some insights on potentially general evolutionary tracks can be deduced. These lines illustrate the evolution of the instellation (stellar irradiation) for the orbits of Earth (dotted yellow: $F_\mathrm{Earth}$) and Venus (dotted orange: $F_\mathrm{Venus}$) around the Sun, and the evolution of a planet around a 0.1 $M_\odot$ star at 0.03 au (comparable to TRAPPIST-1 c/d). The evolution of these instellation lines goes from right to left, which indicates the thermodynamic path young planets take: they start off hot and cool down. For temporal orientation, the points when an irradiation line crosses the steam thermal radiation limit is indicated with a yellow/orange dot. The energy balance of a given planet at a given time is determined by the net balance between in- and outgoing energy. As such, if the planets indicated by the yellow lines were governed by pure steam atmospheres, they would be able to cool when the yellow irradiation lines are below the radiation limit, and they would heat up when they are above it. This would lead to a warming or cooling response of the atmosphere, respectively. This is why these climates are called 'runaway greenhouse' because temperature beyond the critical limit only increases. There is no energetically stable point between $\approx$ 500--2000 K surface temperature if the incoming stellar radiation is above the thermal limit. If a planet starts cool with a surface liquid water ocean (imagine present-day Earth) and the star brightens above the radiation limit, the surface temperature effectively jumps from the starting point temperature (e.g., 500 K) to beyond the limit ($\gtrsim$ 2000 K). Because water vapor is radiatively active, increasing vaporising before the limit drives increasing warming responses, such that the approach toward the radiation limit is (in geologic terms) rapid.

If the indicated planets would be governed by other types of atmospheres the energy balance shifts. For example, \ce{H2}-rich exoplanetary climates would be able to cool when they are right/below the green \ce{H2} lines in Fig. \ref{fig:radiationlimits}, and heat up left/above the green lines for a specific $P$-$T$ pair. Insightful illustrative behaviours of planetary climate emerge from this conceptualisation. Firstly, very young planets ($\lesssim$ Myr) cannot cool down, because their pre-main sequence host stars are luminous. However, there is strong deviation between G-type stars like the Sun and M-dwarf stars: Sun-like stars cool down quickly and reach the stellar main-sequence after a few tens of Myr. M-dwarf stars, on the other hand, take up to several hundred Myr to reach this point. Before this, their bolometric luminosity prevents exoplanets in the later 'habitable' zones to leave the runaway greenhouse phase. Because M stars are the most numerous type of star in the galaxy, and effectively represent the vast majority of exoplanets amenable for atmospheric characterization with transit and direct imaging surveys, understanding this evolutionary phase is crucial for exoplanet climate science. The above considerations related to Figure \ref{fig:radiationlimits} are simplified, as indicated. For example, they do not account for varying planetary albedo, for instance due to clouds and hazes, which may alter the effective radiation limits for varying climate compositions \citep[e.g.,][]{2013ApJ...771L..45Y,2019Icar..317..583P,2019AREPS..47..583H,2021Natur.598..276T,2021JGRE..12606655G,2019ARA&A..57..617M,2021ApJ...918....1M}, and more self-consistent radiative-convective models that can impact the temperature-pressure structure and thus outgoing balance for \ce{H2O}- \citep{Selsis2023} and \ce{H2}-dominated \citep{Innes2023,2024arXiv240106608L} atmospheres. Irrespective of these uncertaintites, planets that are in a runaway greenhouse phase may be observationally distinguishable via their inflated atmospheres: hotter temperature structures lead to larger scales, therefore planets in a runaway greenhouse phase should -- on average -- be larger than planets with their atmospheric constituents condensed \citep{2019A&A...628A..12T,2020A&A...638A..41T,2020ApJ...896L..22M,2021ApJ...914...84A}. \citet{2024PSJ.....5....3S} recently demonstrated that a large-scale transit survey like ESA PLATO may be able to observationally constrain the abundance of steam runaway greenhouse climates, and thus could lead to a first observational test of the habitable zone concept. Because of the abundance of M-dwarf exoplanets, the inclusion of a statistically relevant ensemble of such exoplanetary system across stellar ages is important. Furthermore, the duration of the runaway greenhouse phase and atmospheric differential escape may substantially change the composition of the secondary atmosphere and therefore its net oxidation state. In initially steam-dominated atmospheres, photolytic destruction can overcome the cold trap mechanism (see Section \ref{sec:atmospheres2}) by thermally inflating the atmosphere and differentiating H and O in the outlfow. In result, diffusive separation during escape will enrich residual atmospheres \citep{2013Natur.497..607H,Hamano2015}. This can be particularly effective for M star planets \citep{2016ApJ...829...63S,2018AJ....155..195W,2021AsBio..21.1325B}, but also G star planets may undergo this evolution \citep{2021AGUA....200294K}, creating a potential false-positive biosignature \citep{2018AsBio..18..630M,2022NatAs...6..189K}, but the chemical kinetics of such atmospheres require further attention \citep{Grenfell2018} .

Finally, an important feature of Figure \ref{fig:radiationlimits} is the indicated range for basalt melting between $\approx$1250--1500 K, which is variable and depends on rock composition, but is most sensitive to water content \citep{2003GGG.....4.1073K}. Because of the aforementioned sensitivity of surface temperature on atmospheric composition and total volatile abundance, the phase state of the planetary mantle is strongly coupled to the climate state. Planets with approximately Earth-like water inventories are molten when in a runaway greenhouse phase, as are cloudless sub-Neptunes with $\gtrsim$100--200 bar surface pressure \citep{2021ApJ...918....1M,2021JGRE..12606711L,2022PSJ.....3..127S}. For water- and hydrogen-rich planets, this brings the interaction with the interior into close focus: water-rich terrestrials and super-Earth exoplanets enter the magma ocean regime, which can hide a large fraction of their global water budget in the interior \citep{2021ApJ...922L...4D}, while water-rich sub-Neptunes may enter a supercritical phase state in their deep volatile layers \citep{2023ApJ...944...20P}, with important consequences for their internal mixing and stratification processes \citep{2019ApJ...887L..33K,2020ApJ...891..111K,2024ApJ...962L...8S,2024ApJ...963L...7W}. This leads us towards the tight interconnection between planetary envelope and interior during the formation of planetary atmospheres.

\subsection{Magma Ocean Evolution}
\label{sec:magmaocean}

The magma ocean concept -- the whole Earth being liquid magma instead of solid rock -- was first mentioned by Lord Kelvin \citep{Kelvin1864}, who noted that a whole silicate liquid sphere would "most probably" start crystallizing from the bottom to the top because the silicate adiabat is less steep than silicate melting curves, which leads to the first crystals appearing at the base of the system. The science of liquified planetary mantles really started off after the rocky samples brought back to Earth by the lunar Apollo missions clearly indicated substantial evidence for global melting of the lunar mantle through the composition of the oldest lunar crust \citep{solomon1977magma,1985AREPS..13..201W}. The global distribution of anorthosites and small age spread < 300 Myr on the lunar surface can be formed by the floatation of low-density plagioclase crystals on top of a denser basaltic magma. Since then the magma ocean concept has undergone substantial evolution, has been used to understand atmospheric formation, the chemical segregation of the terrestrial planets and satellites \citep{2012AREPS..40..113E}, and has been extended to rocky exoplanets \citep[e.g.,][]{2017RSPTA.37550394T,2018RSPTA.37680109S}. What defines a magma "ocean"? In its purest form, an ocean implies that the whole of the mantle would be liquid, i.e., above the liquidus of the mineral solution of the mantle. However, for simplicity, we will here describe all mantles with a substantial melt fraction close to or above the rheological transition as "magma oceans". The rheological transition at about 40-60\% melt fraction \citep{1993Litho..30..223A,2007evea.book...91S,2009GGG....10.3010C,2019GeoJI.219..185K} changes the fluid viscosity by about 20 orders of magnitudes, from water-like ($\sim10^{-2}$--$10^{2}$ Pa s) to rock-like ($\sim10^{20}$--$10^{22}$ Pa s). Above this transition (at higher melt fraction), the mantle aggregate behaves rheologically as a liquid, below it more like a solid. The importance of this transition cannot be overstated, as it qualitatively changes the physics of energy transport, as well the redistribution of chemical elements in the planetary mantle, which acts as the intermediary between metal core and atmosphere of rocky planets and exoplanets. Whether a planet is in a full magma ocean state, partially molten, or mostly solid, is a global phase state, and changes how to interpret atmospheric abundances, internal dynamics, and structural relations of exoplanets. In this section, we wil describe the transition from a primordial magma ocean generated by accretion energy and the decay of short-lived radionuclides to solid-like states, and the effect on atmospheric formation of terrestrial-like planets, and lay out some potentially observable implications for super-Earth exoplanets. Permanent dayside magma oceans on tidally-locked rocky exoplanets are discussed in Section \ref{sec:interiors}.

From the viewpoint of atmospheric formation, the mode and timescale of the freeze-out of magma oceans is of primary importance for the build-up of rocky planet atmospheres because atmospheric volatiles can dissolve ('partition') into the magma. We refer the reader to \citet{2018RvMG...84..393S} and \citet{2023FrEaS..1159412S} for comprehensive overviews of these processes from a geochemical perspective.  Most important from an exoplanet perspective is that volatile compounds have diverse chemical affinities for partitioning into silicate magma and binding with metallic phases, and their partition behaviours are strong functions of melt composition (redox state) and the degassing pressure, which, in the case of a global magma ocean, is the atmosphere-interior interface (surface) pressure. Hence, the interplay between a magma ocean and outgassing atmosphere fractionates the relative species present in the atmosphere with respect to the case that no magma ocean is present. This is the physical background to Figure \ref{fig:volatileoutgassing}B. Most importantly, \ce{H2O} is highly soluble; the majority of \ce{H2O} in a chemically equilibrated magma ocean--atmosphere system is dissolved in the magma, and outgasses rapidly once the majority of the mantle starts crystallizing \citep{2013JGRE..118.1155L,2017JGRE..122.1458S}, because the partition coefficients for solids are typically orders of magnitude below those in liquids. Therefore, in Figure \ref{fig:volatileoutgassing}B the amount of \ce{H2O} is a strong function of mantle melt fraction. If most of the mantle is liquid the majority of \ce{H2O} is partitioned into the melt; if most of the mantle is solid it degasses into the atmosphere. The interaction between outgassing and greenhouse effect of radiatively active gases (dominated by \ce{H2O}, \ce{H2}, and \ce{CO2}, perhaps \ce{CO}) creates emergent co-evolution between the planetary mantle and atmosphere. Without dissolution into the magma the atmosphere could be understood in isolation; without the highly non-linear effects of greenhouse gases (Section \ref{sec:atmospheres1}) the crystallization of the mantle could be understood in isolation. Both treatments are pervasive in the literature, but both fall short in capturing the first-order interactions between mantle and atmosphere on a planetary scale. In this respect, super-Earth and sub-Neptune exoplanets open whole new categories of celestial objects that give us direct access to these types of thermodynamic regimes that govern atmospheric formation and dispersion of rocky planets. We will come back to a mid-term outlook on these questions in Section \ref{sec:outlook}.

\begin{figure*}[tbh]
 	\centering
        \includegraphics[width=0.7\textwidth]{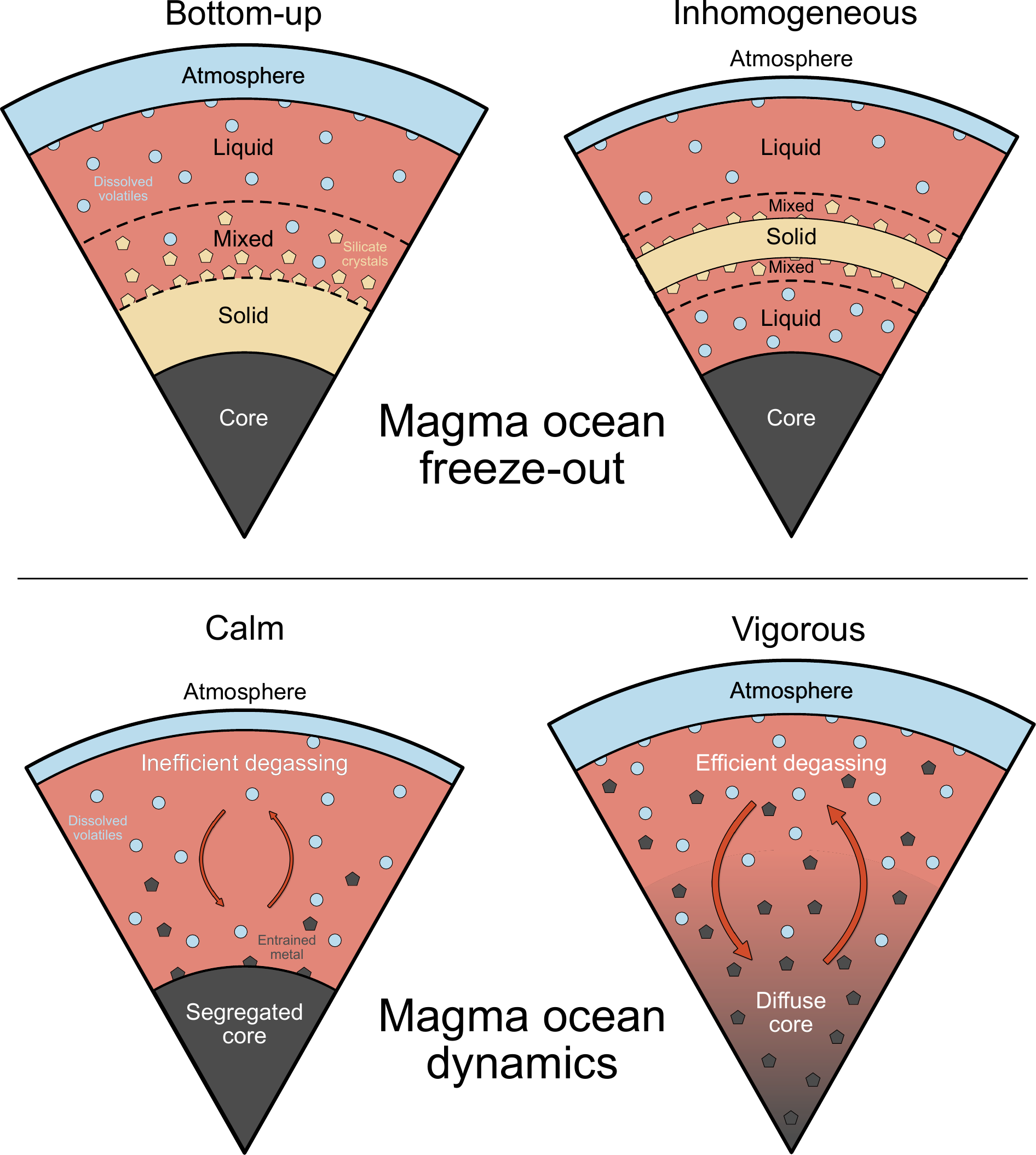}
 	\caption{\textsf{\textbf{Schematic illustration of different modes of mantle crystallization (top) and internal dynamics (bottom) in planetary magma oceans.} Freeze-out of the planetary mantle (top) can proceed in a variety of ways. If the magma ocean crystallizes linearly from bottom to top (top left), atmospheric and liquid mantle are in direct chemical exchange, while the exchange with the core is limited once crystallization has started. Inhomogeneous crystallization (top right) -- here illustrated by freeze-out from the middle-out -- can lead to diverse solidification paths in which the chemical exchange between deep liquid mantle and atmosphere is inhibited, which can store large quantities of volatiles in the deep planetary interior. The strength of turbulent convection in the magma ocean plays a separate non-linear role in mixing chemical reservoirs. Limited convectional vigour in the mantle (bottom left) promotes gravitational settling of metal droplets that form the core. However, in transient, short-lived magma oceans the crystallization timescale can be faster than the transport of volatiles to the near surface, which inhibits the chemical exchange between magma ocean and atmosphere. In highly turbulent magma oceans (bottom right) gravitational settling of metal droplets is inhibited, which may suppress core formation. At the same time, dissolved volatiles are efficiently mixed by turbulent eddies, and thus exchange between mantle and atmosphere should be efficient.}}
    \label{fig:magmaocean}
\end{figure*}

Figure \ref{fig:magmaocean} illustrates two major physical processes operating in planetary magma oceans that can lead to order of magnitude deviation in cooling timescale and atmospheric composition. The top row sketches two end-member scenarios of interaction between the deep planetary interior and the atmosphere. On the top left ('bottom-up') the standard case of magma ocean solidification is illustrated \citep[e.g.,][]{2008ApJ...685.1237E}. In this case, initially the whole of the planet, from the core to the atmosphere, is in causal contact. The thermal Rayleigh number of planetary magma oceans is on the order $10^{28}$ and higher. This means that dominantly molten planetary mantles are highly turbulent: from a fluid dynamics perspective they behave comparable to the atmosphere, with similar timescales for admixing and nucleation processes (clouds and hazes in atmospheres, metal droplets and rock crystals in magma oceans). Mixing is driven by large-scale eddy circulation \citep{2007evea.book...91S} and the magma ocean temperature profile in the standard case can be assumed to be adiabatic. This is the case noted above by Lord Kelvin: the mantle adiabat is less steep than the expected melting curves of mantle rock compositions. During cooling, the adiabat then first intersects at the base of the magma ocean, which drives crystallization. The crystallization front then smoothly propagates upwards until it reaches the surface. Evolution toward the rheological transition in this case is rapid, on the order of $10^3$ to $10^4$ years. Volatile dissolution in this first episode of freezing is high: the mantle volume is large, and volatiles, like \ce{H2O}, can in principle be taken up efficiently. Once the mantle is dominated by the mush regime (melt fraction lower than $\approx$ 50\%), cooling slows down substantially, and mostly depends on the total mass and composition of the atmosphere.  This mode of crystallization is the simplest, and can be captured by boundary layer theory, which is effectively a 0-D approach. 

However, if the mantle does not crystallize in such a smooth fashion, the behaviour becomes non-linear. Experimental evidence indicates that at high pressures rock melting curves can deviate from the above picture, and instead of the bottom the middle or other regions of the mantle can be the location of the first crystals appearing \citep{2007Natur.450..866L,2014RSPTA.37230076S}. In this case the magma ocean would not freeze out from bottom to top, but non-linearly, for example from the middle-out (like shown in the top right of Fig. \ref{fig:magmaocean}, 'inhomogeneous'). Additionally, during late-stage evolution the near-surface region of the magma ocean can crystallize close to the rheological transition, while a large fraction of the interior is still molten \citep{2018PEPI..274...49B,2022PSJ.....3...93B}. Besides the issue of the adiabat crossing the melting curve of a particular mantle composition, there is the important question of the density of silicate melts at high pressure \citep{1983PEPI...33...12O,2010E&PSL.295..435F,2015JGRB..120.6085B,2019E&PSL.516..202C}. FeO is incompatible and decreases the melting temperature of silicate rocks. As crystallization proceeds, melt gets enriched in FeO, denser than crystals and more fusible than Mg-rich silicate. This can promote the formation of a basal magma ocean \citep{2017GGG....18.3385B}. Exacerbating these uncertainties, giant impacts during late-stages of accretion significantly affect pressure-temperature profiles \citep{2019SciA....5.3746L}.  These non-linear crystallization scenarios can have qualitative effects on the expected mass and composition of the atmosphere, because they can disconnect the chemical mixing between planetary sub-reservoirs: core + lower mantle and upper mantle + atmosphere are suddenly disconnected, and volatiles dissolved in the deep interior may not be efficiently mixed upwards. The detailed consequences of such scenarios for atmospheric evolution are yet to be explored, but implications for the secondary atmospheres of M-star rocky exoplanets \citep{2020MNRAS.496.3786M,2023MNRAS.526.6235M} and the bulk volatile fraction of super-Earths and sub-Neptunes \citep{2020ApJ...891..111K,2021ApJ...914L...4L,2021ApJ...922L...4D,2021ApJ...909L..22K,2022PSJ.....3..127S,2023ApJ...954...29P,Kempton2023,2023A&A...674A.224C,2024ApJ...962L...8S,2024A&A...683A.194F,2024arXiv240214072R} may be manifold.

The large-scale circulation of the magma ocean itself influences the chemical redistribution of elements, most importantly volatiles and metal particles that sink toward the core. To start with volatiles, it is unclear which physical effect dominates chemical equilibration (in- and outgassing) between the magma ocean and atmosphere: bubble nucleation or diffusion through the boundary layer. The first scenario depends on the highly non-linear and poorly understood physics of bubble nucleation in turbulent media \citep{,2013Natur.497..607H,2018SSRv..214...76I,2021E&PSL.55316598J,2021SciA....7..406S}, while the second is sensitive to the saturation of the near-surface eddies with deep-mantle volatiles \citep{2018E&PSL.498..418O,2019PEPI..29406294O}. However, as the large-scale dynamics for Earth-sized and larger planets are highly turbulent, the efficacy of volatile redistribution between the deep and shallower mantle should be effective. \citet{2023Icar..39015265S} recently derived scaling laws that relate the convective vigour to degassing efficiency. The resulting picture for whole-mantle magma oceans is straightforward: magma oceans in small protoplanets ($\sim$Mars) can crystalize faster than they can redistribute volatiles between deep and shallow mantle because the overturning circulation is not effective enough to homogenize the whole magma column. In this case, illustrated in the bottom left of Figure \ref{fig:magmaocean} ('calm'), mantle and atmosphere cannot fully equilibrate and thus volatiles can be stranded in the mantle during crystallization. This supports previous results from \citet{2017GGG....18.3078H}, who argued -- based on numerical simulations -- that rapid upward propagation of the crystallization front during cooling can trap a large fraction ($\sim$50 \%) of residual \ce{H2O} and \ce{CO2} in melt pockets that are isolated by small-scale convection \citep{2017GGG....18.2785B,2017JGRE..122..577M,2018E&PSL.491..216B}. The opposing case, potentially valid for Earth-sized and larger exoplanets, operates when planetary-scale convection rapidly equilibrates the magma column. In this case near-surface magma is saturated in volatiles and in- and outgassing can progress on the local equilibration timescale. All else being equal, this suggests that larger super-Earths develop thicker atmospheres while in a magma ocean phase. This is to some extent self-supporting because deeper atmospheres then increase thermal blanketing by the greenhouse effect.

So far, our discussion has given the impression that a given fixed composition supports a specific type of atmosphere, which then can be ingassed and outgassed, depending on the melting state of the mantle. However, mantle and atmosphere are not a closed system toward the metal core, but are chemically open during core formation itself and potentially later on: atmospheric volatiles chemically interact not only with the magma but also with core-forming metals. For discussing this, we first must draw a general picture of the core formation process and its interaction with the mantle.

\subsection{Metal Core Segregation} \label{sec:metalcoresegregation}

All terrestrial planets of the Solar System possess metal cores (Fe, Ni, etc.) of various sizes, and even the icy moons of the outer Solar System show evidence for increased densities and metal cores in their interiors. From a general perspective, core formation is simply gravitational settling and concentration of denser elements deeper into the potential well of rocky planets. However, this process is not as simple as it may seem, because the stresses that must be overcome by small metal grains to sink toward the core through a solid planetary mantle are much larger than those provided by the gravitational force. A theoretical cold and solid planet with a perfectly homogeneous distribution of metals (e.g., Fe) and rocks would not form a metal core. There are three principle ways to overcome this problem that are relevant for the formation of terrestrial and super-Earth exoplanets, all of which require either the silicate mantle or the metal material to be liquid and are sensitive to the main mode of planetary accretion (Section \ref{sec:formation}). Because the liquid phase decreases stresses in the mantle by orders of magnitude, and denser metal particles can sink through the ambient medium efficiently, the main phase of metal core formation is thought to operate during the magma ocean stage.

The first mode of core formation is by mutual giant impacts among already gravitationally segregated protoplanets and planetesimals \citep[e.g.,][]{2017E&PSL.474..375J,2017E&PSL.458..252F,2019Tectp.760..165T}. In this picture, direct core-core merging minimizes the physical and chemical interaction between the protocores and the mantle itself, but induces a fully molten mantle, potentially leading to complete vaporizing of the mantle and atmosphere of the planet \citep{2023E&PSL.60818014C}. The core composition after perfect core-core mergers is the mixed composition of both. However, the smaller the impacting protoplanet, the more likely is the physical destruction of the impacting protocore and the chemical dissolution of it in the planetary magma ocean by fluid dynamical instabilities at the protocore-magma interface \citep{2011E&PSL.310..303D,2014E&PSL.391..274D,2016NatGe...9..786L,2021E&PSL.56416888L}. A high degree of metal-silicate equilibration is reached when the core fragments completely, the degree of which is dependent on the mode of accretion and the energetics of protoplanet mergers \citep{2018SSRv..214..101N}. Variations in the type of impact, geometry, and angular momentum of an individual accretion event between two protoplanets drastically change the degree of metal-silicate and volatile-mantle mixing \citep{2018E&PSL.487..117N,2020AIPC.2272h0003S,2021arXiv210302045C}.

The second mode of core segregation is by gravitational settling of metal droplets entrained in the magma ocean \citep{1990orea.book..231S}, which -- analogous to precipitation in the atmosphere -- is a competition between shear stresses in the magma flow and gravity. Typically, gravitational settling in protoplanets is thought to be fast and efficient; rain-out of metal droplets operates on a timescale of days to weeks \citep{2012AREPS..40..113E}. If magma ocean also crystallize from the bottom to the top, this would mean that core segregation and atmospheric degassing can be understood as a single, linear process. This type of reasoning supports approaches that postulate the chemical equilibration between core and mantle, followed by isolation of the core in super-Earth exoplanets \citep{2008ApJ...685.1237E,2017ApJ...835..234S}. However, analogous to the admixture of volatiles into shallower parts of the mantle, the large-scale dynamics of the magma ocean influence this type of core formation \citep{1993JGR....98.5375S}: vigorous convection at extreme Rayleigh numbers can suppress gravitational settling by dispersion of growing metal droplets, which may suppress efficient core formation in super-Earths and sub-Neptunes \citep{2021ApJ...914L...4L}. These scenarios are illustrated at the bottom of Figure \ref{fig:magmaocean}: in the 'calm' scenario the Rayleigh number of the system is low, hence entrained metal droplets efficiently coagulate and settle toward the core. In the 'vigorous' scenario the convection in the magma ocean is at extreme Rayleigh numbers and hence the whole of the planetary interior is mixed together.

The third potential mode of core segregation is debated for rocky and terrestrial planetesimals and icy moons in the Solar System: high bulk sulfur contents and intermediate levels of internal heating (e.g., by radioactive decay or tidal forcing) may result in an interconnected network of molten Fe-FeS metal that can gravitationally segregate by porous flow through the solid mantle matrix   \citep{2003Natur.422..154Y,2017PNAS..11413406G,2023SciA....9F3955T}. From an experimental perspective, this mechanism is criticized because the micro-structural properties of silicate crystals may efficiently close available pore space and thus hinder metal connectivity \citep{2009PEPI..177..139B,2015E&PSL.417...67C}, in particular at high pressure. Mantle-core differentiation in this third mode may operate on Gyr timescales, while the first two modes of metal-silicate separation would operate on the accretion timescale.

In reality, rocky exoplanets will undergo a mixture of these differentiation regimes, in particular during accretion. Whether permanent dayside magma oceans are comparable to these regimes will be explored in Section \ref{sec:interiors}.

\subsection{Volatile redistribution between core, mantle, and atmosphere} \label{sec:volatileredistribution}

Whether the metal part of a planet segregates to the core or remains entrained in the magma will lead to substantially different atmospheric compositions. This is because, first, redistribution of redox-active elements, such as Fe and H, in the planet will change the redox state of the mantle, and hence the atmosphere (Figure \ref{fig:volatileoutgassing}). Second, atmospheric volatiles, in particular H, N, C, and S, chemically bind to core-forming metals, which is sensitive to ambient pressure and temperature \citep[e.g.,][]{2016AmMin.101..540H,2017E&PSL.469...84S,2023FrEaS..1159412S,2020PNAS..117.8743F,2022E&PSL.59817847G,2022E&PSL.59317650Z}. Therefore, metal core formation and atmospheric build-up are tightly coupled during magma ocean stages. This can be particularly important if the core and mantle material can be homogenized over longer time spans by vigorous convection \citep{2021ApJ...914L...4L}. Additionally, interaction of water with core-forming metals can lead to divergence between the structural states of volatile-rich exoplanets and the terrestrial planet population. Planets that accrete very high water-mass fractions, so-called water worlds (cf. Section \ref{sec:formation}) may escape metal core formation by reacting most available metal with \ce{H2O} \citep{2008ApJ...688..628E}. Interaction between primordially-accreted hydrogen and mantle-derived iron can generate water by the reaction \ce{FeO + H2 -> H_2O + Fe^0} \citep{2021ApJ...909L..22K,2020MNRAS.496.3755K,2022NatAs...6.1296K}. The detailed consequences of these interactions are currently coming closer into the focus of the community, as a whole sequence of reactions between metal, silicates, and atmospheric volatiles is possible, which can shift both the abundances of volatile elements in the core, and the mixing ratios of major atmospheric compounds in the atmosphere \citep{2022PSJ.....3..127S}.

The redox state of magmas is typically measured through the abundance of different iron redox states, which act as a measure of the global redistribution of valence electrons. Fe can be present in the mantle in metallic form (\ce{Fe^0}), or as ferrous (\ce{Fe^2+}, for example in \ce{FeO}) or ferric (\ce{Fe^3+}, for example in \ce{Fe2O3}) oxides. Silicate magmas and solids at different pressures and temperatures, however, have different affinities for hosting metal in its ferrous or ferric forms, which 'disproportionates' the abundance of Fe and O between magma and rock. For example, the redox reaction \ce{3 FeO + Al2O3 <-> 2 FeAlO3 + Fe^0} creates perovskite and splits off iron metal. During core formation \ce{Fe^0} would be expected to sink toward the core and become isolated from the mantle. On a global planetary scale, this phenomenon tends to oxidize the mantle of planets as a function of pressure: larger planets with higher internal pressures tend to develop more oxidized mantles, because the respective redox reactions are sensitive to pressure \citep{2020NatCo..11.2007D}. Due to the rapid mixing timescales and high temperatures, mantle oxidation via iron disproportionation is expected to be rapid during the magma ocean stage \citep{2005E&PSL.236...78W,2019Sci...365..903A,2012E&PSL.341...48H}, but can potentially proceed on a slower timescale after solidification \citep{2004Natur.428..409F}. Even in the absence of metal segregation to the core, iron disproportionation is expected to alter the atmospheric composition by changing the redox gradient in the mantle \citep{2022GeCoA.328..221H,2023PSJ.....4...31M}. The differing timescales and efficacies of the suggested disproportionation effects have not been quantified in a general picture as of yet. However, exoplanet science in principle has the means to testing first-order predictions of the timescales of planetary oxidation and core-mantle-atmosphere equilibration. For example, redox disproportionation and hydrogen ingassing during planetary accretion predicts that all rocky planets that form with a magma ocean in the disk should develop a finite amount of water by the reaction of \ce{H2O} with mantle \ce{FeO} \citep{2020PNAS..11718264K,2021ApJ...909L..22K,2020MNRAS.496.3755K,2022NatAs...6.1296K}. On the other hand, various degrees of metal-silicate equilibration due to magma ocean dynamics may shift the atmospheric abundances \citep{2021ApJ...914L...4L,2022PSJ.....3..127S}. Reduced secondary atmospheres produced by incomplete mantle differentiation or other mechanisms may be distinguishable through trace species in super-Earth atmospheres \citep{2014ApJ...784...63H,2021AJ....161..213S,2021ApJ...921L..28R,2023ApJ...948L..20H}


\section{Atmospheric Structure and Dynamics} \label{sec:atmospheres2}
Every new exoplanet discovered shows the stunning diversity of worlds in our galaxy. As we enter an era of planetary characterization, we expect the atmospheres of rocky exoplanets to exhibit equally diverse characteristics. However, certain trends are expected to emerge in relation to the effect of the host star on the planet's surface and the potential outgassing of magma oceans. In this section, we discuss the expected trends based on theoretical calculations and describe the composition, structure and dynamics focused on observations prior and shortler following the launch of JWST.

\subsection{Atmospheric Compositions}
\begin{figure*}[tbh]
 	\centering
        \includegraphics[width=0.99\textwidth]{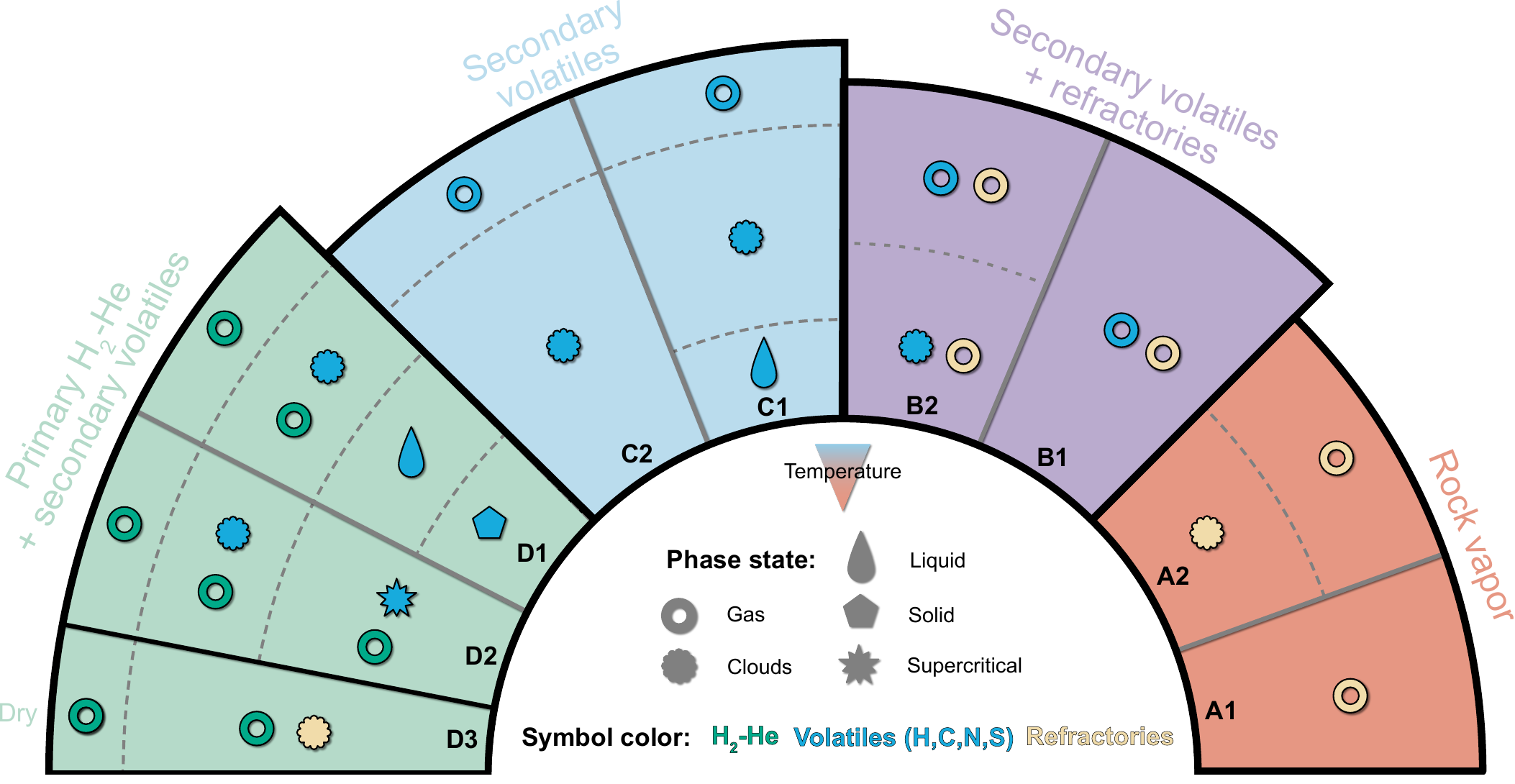}
 	\caption{\textsf{\textbf{Illustration of plausible classes of atmospheres on super-Earths and Earth-like exoplanets expected from theoretical calculations and observational constraints.} The different radii in each case represent the increase in mean molecular weight expected from H- (left) to rock-dominated (right). Different symbols show the presence of gases or  condensates in the atmospheres. Information on liquids, solids or supercritical fluid at higher pressures is also shown. Temperatue generally increases from top (blue, purple) to bottom (green, red) classes, either driven by stellar irradiation (red) or greenhouse forcing (green). Note that the separation in different layers is schematic and does not represent the actual depth of such layers in the planet.}}
    \label{fig:atmospheres-schematic}
\end{figure*}

The four rocky bodies in our Solar System with substantial atmospheres--Venus, Earth, Mars and Titan--each have their own unique atmospheric compositions. Venus has a thick atmosphere made up mainly of carbon dioxide, with smaller amounts of nitrogen, sulfur dioxide, and other trace gases, which creates a strong greenhouse effect and extremely high surface temperatures \citep{TAYLOR2014305}. Earth's atmosphere is predominantly nitrogen and oxygen, with trace amounts of other gases, such as carbon dioxide, argon, and neon, that together create the right conditions for life to exist on our planet \citep{SHOWMAN2014423}. Titan's atmosphere is primarily composed of nitrogen but also has some methane, with trace amounts of other hydrocarbons and nitrogen-rich organic compounds \citep{COUSTENIS2014831}. Finally, Mars has a thin atmosphere consisting mostly of carbon dioxide and its atmospheric pressure is much lower than Earth's \citep{CATLING2014343}. This diversity is an outcome of accretion, delivery and loss of material to space, internal chemistry and differentiation processes, and exchange between the atmospheres with the surface and interior, as discussed in the previous sections. Given the diverse range of exoplanetary densities and physical processes operating on them, we can anticipate a substantial diversity in low-mass exoplanet atmospheres. Figure \ref{fig:atmospheres-schematic} shows a schematic figure with the different types of atmospheres expected in low-mass exoplanets up to sub-Neptune sizes. In the figure, we can see different cases in different colors, and within each case the dashed lines separate different layers of the planet, including potential cloud layers, oceans (labeled as liquid), high pressure ice layers (solid), and supercritical fluids. In the following discussion, we will describe each one of these potential atmospheres as inferred from theoretical studies and observations.

\paragraph{Lava planets} Because of their extreme proximity to the host stars, these planets (Fig. \ref{fig:atmospheres-schematic}, A1--2) are typically tidally locked and the strong irradiation on the perpetual dayside can make the planetary surface to reach very high temperatures, high enough to melt their surfaces into an enduring magma ocean. This magma ocean is expected to vaporize and form a silicate-rich outgassed atmosphere, mainly composed of vaporized rock material \citep{Schaefer2009, Leger2011, Miguel2011}. For the most extreme cases in this population, corresponding to an orbital distance closer than 0.1 AU for a planet around a solar-type star, it is expected that if their atmospheres had any volatile species at some point in their history, those were lost due to the strong interaction with the star, leaving the planets with a tenuous and heavy atmosphere. Several papers over the years have studied the potential composition of these atmospheres \citep{Schaefer2009, Miguel2011, Ito2015, Kite2016, 2020ApJ...891..111K, Zilinskas2022}.

\begin{figure}[tbh!]
 	\centering
 	\includegraphics[width=0.47\textwidth]{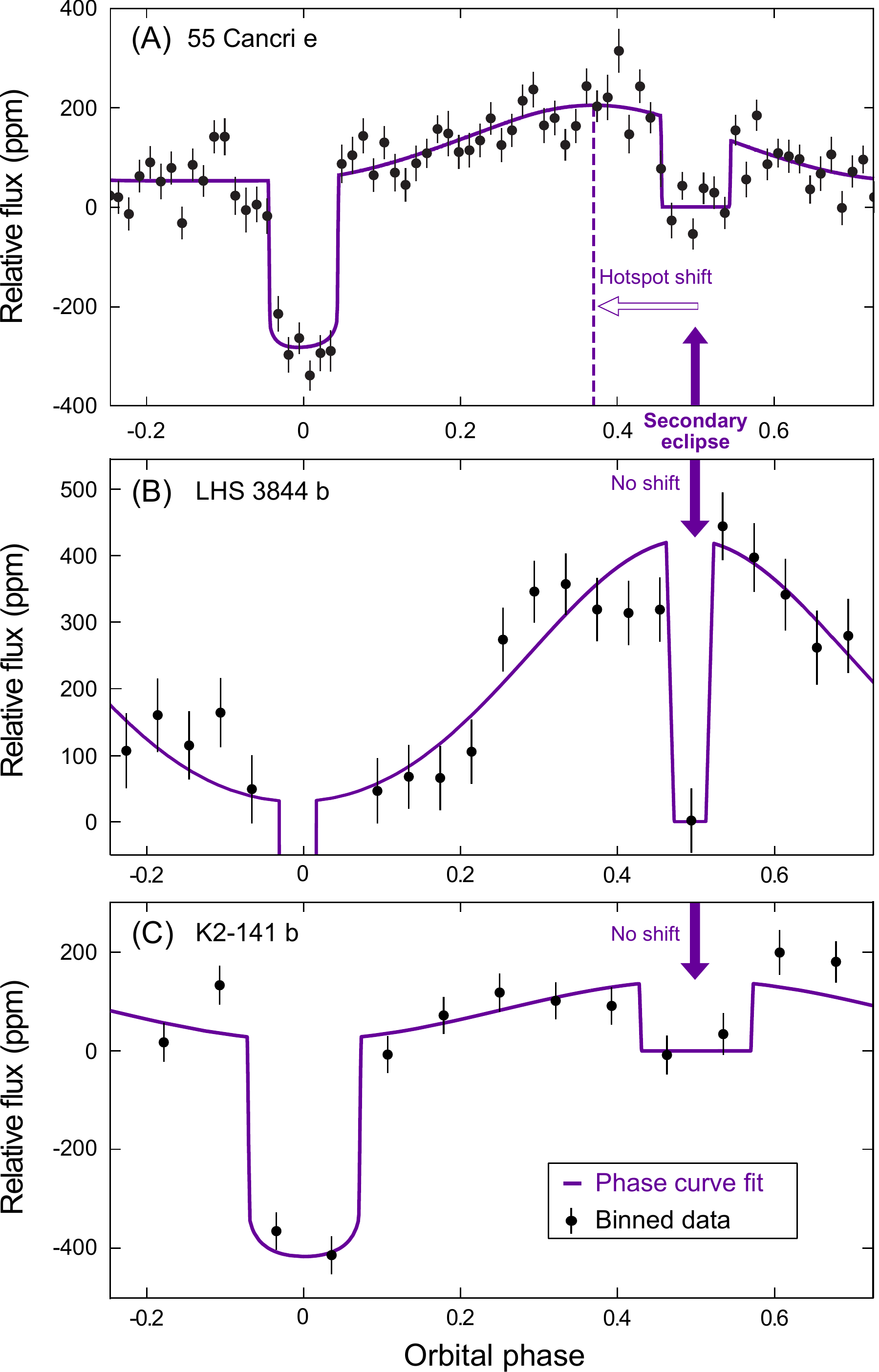}
 	\caption{\textsf{\textbf{Measured phase curves of the ultrashort-period super-Earth exoplanets 55 Cancri e, LHS 3844 b, and K2-141 b.} Binned observational data is displayed in black, the magenta lines represent fits to the data. The phase curve of 55 Cancri e displays an offset in the expected hotspot from the orbital location of the secondary eclipse, presenting evidence for a redistribution of heat. The LHS 3844 b and K2-141 b data shows no measurable offset, indicating no heat redistribution and therefore providing evidence for the absence of a thick volatile atmosphere. Data from \citet{2016Natur.532..207D}, \citet{2019Natur.573...87K}, and \citet{2022A&A...664A..79Z}, reproduced with permission.}}
    \label{fig:phasecurves}
\end{figure}

The most common approach to study this problem is to start from the bottom up and estimate the composition of the atmosphere as it degasses from the magma ocean. This can be done by estimating the composition of the gases that will be in chemical equilibrium with the melt \citep{Fegley1987, Schaefer2004,2020ChEG...80l5594F}, which has been adapted in recent open-source variants \citep{2023M&PS...58.1149V,2023ApJ...947...64W}. These calculations find that the most abundant gases in these atmospheres are \ce{SiO}, \ce{SiO2}, \ce{O2}, \ce{Na}, and \ce{Fe}, among other gases \citep{Schaefer2009}. Condensables of species such as \ce{MgSiO3} and \ce{Mg2SiO4} are expected \citep{Mahapatra2017} when we go farther away from the substellar hot-spot and the planet cools down. The abundance of the outgassed atmospheres primarily depends on the temperature of the melt and its composition. Because the temperature of the surface is mostly governed by stellar irradiation, the stellar type and semi-major axis of the planet are very important when determining the temperature of the magma and thus, the atmospheric composition \citep{Miguel2011}. On the other hand, variations in stellar refractory abundances and magmatic evolution and differentiation determine the mantle composition, and so the vapour composition is a major uncertainty. Therefore, most efforts so far use known compositions from the Solar System. As discussed in Section \ref{sec:observations}, there are observations made with Spitzer and K2 of a notable example in this category: K2-141 b \citep{2022A&A...664A..79Z}, shown in the bottom panel in figure \ref{fig:phasecurves}. While the K2-141 b data might be consistent with a vaporized rock atmosphere, another planet in this class might have no atmosphere at all: GJ 1252 b \citep{2022ApJ...937L..17C} and therefore, finding if these planets have atmospheres is a crucial question to ask in order to better understand the nature of these worlds and the interaction with the parent star. While there are currently not many observations of planets in this class, many are expected to be made with the JWST (Section \ref{sec:futureObs}). 

At the farthest extreme of this population, the small class of disintegrating lava planets merits their own explanation. These are planets detected through the dust tails produced by evaporative outflows from their molten surfaces. So far we have only three catastrophically evaporating planets discovered by the Kepler/K2 missions: Kepler 1520b \citep{2012ApJ...752....1R}, KOI-2700b \citep{2014ApJ...784...40R} and K2-22b \citep{2015ApJ...812..112S}. What is intriguing about these planets is that, despite being capable of destruction within a fraction of the stellar lifetimes, their characterization can offer additional insights into the interiors and the interaction between their interiors and atmospheres of small rocky worlds. \citep{2023MNRAS.518.1761B, 2024MNRAS.528.1249C,2024MNRAS.528.4314C}. 

\paragraph{Rocky exoplanets with volatiles and refractory atmospheres} However, not all hot rocky exoplanets are expected to lack volatiles in their atmospheres. It is possible that some planets could retain volatile gases from their secondary or primary atmospheres, either because they are located farther from their host star and are less exposed to stellar activity, or because they initially had very massive atmospheres that were not completely lost (Fig. \ref{fig:atmospheres-schematic}, B1--2). Such planets might still be strongly irradiated and have surface temperatures high enough to melt the surface, which can explain the unusual underdensities of some exoplanets in Fig. \ref{fig:M-R}, e.g., 55 Cnc e, TOI-561 b, and HD 3167 b. It is also possible that some of these planets are located further away from the star but have a very thick atmosphere with greenhouse gases that will heat the surface and drive it to the melting point (Section \ref{sec:atmospheres1}). In these cases, the atmospheres of these planets are composed of volatile species mixed trace elements from surface vaporization \citep{2022PSJ.....3..127S, Zilinskas2023, 2023A&A...678A..29M} (purple B1 and B2 cases in figure \ref{fig:atmospheres-schematic}). These atmospheres are expected to be dominated by gases like \ce{H2O}, \ce{CO}, \ce{CO2}, \ce{N2} or \ce{SO2} and hydrocarbons such as \ce{CH4}, \ce{C2H2} and \ce{HCN} \citep{Miguel2019, Zilinskas2020, Herbort2020}, but they might also have non-negligible concentrations of \ce{SiO} and \ce{SiO2} depending on their metallicity and C/O ratios \citep{2022PSJ.....3..127S,Zilinskas2023,2023ApJ...954...29P,2023A&A...674A.224C}. These atmospheres might also present cloud condensates of species such as \ce{KCl}, \ce{NaCl}, \ce{FeS}, \ce{FeS2}, \ce{FeO}, \ce{Fe2O3}, and \ce{Fe3O4} \citep{Herbort2022}. A notable example in this class is 55 Cnc e (Figure \ref{fig:phasecurves}, top panel), whose recent JWST observations suggest the presence of an atmosphere \citep{Hu2024}, which we will discuss in more detail in Section \ref{sec:3d}. 

\paragraph{Rocky exoplanets with volatile atmospheres} These planets are located further away from the star compared to the previous cases (Fig. \ref{fig:atmospheres-schematic}, C1--2). As a result, they were less exposed to stellar activity during their formation and evolution, and they might have retained secondary atmospheres primarily composed of volatile species that were degassed from the interior (see Section \ref{sec:atmospheres1}). Nevertheless, we note that there are at least a few examples in this class that are consistent with no atmosphere at all: LHS 3844 b \citep{2019Natur.573...87K} and TRAPPIST-1 b+c \citep{Greene2023,Zieba2023}. LHS 3844 b is hot enough (1040$\pm$40 K) to be in between the B and C classes. Motivated by the atmospheric composition of the Solar System terrestrial planets, the reservoir budget of volatiles available from protoplanetary disks, and information obtained from the composition of meteorites (see sections \ref{sec:planetesimal_evolution} and \ref{sec:atmospheres1} for a critical discussion of these assumptions), many studies have performed calculations assuming that warm rocky exoplanets have oxidized atmospheres mainly composed of gases such as  H$_2$O, CO$_2$ and N$_2$ \citep{2008ApJ...685.1237E, Forget2014, Morley2017, 2021NatAs...5..575T, Jordan2021}. This type of atmosphere is represented with light-blue in Figure \ref{fig:atmospheres-schematic}, where we also see that they might posses a layer where some of the volatiles condense and form a cloud layer. Some examples of potential exoplanets in this class are LHS 1140 b, GJ 1132 b, GJ 486 b, and TRAPPIST-1 d--h (Section \ref{sec:observations}). 

\paragraph{H-He-dominated atmospheres} These planets either retain some of the primordial H$_2$-He in their atmospheres or generate secondary atmospheres by internal processes (Fig. \ref{fig:atmospheres-schematic}, D1--3). The former category, building a bridge to the larger sub-Neptune class, might also have large amounts of volatile ice directly accreted from the planet formation process, although their origin is still under debate (Section \ref{sec:atmospheres1}). One of the ideas to explain the origin of H-He atmospheres is that these planets were formed as Neptune-like planets and lost most of their atmospheres due to their interaction with the star \citep{2020SSRv..216..129O}. Another idea is that they have never been big enough to start the runaway gas accretion and only retained some H$_2$-He from the nebula that mixed with large amounts of volatiles such as water that are expected to be degassed from their interiors \citep{2016ApJ...817...90L}. A third option is that these planets are chemically reduced and degas H-rich atmospheric compounds, counteracting atmospheric escape (Section \ref{sec:volatileredistribution}). In either case, these planets are expected to have non-negligible amounts of H$_2$-He in their atmospheres, and therefore a lower mean molecular weight than the previous cases, making them good targets for transit spectroscopy. In Figure \ref{fig:atmospheres-schematic} we show the three possible cases for this category: water-rich planets with large amounts of H$_2$-He in their atmospheres (D1 and D2), and dry planets whose atmosphere is dominated by H$_2$-He with the potential formation of clouds of refractory species (D3).  

\subsection{Thermal structure and spectra of short-period exoplanets}
The thermal structure and spectrum of an exoplanet's atmosphere depends on its composition, the opacity of the different species, the presence or absence of clouds and hazes, the irradiation received from the star, the circulation of its atmosphere, and the interaction of the atmosphere with the interior and surface \citep{Heng2018,2022ARA&A..60..159W}. From a theoretical perspective, in order to obtain the temperature structure in an atmosphere, the radiative transfer equation needs to be solved, which is an energy conservation equation that describes the change in the intensity of radiation traveling through a medium over a certain distance under the effects of absorption, emission and scattering, taking into account both the radiation coming from the planet interior and from the host star \citep{2010ppc..book.....P}. The fact that radiation is travelling in all directions and the potential variety of potential different species presents a complex numerical problem. While there are detailed models that calculate this considering the properties of the different species present in an atmosphere \citep[e.g.,][]{Malik2017}, semi-analytical approximations are also valuable because they allow for fast calculations that can aid retrieval models and are a good first order approximation to understand the problem. Available semi-analytical solutions to the radiative transfer problem \citep[e.g.,][]{Guillot2010, Robinson2012, Heng2014} assume that the radiation field is dominated by two directions (from the star and the planet interior) and usually also make simplifications on the wavelengths of this radiation, assuming grey atmospheres where the maximum of the energy distribution from the stellar radiation is in the visible/UV (shortwave) and that coming from the planet is in the infrared (longwave) \citep{Guillot2010}. For old rocky planets that have escaped their primordial magma ocean phase, the radiation coming from the star is the dominant one and different wavelengths are absorbed at different heights or pressures in the atmosphere: short wavelength radiation is absorbed at low pressures while longer wavelengths can reach the planetary surface. The light absorbed by different molecules heats the atmosphere: species that absorb strongly at short wavelengths can cause heating of the atmosphere at low pressures and might result in temperature inversions; species that absorb strongly in the infrared will cause heating of the planetary surface, i.e., the greenhouse effect (Section \ref{sec:atmospheres1}). 

The presence of molecules dominating atmospheric composition makes the calculation of radiative transfer challenging, as they have hundreds of times more absoprtion lines than atoms \citep{Sharp2007}. Furthermore, many of the chemical constants and parameters needed for these calculations were originally measured at low (room) temperature \citep{Gordon2017} and extrapolating these parameters to the extreme conditions observed in hot exoplanets can introduce significant errors in the calculations. Nevertheless, there is an increasing effort in the community to obtain better absorption and scattering data for calculations in exoplanet atmospheres. Some examples of this are the line-lists calculated from numerical calculations \citep{Tennyson2016}, and increasing efforts to obtain measurements in the laboratory at high temperatures \citep{Rothman2010, Hargreaves2020, Gordon2022}. Furthermore, even laboratory studies on ion chemistry have become increasingly important for rocky exoplanets \citep{Bourgalais2020, Bourgalais2021}, as they might induce the presence of hazes in the atmospheres, suppressing spectral features \citep{Arney2017,2018NatAs...2..303H}. 

For hot rocky planets, a potentially important aspect to determine is the fraction of light that is expected to be reflected from their surfaces and magma oceans. The few observations available \citep{Demory2014, Jansen2018} show values that can be contrasted with laboratory and theoretical predictions \citep{Rouan2011, Essack2020} and can be explained by the presence of exotic magmas \citep{Rouan2011}, mineral silicate-dominated atmospheres \citep{Hamano2015}, \ce{H2O} and \ce{CO2} atmospheres \citep{2019Icar..317..583P} and potentially even waves on magma ocean surfaces \citep{Modirrousta-Galian2021}, but there is a prevalent dichotomy due to the limitations of the data. Laboratory data and theoretical predictions are also being developed for the spectra of rocky planet surfaces \citep{2012ApJ...752....7H} and magmas \citep{Essack2020,Fortin2022} that could aid future observations on potentially airless hot rocky planets. 

\begin{figure*}[tbh]
 	\centering
 	\includegraphics[width=0.99\textwidth]{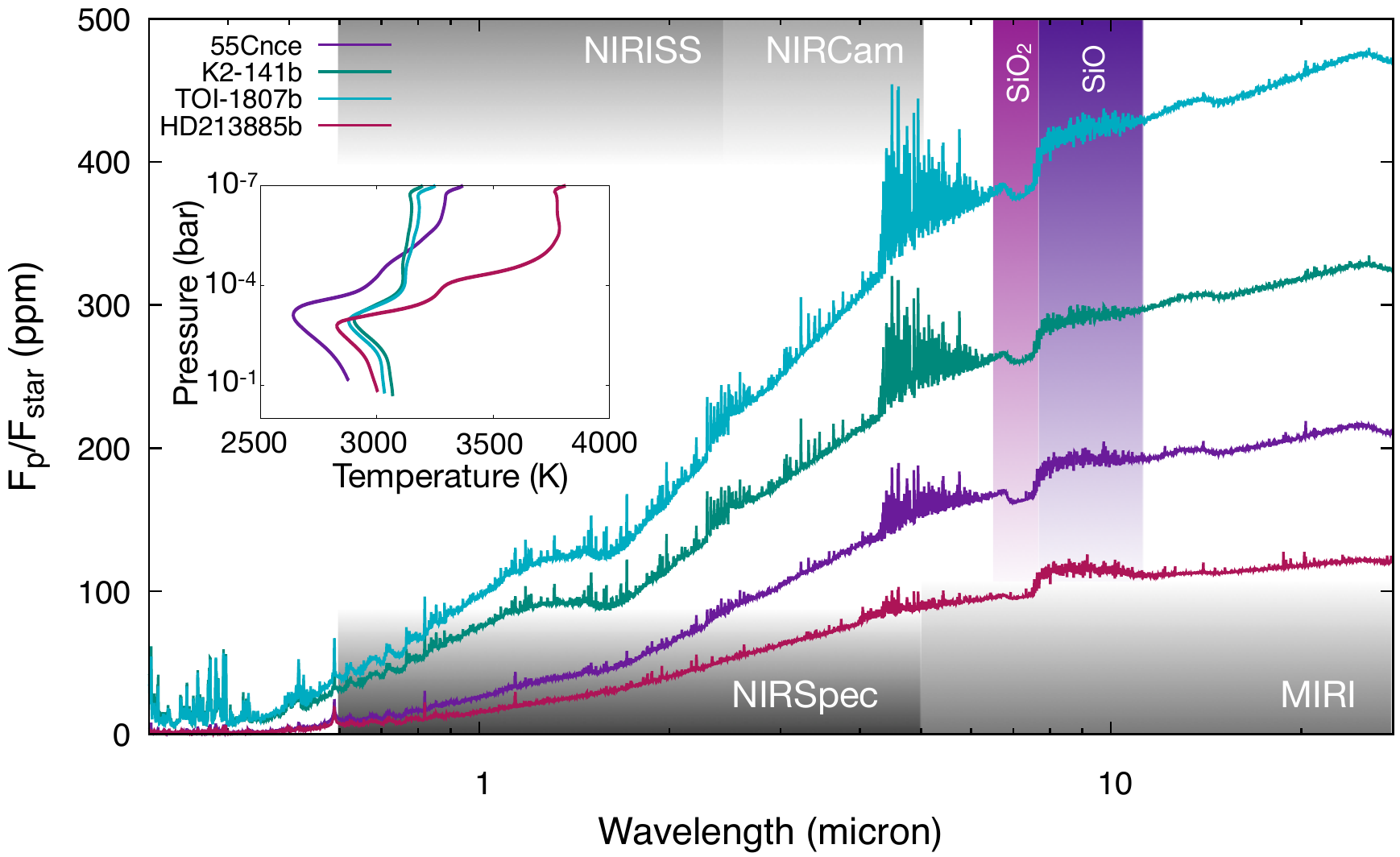}
 	\caption{\textsf{\textbf{Theoretical emission spectra of four hot rocky exoplanets.} The calculations were made assuming 1D atmospheres composed of vaporised lava. The four different instruments of JWST and the regions of interest for each one are indicated as reference. Extracted from \citet{Zilinskas2022} with permission from the authors.}}
    \label{fig:rockvapour}
\end{figure*}

For lava planets with outgassed silicate-dominated atmospheres, theoretical studies show that the presence of molecules that cause strong shortwave absorption could cause temperature inversions in their atmospheres \citep{Zilinskas2021}, as is shown in some cases in the small sub-panel in Figure \ref{fig:rockvapour}. These inversions could be due to molecules such as \ce{CN}, if volatiles are present, but also due to \ce{SiO} in pure vaporised lava atmospheres \citep{Zilinskas2021, Zilinskas2022}. The way that these inversions affect observations is by the presence of emission instead of absorption features in the observed spectra, as can also be seen in the case of the \ce{SiO} features shown in figure \ref{fig:rockvapour} (dark-purple highlight in the spectra). Other features in these atmospheres that can be potentially detectable with JWST are \ce{SiO2} (purple highlight in the spectra). The \ce{SiO} to \ce{SiO2} ratio may be a primary determinant to distinguish the magma redox state \citep{2023ApJ...947...64W} (cf. Section \ref{sec:atmospheres1}).

\subsection{Atmospheric dynamics of short-period exoplanets}\label{sec:3d}  Because dynamics affects the redistribution of energy and the transport of species from one region to another, the circulation of the atmosphere can have a significant effect on the thermal structure and vertical abundances of species in the atmosphere. Additionally, due to the change in the temperature and vertical distribution of the species, dynamics can also affect the opacity of the atmosphere and the emitted spectrum \citep[see review by][]{pierrehumbert2019}. 

Short-period exoplanets are expected to be tidally locked. This means that the tidal stresses from the central star have caused the planet to spin down its rotation to the point where the length of the day equals the length of a year \citep{Barnes2017,pierrehumbert2019}. Because of this, one way of characterizing these planets is to observe the flux of the planet during its entire orbit around the star and obtain its phase curve. These observations are important to observe the entire planet and obtain brightness temperatures, study the redistribution of heat and circulation of the atmospheres. Models on the circulation of short-period rocky planets were developed for exoplanets based on the rocky planets in the Solar System \citep{Lora2018, Kane2022} and by the first phase curve observations \citep{Castan2011, Hammond2017, Nguyen2020, Nguyen2022}. Some of these observations are shown in Figure \ref{fig:phasecurves} for 55 Cnc e \citep{2016Natur.532..207D}, LHS 3844 b \citet{2019Natur.573...87K}, and K2-141 b \citet{2022A&A...664A..79Z}. In Figure \ref{fig:phasecurves}, purple arrows indicate the secondary eclipse during the planetary orbits around their host stars. In some cases, the maximum flux of radiation or hot-spot coincides with the secondary eclipse, as is the case of LHS 3844 b and K2-141 b, but in the case of 55 Cnc e, the maximum flux is not on the substellar point, but it is shifted. This shift was also found in Hot Jupiters \citep[e.g.,][]{2020SSRv..216..139S,2021MNRAS.501...78P}, and is a potential indication that the planet has a thick atmosphere that is re-distributing the heat, a mechanism that was also found with general circulation models, when using \ce{N2} and \ce{CO2} atmospheres \citep{Hammond2017}, although a recent re-analysis of 55 Cnce data shows a much smaller effect than noticed before \citep{Mercier2022}. 55 Cnc e in particular keeps adding new nuances to the multi-dimensional nature of super-Earth atmospheres. Recent observations with CHEOPS and ground-based surveys \citep{2023A&A...677A.112M,2023A&A...669A..64D} provide strong evidence for rapid variation of the planetary signal, suggesting short-term variability. Based on models of the internal geodynamic and magmatic evolution of 55 Cnc e \citep{2023A&A...678A..29M}, potential explanations for the transit variability are intermittent outgassing \citep{2023ApJ...956L..20H} of either volatile or refractory compounds, or the presence of a circumstellar torus of dust driven by radiation pressure and gravity \citep{2023A&A...677A.112M}. The first explanation would fit well with a hydrous magma ocean as explanation for 55 Cnc e's significant underdensity relative to Earth \citep{2021ApJ...922L...4D}. JWST thermal emission observations suggest the presence of a volatile atmosphere on 55 Cnc e \citep{Hu2024}, which lends further support to the hydrous magma ocean interpretation.

\begin{figure*}[tbh]
 	\centering
 	\includegraphics[width=0.99\textwidth]{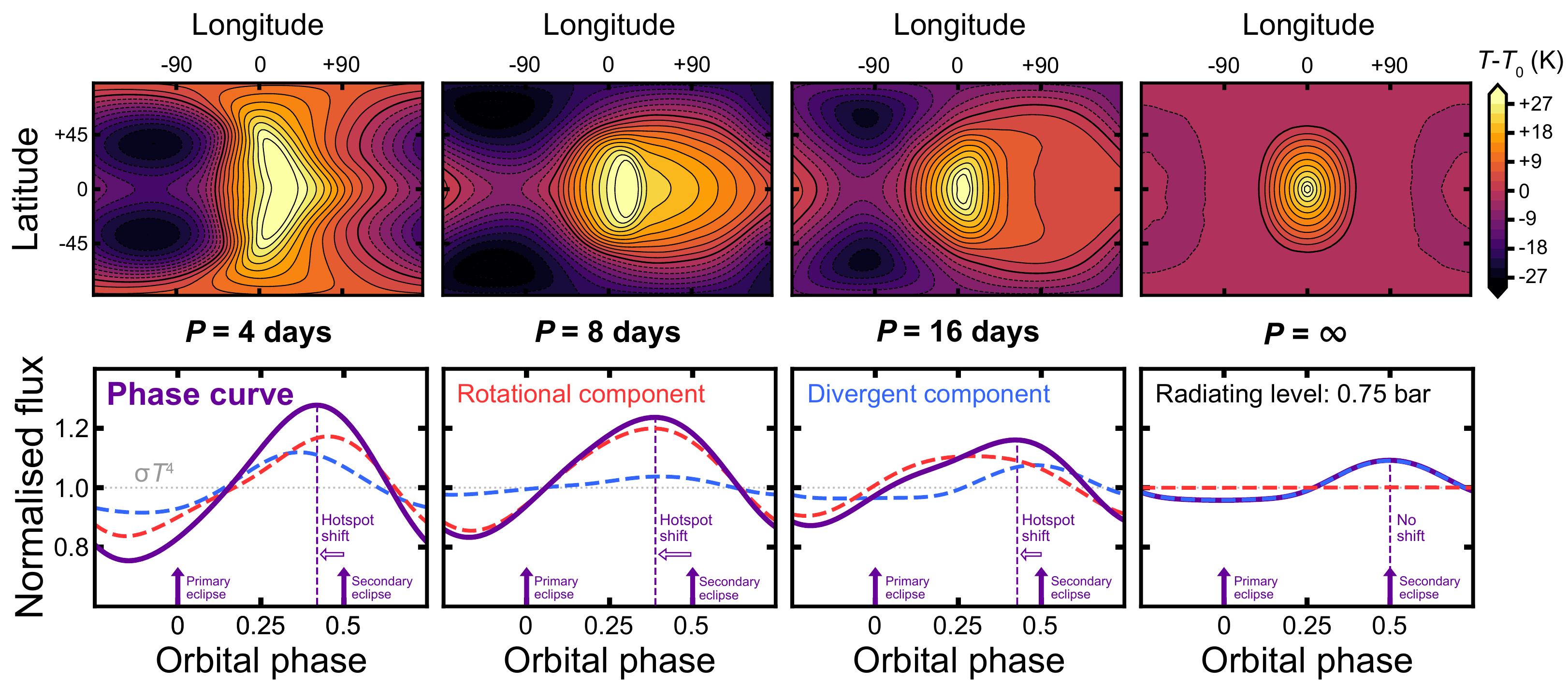}
 	\caption{\textsf{\textbf{Temperature maps from a global circulation model (top) and simulated phase curves (bottom) from broadband thermal emission at a radiating level of 0.75 bar for tidally locked terrestrial planets at various orbital periods.} The temperature field in the top row shows the deviation from global mean temperature ($T_0 \approx$ 280 K, 281 K, 282 K, and 285 K from left to right), illustrating the change in hotspot shift with varying orbital period. The phase curve (solid purple line) is split into its contributions from the rotational (red-dashed: stationary jet and hemispheric eddy circulation) and divergent (blue-dashed: isotropic flow from day- to nightside). Surface pressure, planetary radius, mean molecular weight, and gravity acceleration in the simulation are set to be those of Earth. Simulation data from \citet{2022ApJ...941..171L}, reproduced with permission.}}
    \label{fig:circulation}
\end{figure*}

Figure \ref{fig:circulation} shows an example of theoretical temperature maps (top panels) and phase curves (bottom panels) calculated for a generic rocky exoplanet. Different panels (from left to right) show the results obtained under the assumption that the planet rotates with a period of 4 days, 8 days, 16 days and infinite days correspondingly. In all but the infinite-period case there is a hot-spot shift from the sub-stellar point, which is dominated by rotational standing Rossby waves (red lines in the bottom panels), which are shifted by the zonal jet \citep{Tsai2014, Hammond2018, 2022ApJ...941..171L}. Bottom panels also show the contribution to the phase curve of different circulation components obtained by separating the horizontal velocity into its constituent parts: rotational and divergent \citep{2021PNAS..11822705H,2022ApJ...941..171L}. Fig. \ref{fig:circulation} emphasizes the effects of planetary rotation rate on the atmospheric dynamics, but stellar flux, atmospheric mass, surface gravity, optical thickness, and planetary radius similarly affect the atmospheric circulation and temperature distribution on rocky exoplanets \citep{Kaspi2015,Guendelman2020}. Even parameters such as cloud particle size \citep{Komacek2019} and interaction between the surface and the atmosphere might have an important effect on the circulation and eclipse spectra observed for these planets \citep{May2020}. Because many of these parameters are unknown, all the cited papers exploring the parameter space are relevant to know the uncertainties and the most relevant parameters that we need to obtain from observations in order to better characterize dynamics in these atmospheres.

\subsection{Observational challenges, and the case of planets around M stars}    \label{sec:futureObs}
Condensates can form as clouds and hazes in an exoplanet's atmosphere, and thick layers of these can block light and flatten observed spectra \citep{Fortney2005, Berta2012, Morley2013, Kreidberg2014}. Therefore, studying condensates is crucial for interpreting observations and understanding exoplanet atmospheres. However, not all planets might host clouds, and these can be avoided to some extent by observing the planet in emission spectroscopy that probes deeper in the atmosphere \citep{Lustig-Yaeger2019}. The study of condensates is also important to determine atmospheric stability, and is especially relevant for hot planets, where thin atmospheres may not redistribute heat efficiently, leading to the formation of condensates in colder regions. In some cases, these cold regions can have temperatures even lower than the planet's equilibrium temperature and act as cold traps \citep{Leconte2013, 2022ARA&A..60..159W}. As an example, in the atmospheres of terrestrials planets in the Solar System, there are regions in perpetual shadow that act as cold traps, such as the polar craters on Mercury \citep{Neumann2013}. Cold trapping becomes particularly important for global exoplanetary climate when the main component of the atmosphere is condensing, which can lead to a runaway process and atmospheric collapse on the nightside \citep{1997Icar..129..450J,2011ApJ...726L...8P,Heng2012,Wordsworth2015,2016ApJ...825...99K,2015MNRAS.453.2412C,2016MNRAS.461.1981C,2020A&A...638A..77A}. Solar System examples of this process are the atmospheres of Mars, Triton and Pluto \citep{2022ARA&A..60..159W}, and this might be the case for lava planets with atmospheres dominated by Na, SiO and Mg, which can condense on their nightsides \citep{Kite2016}.  

When considering transiting signals to characterize exoplanet atmospheres, one of the biggest challenges is the signal-to-noise ratio of the planet's atmospheric features, as well as the long observation times required for cooler planets located farther from the star. To overcome these obstacles, the exoplanet community has focused on characterizing rocky planets around small, cool M stars. These stars are the most common in the solar neighborhood, have smaller radii that improve the signal-to-noise ratio, and are cooler, meaning that temperate planets orbit on relatively short orbital periods, on the order of tens of days instead of a year. As a result, nearly all characterized rocky exoplanets orbit these stars. M stars do have some characteristics, however, that may impact their planets' atmospheres to retain volatile envelopes. In particular their high activity levels and UV fluxes, which exceed what is predicted by common stellar models. The frequent and intense flares and winds of these stars may have a significant impact on the planets around them, affecting their atmospheric composition, global climate, and potential habitability \citep{Segura2010, 2023MNRAS.521.3333L, Ridgway2023, 2023MNRAS.521.5880R}. To fully understand the effects that M stars may have on their planets' atmospheres, it is imperative to better characterize their (X)UV emission. The MUSCLES and Mega-MUSCLES treasury surveys have made a significant effort in this regard, observing UV fluxes of many stars that host small planets using the Hubble Space Telescope, including planets whose atmospheres will be priority for characterization with JWST \citep{France2016, Youngblood2016, Loyd2016, 2019ApJ...871L..26F,Wilson2021}.


Despite the challenges faced in characterizing exoplanet atmospheres, the future looks promising for this field. Facilities such as JWST, CHEOPS, and TESS, as well as future missions like PLATO and Ariel \citep{2021arXiv210404824T}, and ground-based telescopes like the Extreme Large Telescope (ELT), promise to provide new data that will revolutionize the field in the coming years. Currently, there are only a few exoplanets with less than 2.5 Earth radii and/or a mass less than 10 Earth masses with atmospheres that have been characterized through either emission or phase curve observations from space (with no conclusive atmospheric features so far found, see Section \ref{sec:observations}). However, the upcoming years of JWST operations will lead to unprecedented data on $\gtrsim$tens of small, rocky exoplanets in this category, some of which will be observed by multiple instruments. The next decade of atmospheric characterization will radically change our knowledge of these planets and help us put the terrestrial planets in our own Solar System into context within the galaxy.

\FloatBarrier
\section{Interior Structure and Dynamics} \label{sec:interiors}

\begin{figure*}[tbh]
 	\centering
        \includegraphics[width=0.99\textwidth]{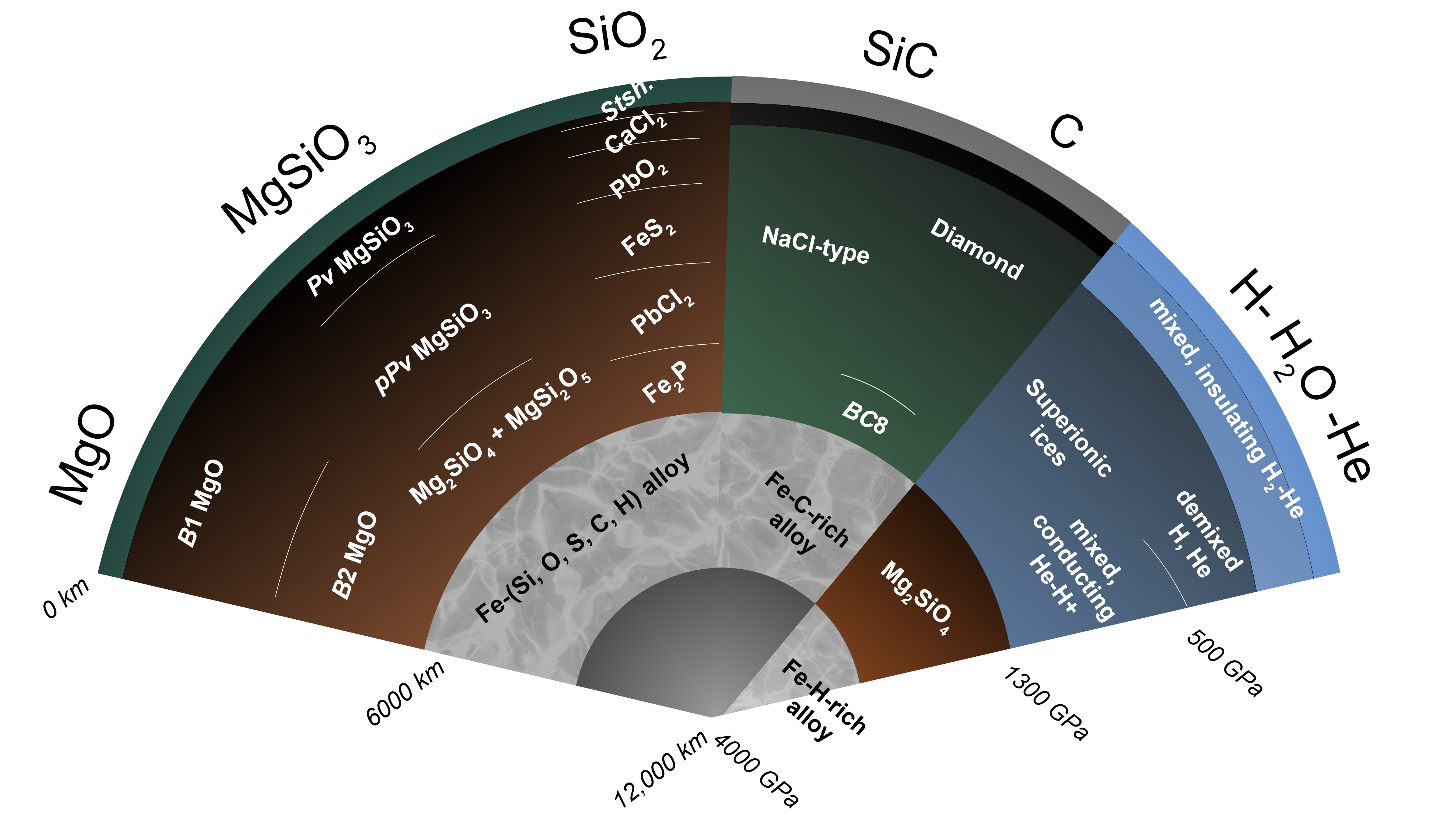}
 	\caption{\textsf{\textbf{Experimentally predicted geochemistry of temperate super-Earth exoplanets with varying elemental composition.} Schematic illustration by Claire Zurkowski; extended from \citet{duffy20152} with experimental data for MgO: \citet{2012Sci...338.1330M,2012PhRvL.108k1101W,2013NatGe...6..926C,2021PhRvB.104a4106H}; \ce{MgSiO3}: \citet{2004Sci...304..855M,2006RvGeo..44.3001H,2017E&PSL.478...40U,2018PhRvB..97u4105F}; \ce{SiO2}: \citet{1995Natur.374..243K,1989Natur.340..217T,1997AmMin..82..635K,2003GeoRL..30.1207M,2005Sci...309..923K,2005PhRvB..71f4104O,2006Sci...311..983U,2011PNAS..108.1252T,2011PhRvB..83r4102W,2011PhRvL.107d5701D,2011PhRvB..83m4114D}; \ce{SiC}: \citet{1997PhRvB..55.8034S,1993PhRvB..4810587Y,2018JGRE..123.2295M,2022NatCo..13.2260K}; C: \citet{2014PhRvB..89v4109B,2019AsBio..19..867H}; H--\ce{H2O}: \citet{2019A&A...621A.128M,2020A&A...643A.105H,2019Natur.569..251M,2021Natur.593..517B}.}}
    \label{fig:geochemistry}
\end{figure*}

So far we have discussed the physical and chemical processes operating in exoplanets with a focus on atmospheric volatiles and their processing in exoplanetary atmospheres. However, as outlined in Section \ref{sec:atmospheres1}, the atmosphere and interior of super-Earth exoplanets are in tight connection with each other following accretion. After the initial magma ocean epoch is over, atmosphere and interior continue to energetically and chemically equilibrate, but on longer timescales, which are primarily dependent on the thermodynamic and phase state of the planetary interior. We here define 'interior' as any planetary layer that is not mainly gaseous. This goes from the deep interior of volatile-rich super-Earths, which may host high-pressure ice phases, to the rocky mantles and metal cores of potentially Earth-like worlds. Composition, structure, and dynamics influence each other, but qualitatively different regimes may emerge if exoplanets are very rich ($\gtrsim$ wt\% levels) in volatiles, compared to when they are similarly depleted in H-C-N-S compounds as the terrestrial planets of the Solar System. In this section we focus on the interior mineralogy and structure of solidified exoplanets (Fig. \ref{fig:geochemistry}) and the geodynamics of dominantly rocky exoplanets (Fig. \ref{fig:geodynamics}). These two figures illustrate key concepts of the following discussion.

\subsection{Volatile-rich super-Earths}

At the time of writing, observational data on the existence of extrasolar 'ocean planets' \citep{2003ApJ...596L.105K,2004Icar..169..499L,2007Icar..191..337S} -- Earth-sized or super-Earth-sized planets with $\gtrsim$ 1 wt\% levels of water and other volatiles but without a massive hydrogen-dominated atmosphere -- is inconclusive \citep{2022ApJ...933...63N,2023ApJ...947L..19R}. The principal problem behind this is that most current observational data is limited to mass and radius, without compositional information from, for example, transit spectroscopy. All attempts of obtaining direct compositional information from exoplanets in the density-size regime compatible with Earth-like compositions has been hampered by observational limitations (Section \ref{sec:observations}). With mass and radius being the only two constraints, the mean density of an exoplanet can be fitted with a variety of interior structure models \citep[e.g.,][]{2008ApJ...673.1160A,2010ApJ...712..974R,2015A&A...577A..83D,2017A&A...597A..37D,2022MNRAS.513.5256H,2023ApJ...944...42U,2024A&A...681A..96H}, prediciting a maximum range of pressures and temperatures of $\approx$2400 GPa and $\approx$6500 K, respectively \citep{2007Icar..191..337S}. However, as outlined in sections \ref{sec:observations} and \ref{sec:formation}, several key predictions from planet formation theory \citep[e.g.,][]{2019NatAs...3..307L,2019PNAS..116.9723Z,2020A&A...643L...1V,2022ApJ...939L..19I} hint toward the likely existence of these worlds, and observational surveys close in on their distribution and prevalence in exoplanetary systems \citep{2022Sci...377.1211L,Diamond-Lowe2022,2023NatAs...7..206P,2023AJ....165..167C,2024MNRAS.52711138O}.

From a Solar System point-of-view, the existence of such ocean worlds is uncontroversial: all minor planetary bodies of the outer Solar System, such as Ganymede, Europa, or Pluto, are dominated by mantles primarily composed of volatile ices. However, super-Earths with such large volumes of volatiles ices would experience much higher pressures and temperatures than the icy planetary bodies of the outer Solar System. Often \ce{H2O} is discussed as the primary volatile because of its cosmochemical abundance and phase properties, which make it the dominant contributor to greenhouse warming in hot atmospheres (Section \ref{sec:atmospheres1}), and to atmospheric albedo in colder planetary environments. The theoretical maximal water content of volatile-rich worlds is about 50 wt\%, the mixing ratio of water to other species beyond the snowline in a Solar-composition protoplanetary disk \citep{2003ApJ...591.1220L}. At such extreme water-to-rock ratios (Fig. \ref{fig:geochemistry}: \ce{H}-\ce{H2O}-\ce{He}), the pressure in the interior very quickly solidifies the volatile ices with increasing depth and reaches temperatures and pressures that are dominated by high-pressure forms of water ice \citep{2019AdPhX...430316V,2020SSRv..216....7J}. In a chemically-differentiated body -- which is the typical assumption of exoplanet structure models \citep{2007Icar..191..337S,2007ApJ...659.1661F,2007ApJ...669.1279S} -- the high-pressure phase physically separates the underlying mantle from the overlying atmosphere. At mantle conditions, ice VI occurs between $\approx$0.6--2.2 GPa, ice VII until $\approx$70 GPa. Beyond that, relevant for the deep mantles of ice-rich super-Earths, insulating and superionic states of ices form (ice X, ice XVIII), which have extreme viscosity and behave like a solid \citep{2019Natur.569..251M}.  As is outlined in Section \ref{sec:interior_cycling},  climate stability, stable temperatures, and the chemical availability of nutrients may be important for the potential habitability of exoplanet surfaces. Therefore, densely-packed ice phases potentially inhibit these conditions \citep{2016Icar..277..215N,2018ApJ...864...75K}. However, solid-state convective transport in the ice may overcome this barrier, exchanging electrolytes between solid deep and shallow liquid ocean layers \citep{2018ApJ...857...65L}, which finds tentative experimental support \citep{2022NatCo..13.3304J,2022NatCo..13.3303H}.

If no additional pressure from an overlying H-He-rich atmosphere (like in sub-Neptunes) is created, comparison of the phase diagram of \ce{H2O} with mantle adiabats suggests that $\gtrsim$3 Gyr-old exoplanets on temperate orbits are most often solidified \citep{2014ApJ...784...96Z}. In such a scenario, the effect of internal temperature variations on structural models with similar compositional assumptions is minor \citep{2007ApJ...670L..45V,2007ApJ...669.1279S}. However, as illustrated in Figure \ref{fig:M-R}, the majority of currently known super-Earths are subject to high irradiation from their host star, potentially driving these planets into a runaway greenhouse regime (Section \ref{sec:atmospheres1}, Figure \ref{fig:radiationlimits}). For volatile-rich planets in a runaway greenhouse, the molten interior can dissolve large ($\gg$ 1 wt\%) quantities of volatiles, in particular water \citep{2021ApJ...922L...4D}, and under reducing conditions nitrogen \citep{2024ApJ...962L...8S} and sulfur \citep{2016E&PSL.448..102N} compounds. For a recent comprehensive review on volatile partitioning into melt and metal phases in planetary interiors see \citet{2023FrEaS..1159412S}. Phase transitions in the interior of volatile-rich exoplanets can therefore mask substantial compositional variations. While there is certainly room for detailed experiments on solid-state phases of volatile-rich compositions, a major critical question surrounding the composition and evolution of observable super-Earths is therefore the admixture of atmospheric volatiles into melted planetary interiors, and the transition of molten planetary regimes toward partial crystallization. In order to better understand this problem, we will require compositional information on a diverse set of super-Earths in the critical irradiation-size regime. Possible candidates to provide observational insight into this question are, among others, TOI-561 b, HD 3167 b, 55 Cnc e, K2-131 b, and the Kepler 138 and K2-3 systems. From an experimental perspective, it will be crucial to investigate the dissolution of atmospheric volatiles into magmas at increasing pressures and diverse compositions, for example with varying Fe/Si and Mg/Si ratios \citep{2010ApJ...715.1050B} and redox states (Section \ref{sec:atmospheres1}). On the path toward understanding the interior and climate conditions of lower-mass planets, sub-Neptune exoplanets have a crucial role to play. 

Sub-Neptunes that straddle the boundary between super- to sub-runaway climates, such as K2-18 b \citep{Innes2023,2024ApJ...963L...7W}, may either host metallic \citep{2019Natur.569..251M} or supercritical \citep{2021ApJ...914...84A,2023ApJ...944...20P,2024arXiv240303325B} water phases in their deep interior, depending on the thermal evolution of H-rich planets after their formation. At supercritical conditions, gases can become fully miscible \citep{2014Life....4..331B,2016ScChD..59..720N,2017SciA....3E0240P}, suggesting that supercritical sub-Neptune interiors can, in essence, take up substantial amounts of water in their deep interiors. Both the transition from high-pressure ices to liquid water at their atmosphere-interior interface \citep{2021MNRAS.505.3414N,2019ApJ...887..231L,2021ApJ...914...38Y,2021ApJ...921L...8H,2021ApJ...922L..27T} and the gas to supercritical to magma transition have been proposed to have observable diagnostics \citep{2024ApJ...962L...8S}. Recent high-pressure experiments confirm that at temperatures above the melting point of \ce{MgSiO3} water and rock become miscible \citep{2022ApJ...926..150V,2022NatSR..1213055K}, thus questioning the validity of simplified shell models for the structural modeling of water-rich sub-Neptunes \citep{2024ApJ...962L...8S} and super-Earths \citep{2021ApJ...922L...4D}.

\subsection{Rocky super-Earths and Earth-sized exoplanets}

So far it is unclear what constitutes the boundary between 'rocky' and other potential compositions and interior structures of super-Earths and Earth-sized exoplanets. In the past years, therefore, guidance has been found in using the Earth and terrestrial planet-like materials as a starting point, and investigate compositional and dynamical variations when the system parameters are changed. Comparison with direct observations are non-existent to sparse aside from the TRAPPIST-1 planets, which are robustly constrained to be underdense relative to a pure Earth-like composition. However, from an experimental and modeling perspective, key results have emerged that can guide telescopic exploration. We distinguish here developments in three directions: compositional (\ref{sec:interior_structure}) and dynamical (\ref{sec:interior_dynamics}) aspects of the interior, and how these interplay (\ref{sec:interior_cycling}) with the climate system on geologic timescales.

\subsubsection{Interior structure and composition} \label{sec:interior_structure}

As a reference point, the Earth's mantle extends to a depth of $\approx$2900 km, with a pressure of $\approx$135 GPa. The temperature and pressure at the center of the Earth are $\approx$5500 K and $\approx$365 GPa, respectively \citep{nimmo2015energetics}. Pressures and temperatures inside solidified rocky super-Earths ($\leq$10 M$_{\rm{Earth}}$) reach up to $\approx$10000 K and $\approx$4 TPa \citep{2012A&A...541A.103W}. At such extreme pressures, nominally Earth-like mantle materials undergo phase transitions that are unknown from our own planet, but can be explored using synthetic materials in high-pressure laboratories. A particular challenge hereby is that the elemental abundances of refractory materials (such as Mg, Si, or Fe compounds) are likely fractionated from their terrestrial counterparts. The elemental abundances of refractory (but not volatile) species in the Sun and planetary materials are closely related, which is similarly expected for rocky exoplanets \citep{2010ApJ...715.1050B,2012ApJ...758...36M,2022AJ....164..256H,2019MNRAS.482.2222W,2022MNRAS.513.5829W,2023ApJ...944...42U}. The composition of Earth is thus dominated by O, Fe, Mg, and Si, with some amounts of Ca and Al \citep{1995ChGeo.120..223M}, segregated into a structure of Mg-rich silicates and oxides on top of a metallic, Fe-rich core. Similarly to the modeling of volatile-rich planets, structural models of super-Earth and Earth-like exoplanets employ equations of hydrostatic equilibrium with a fixed composition in each layer and a symmetric planet
\citep[e.g.,][]{2006Icar..181..545V,2007ApJ...656..545V,2007ApJ...669.1279S,2007ApJ...659.1661F,2012A&A...541A.103W,2016ApJ...819..127Z,2015A&A...577A..83D,2023ApJ...944...42U}. The temperature distribution in the interior is usually assumed to be adiabatic. To relate temperatures and pressures to density, empirically determined material properties are fitted to an equation of state model (EOS), such as the Birch-Murnaghan or Vinet equations.

To start with the innermost part, rocky exoplanets will contain a core of metallic alloy due to chemical differentiation during and after planetary accretion (Section \ref{sec:atmospheres1}). Under high pressure, solid iron and its alloys undergo several phase transitions, from body-centered-cubic (bcc), to hexagonal-close-packed (hcp), and face-centered cubic (fcc) \citep{2018Icar..313...61H,2018SciA....4.5864W,2020Mine...10..100M,2017NatSR...741863D}. The melting curves of iron alloys at core conditions \citep{2013Sci...340..464A,2020Mine...10...59I} depend critically on the presence of light elements, such as H, S, O, Si, and others \citep{2014RSPTA.37230076S}. In addition to Fe-rich metals, the Earth's core contains several wt\% of light elements, best recent estimates suggest the likely range of compositions for the outer core to be: Fe + 5\% Ni + 1.7\% S + 0–4.0\% Si + 0.8–5.3\% O + 0.2\% C + 0–0.26\% H by weight; and for the inner core: Fe + 5\% Ni + 0–1.1\% S + 0–2.3\% Si + 0–0.1\% O + 0–1.3\% C + 0–0.23\% H by weight \citep{hirose2021light}, with a recent community debate on whether the light element budget leans more toward an O- \citep{2013Sci...339.1194S,2015PNAS..11212310B}, Si- \citep{2015GeCoA.167..177F,2015Icar..248...89R}, or H-dominated \citep{2020GeoRL..4788303Y,2020NatGe..13..453L,2021NatCo..12.2588T,2022GeoRL..4996260T} composition. Light element incorporation is important in the context of planet formation models \citep{2023ASPC..534.1031K,2023ASPC..534..907L}, as the significant abundance of H in the core provides strong evidence for (i) accretion of the Earth in the presence of a hydrous magma ocean \citep{2018SSRv..214...76I,2023Natur.616..306Y}, and/or (ii) rapid incorporation of water-rich planetesimals during the initial accretion phase \citep{2021Sci...371..365L,2021NatGe..14..369G,2024NatAs.tmp...13G}, suggesting early chemical equilibration between core, mantle, and (proto-)atmosphere. 0.26 wt\% H in the outer core corresponds to $\gtrsim$28 Earth oceans that are locked up in the core, a number that dwarfs the amount of water in the mantle (3--10 oceans) and on the surface (1 ocean) \citep{2017SSRv..212..743P,2021AGUA....200323D}. Planetary cores start out liquid and solidify over time due to the core adiabat intersecting the melting curves of iron alloys (Fig. \ref{fig:geochemistry}). The energetics of this evolution is discussed in Section \ref{sec:interior_dynamics}.

Inside the mantle of Earth the minerals undergo a series of phase transitions, with an important discontinuity at 660 km depth, forming perovskite, (Mg,Fe)\ce{SiO3}, and ferropericlase, (Mg,Fe)O. \ce{MgSiO3} further transforms to postperovskite in the lower mantle, leading to a composition of about 70\% Mg-perovskite, 20\% ferropericlase, and 10\% Ca-perovskite in the lower mantle. The internal mineralogy of super-Earths will differ depending on composition, pressures, and temperatures, forming solid solutions of refractory species. From an experimental point-of-view, the main constituents of such mantles are understood, even though major uncertainties remain with regards to physical properties, dissolution of light elements within individual mantle and core phases, and the thermal evolution of the planet. For providing a baseline of the experimentally verified mantle compositions, we here discuss the major phases of simple oxides and silicates that have important implications for density, electrical conductivity, and mechanical strength, following \citet{duffy20152}, including recent updates on important mantle constituents (Fig. \ref{fig:geochemistry}).

For the mantle, Mg/Si, Fe/Si, and C/O ratios are important characteristics that change the internal mineralogy, and thus physical properties that influence volatile cycling and geodynamic regimes. Modeling suggest that varied Mg/Si ratios during planet formation  from different stellar abundances translate directly into planetary compositions \citep{2012ApJ...760...44C,2012ApJ...747L...2C}, changing the relatve fraction of pyroxene, olivine, and feldspars in planetary mantles. Mg/Si thus may yield a first-order indication of mantle mineralogy, with high-Mg/Si stars leading to weaker, ferropericlase-rich mantles, and low-Mg/Si stars leading to mechanically stronger mantles \citep{2023ApJ...948...53S}. However, from the Solar System itself it is clear that this relation is not directly translatable, and intra-system variation is found on meteorites and asteroids \citep{2002Natur.416...39D,2019E&PSL.507..154L,2020GeCoA.277..334C}. Fe/Si is expected to directly influence metallic core sizes \citep{2023ApJ...948...53S}, which finds tentative (and contested) observational evidence \citep{2021Sci...374..330A}. The C/O ratio is most easily affected by differences in planetary formation scenarios and mantle-core partitioning of light elements \citep[e.g.,][]{2005astro.ph..4214K,2010ApJ...715.1050B,2020PNAS..117.8743F,2021ApJ...913L..20L}, but photospheric abundances of exoplanet host stars similarly show a wide spread in C/O \citep{2021A&A...655A..99D}.

Periclase \citep[MgO,][]{2012Sci...338.1330M,2012PhRvL.108k1101W,2013NatGe...6..926C} has one important phase transition from B1 to B2 at around 400--600 GPa, where MgO transforms from a rock salt to a CsCl-type structure. The melting curve of MgO, however, has been under substantial debate. Recent shock compression experiments indicate that MgO is likely solid in the deep mantle of super-Earths \citep{2021PhRvB.104a4106H}. As discussed above, \ce{MgSiO3} \citep{2004Sci...304..855M,2006RvGeo..44.3001H,2017E&PSL.478...40U} is present in the Earth's mantle in the perovskite and postperovskite forms, with the latter likely playing a dominant role in setting the geophysical parameters in the interior of rocky super-Earths \citep{2007ApJ...670L..45V,2012A&A...541A.103W,2013Icar..225...50T,2020JGRE..12506124B}. Possibly further phases beyond $\approx$500 GPa pressures, such as formation of \ce{Mg2SiO4} and \ce{MgSi2O5} is predicted by theoretical calculations. Recent experimental results are supporting the stability of \ce{Mg2SiO4} at super-Earth mantle conditions \citep{2022JGRE..12707344Z}, which is suggestive of an oxidative trend with increasing mantle depth. While silica are the most abundant oxide in Earth's mantle, \ce{SiO2} is only prevalent in localized regions. However, wider ranges in $P$--$T$ for super-Earths and fractionation in refractory species make it seem likely that \ce{SiO2} can become a dominant mineral in rocky exoplanetary mantles. With increasing pressure, \ce{SiO2} undergoes several phase transitions \citep{1995Natur.374..243K,1989Natur.340..217T,1997AmMin..82..635K,2003GeoRL..30.1207M,2005Sci...309..923K,2005PhRvB..71f4104O,2006Sci...311..983U,2011PNAS..108.1252T,2011PhRvB..83r4102W,2011PhRvL.107d5701D,2011PhRvB..83m4114D}, including quartz, coesite, stishovite, and further high-pressure phases. Variations in C/O can lead to very carbon-rich planets, dominantly made of \ce{SiC} \citep{1997PhRvB..55.8034S,1993PhRvB..4810587Y,2018JGRE..123.2295M,2022NatCo..13.2260K} or pure diamond \citep{2014PhRvB..89v4109B,2019AsBio..19..867H}.

These mineralogical phase transitions in the deep mantles of super-Earths likely have a small effect on mass-radius relations of temperate, solidified worlds \citep{2019JGRE..124.1704U}, however, they crucially affect geophysical parameters, such as the melting temperature, thermal expansivity, the Grüneisen parameter, thermal and electrical conductivities, and rheological properties \citep[e.g.,][]{2023Mine...13..885G,2023PhRvR...5c3194G} that affect the geodynamics and melting of rocky exoplanets to first order. In the following, we will discuss some of these aspects.

\subsubsection{Interior dynamics} \label{sec:interior_dynamics}

\begin{figure*}[tbh]
 	\centering
 	\includegraphics[width=0.99\textwidth]{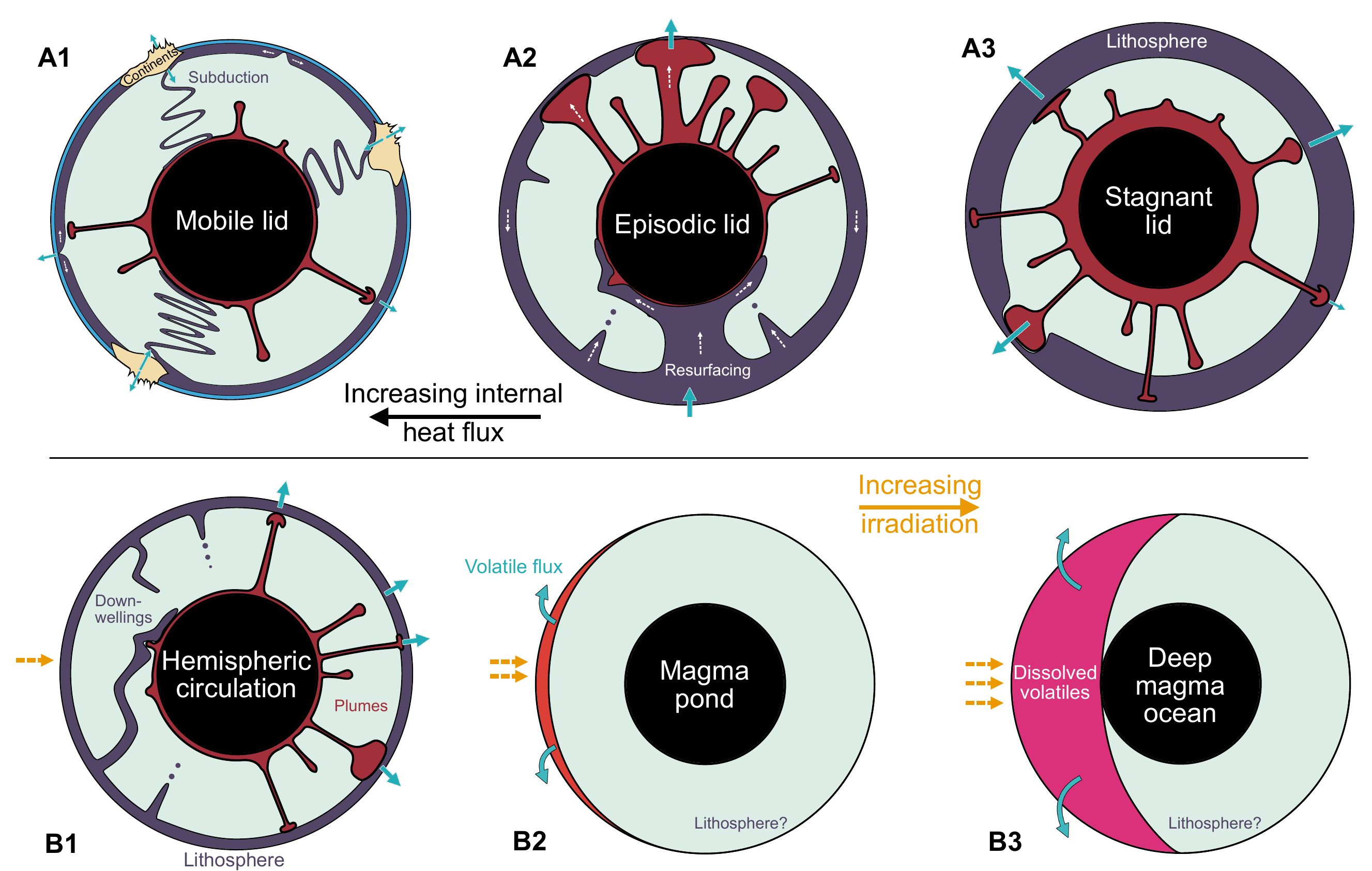}
 	\caption{\textsf{\textbf{Interior dynamics of Earth-like exoplanets on temperate orbits (A1--3) and tidally locked rocky exoplanets (B1--3).} The indicated regimes illustrate the main drivers of lithospheric renewal and thickening, and the main volatile fluxes. Description of the scenarios and physical processes can be found in the main text.}}
    \label{fig:geodynamics}
\end{figure*}

The thermal evolution of super-Earths and Earth-like exoplanets will depend on the material properties in the deep interior, as well as ambient stellar and atmospheric conditions. In particular the mantle viscosity -- its resistance against shear stress -- is sensitively affected by the above described mineralogical properties. At mantle pressures, viscosity is dominated by the thermally-driven disclocation of defects in the crystal structure, which is typically described using an Arrhenius-type functional form \citep{hirth2003rheology,2012A&A...541A.103W}, but strongly dependent on the exact solid solution and conditions in the mantle. Experimental and modeling uncertainties therefore have generated a diverse array of proposed geodynamic regimes of rocky exoplanets that differ qualitatively from the regimes present in the modern Solar System. These dynamical regimes differ widely in associated chemical exchange across boundary layers, such as between the core and mantle, and mantle and atmosphere, including thermal transport properties, which ultimately drives the cycling of volatiles and thus climate evolution. Planetary heat flux also affects mantle and core crystallization and therefore the generation of a dynamo.

Fig. \ref{fig:geodynamics} illustrates end-member cases of interior geodynamic regimes that may govern the long-term evolution of rocky exoplanets. Sub-figures A1--3 focus on Earth-like regimes with high rotation rates, while sub-figures B1--3 display tidally-locked regimes, which may govern the global tectonics of short-period exoplanets. To set the stage, we will first discuss exoplanets with high rotation rates and at more temperate conditions, i.e., further away from their central star (A1--3). A1 illustrates the geodynamic regime of an Earth-like planet: the surface is divided in a number of plates that are vertically displaced relative to each other by horizontal motion, a mobile lid or plate tectonics regime. The stiff plates are generated at divergent margins (mid-ocean ridges) and are subducted into the mantle at convergent margins. The continents are held up at the surface due to buoyancy. In this geodynamic regime the heat flux across the lithosphere, the uppermost stiff part of the mantle, is high, as heat can escape by convection. In addition, volatiles can actively cycle between mantle and atmosphere due to continuous exchange through the surface layer. A3 showcases the most dynamically inactive geodynamic regime, a stagnant lid. In this regime the internal parameters are not sufficient to ablate or break the upper crust and heat transport is governed mainly by conduction through the lid. A2 represents an intermediate, episodic regime, where planets may cycle between phases of stagnant and mobile lid tectonics. Importantly, these three regimes are not linearly related to specific parameters, but a flurry of additional convective regimes and transitions between them are possible. Other recent developments --  especially relevant for Venus past and present and early Earth, include plume-lid or squishy-lid regimes \citep{2015Natur.527..221G,2017NatGe..10..349D,2018NatGe..11..322L}, and heat-pipe tectonics \citep{2015GeoRL..42.9255M,2017E&PSL.474...13M}, which is represented by Io in the current Solar System 

The discovery of the first exoplanets in the super-Earth regime in the middle- to late 2000s triggered a theoretical debate on the likelihood of plate tectonics among super-Earths \citep[e.g.,][]{2007ApJ...670L..45V,2007ApJ...665.1413V,2007GeoRL..3419204O,2009ApJ...700.1732K,2010ApJ...725L..43K,2011Icar..212...14K,2011E&PSL.310..252V,2012ApJ...748...41S,2012E&PSL.331..281F,2012ApJ...755..132L,2013Icar..225...50T,2014P&SS...98...41N,2016GeoRL..43.9469W,2016PEPI..255...80O}. Plate tectonic motion on the surface of the Earth is crucially important for volatile fluxes and likely affects long-term surface habitability via the carbonate-silicate cycle (Section \ref{sec:interior_cycling}). In the absence of data, theoretical models focused on scaled-up versions of the Earth, with explorations and extrapolations of different mechanisms, such as the influence of water and mantle viscosity, the internal heat budget due to variations in radioactive elements, tidal forcing, changes in viscosity due to different mineralogy, or increased internal convectional vigor due to larger Rayleigh numbers in super-Earths \citep[e.g.,][]{2014ApJ...789...30H,2015AsBio..15..739D,2024ApJ...961...22S}, not reaching convergence \citep{2018RSPTA.37670416L}. A major uncertainty in modeling mobile lid planets and thus Earth-like volatile cycles is the very onset of subduction, which may require special circumstances \citep[e.g.,][]{2018Tectp.746..173S,2021PreR..359j6178K}.

In the Solar System, Earth is the only planet that undergoes mobile lid tectonics at present day, all other planetary bodies exhibit a form of geodynamics closer to a stagnant lid. Therefore, stagnant lid geodynamics may be a better approximation for exoplanets, and recent modelling approaches have developed in this direction \citep{2017A&A...605A..71T,2019A&A...625A..12G,2018AsBio..18..873F,2019ApJ...875...72F,2017Natur.545..332R,2018NatGe..11..322L,2017PEPI..269...40N,2018A&A...614A..18D}. Particular emphasis has been placed on the interaction between the planetary interior and atmosphere, which on Earth is enabled mainly by plate tectonics. Aside from the effect on volatiles, variations in internal composition may affect other geophyiscal parameters, such as core evolution. For instance, highly carbon-enriched planets with a graphite shell have been suggested to be highly conductive, loosing heat much faster than a representative Earth-like composition in a stagnant lid regime \citep{2019A&A...630A.152H}. Varying levels of radiactive heating, e.g., due to declining rates of stellar nucleosynthesis over galactic evolution, will have an intrinsic effect on planetary heat production, thus changing the vigour of convection in the planetary interior \citep{2011E&PSL.310..252V,2014Icar..243..274F,2015ApJ...806..139U}.

On fast rotating planets on temperate orbits, the geodynamic mode governs heat flux through the planetary surface. Planets with more mobile regimes experience higher heat flux, while those closer to stagnant lid regimes experience lower heat flux \citep{nimmo2015energetics}. This is important for the thermal and compositional evolution of exoplanetary cores \citep{2020JGRE..12506124B,2021JGRE..12606724B} because crystallization of an inner core (as in the Earth) is driven by cooling. It is sometimes stated that rotation rate is the dominant factor governing magnetic field generation, however, in the Earth the magnetodynamo is primarily governed by inner core crystallization \citep{2015PEPI..247...36L}. Based on a combination of geodynamic modeling and mineral physics data \citet{2010ApJ...718..596G} and
\citet{2013JGRE..118..938V} suggested that super-Earths do not form an inner core. However, recent experimental determination of the melting curve at super-Earths pressures refutes this argument \citep{2022Sci...375..202K}. Magnetodynamos can also be generated by strong convection in very viscous mantle layers, such as internal magma oceans \citep{2021JGRE..12606739B,2022ApJ...938..131Z}, or through compositional effects. The latter requires either liquid metal enriched in light elements due to the solidification of pure Fe at the inner core boundary, which can generate convective motion in the outer core due to their compositional buoyancy. Alternatively, Fe-rich liquid can also sink from the core-mantle boundary to the inner core due to light element exsolution, dominated by MgO  \citep{2016Natur.529..387O,2016Natur.536..326B} or \ce{SiO2} \citep{2017Natur.543...99H}.

In the last few years, advances in telescopic observations of exoplanets have started to reveal details on individual super-Earths, such as contraints on the surface temperature and likely presence of an atmosphere (sections \ref{sec:observations} and \ref{sec:atmospheres2}). If (some) ultra-short period super-Earths and terrestrial-sized planets indeed have no volatile envelope \citep[e.g.,][]{2019Natur.573...87K,2022A&A...664A..79Z,2022ApJ...937L..17C,Zieba2023,Greene2023,2024ApJ...961L..44Z}, this enables direct constraints on surface geochemistry and potentially internal dynamics. For example, geodynamic models of tidally-locked regimes suggest hemispheric flow patterns hat redistribute magma and volatiles qualitatively different than in the regimes discussed above \citep{2011ApJ...735...72G,2011ApJ...736L..15V,2012Sci...338.1330M,Kite2016,2021ApJ...908L..48M,2023A&A...678A..29M}. Figures \ref{fig:geodynamics} B1--3 illustrate these tidally-locked regimes. A1 showcases a cooler scenario, with an atmosphere-less planet that is tidally-locked, but not molten. For such a scenario with an Earth-sized planet, \citet{2011ApJ...736L..15V} found that a hemispherically-split geodynamic regime (degree-1 convection) would take place, with the antistellar side of the planet undergoing thickening of the lithosphere and downwellings, and the opposite on the hotter substellar side. However, \citet{2021ApJ...908L..48M} modelled the specific parameters of the super-Earth LHS 3844 b with temperature maps inferred from the Spitzer phase curves from \citet{2019Natur.573...87K}, and found that both dayside and nightside can in principle undergo up- or downwellings in a hemispheric geodynamic regime, depending on the yield stress parameter of the lithosphere. The yield stress parameterises the strength of the lithosphere to break and participate in foundering or subduction and is, among other parameters, affected by the water content of the upper mantle. \citet{2020PSJ.....1...36K} argued for a volatile-poor formation of LHS 3844 b, but exclusively modelled the planet to be in a symmetric stagnant lid regime and based their conclusions on non-thermal atmospheric escape processes, thus not recovering the hemispheric flow patterns observed in multi-dimensional geodynamic models. In general, it is unclear if such super-Earths are rigidly tidally-locked or undergo true polar wander \citep{2015Sci...347..632L,2018NatGe..11..168L,2017A&A...603A.108A}, which would rotate their dayside over time. In the upcoming years this question may be resolved by constraining the surface mineralogy of ultrashort-period super-Earths \citep{2024ApJ...964..152L}, as tidally-locked and rotating dynamics should evolve different surface mineralogy. More advanced, dimensionally-resolved models of specific exoplanets are required that can be compared against upcoming JWST measurements of surface spectroscopy and phase curve data.

Hotter regimes of tidally-locked exoplanets (Fig. \ref{fig:geodynamics} B2--3) then give us novel access to the distribution of molten and vaporized mantles. The latter is described in Section \ref{sec:atmospheres2}, and here we focus on atmosphere-less planets with a surface temperature  $\lesssim$2000 K. In this thermal regime, the surface is not mainly vaporized, but in liquid phase state, potentially close to the rheological transition (Section \ref{sec:magmaocean}), thus yielding insights into the fluid dynamics of planetary magma oceans, opening a novel, observationally-driven field: planetary magma oceanography. An important question is the depth and geometric extent of a dayside magma ocean, which sensitively depends on the composition of the mantle in volatiles and refractory species, and the thermal trajectory and fluid dynamics of the magma ocean. For instance, \citet{2011ApJ...735...72G} suggested dayside magma oceans to be shallow (ponds), but \citet{2022ApJ...936..148B} argued for a deep mantle extent of dayside magma oceans based on a different parameterization of the mantle melting curve \citep{2010Sci...329.1516F}. \citet{2023A&A...678A..29M} demonstrated for the specific of 55 Cnc e that this discrepancy is exacerbated by choices related to the parameterisation of eddy diffusion in turbulent convection. In particular, the flow regime of lateral convection (where thermal and gravitational forces are misaligned with an angle other than 180$^\circ$; in Earth-like buoyancy-driven convection thermal and gravitational force are exactly anti-aligned). Since liquid magma can store orders of magnitude more water than solids (Section \ref{sec:atmospheres1}), the difference between magma ponds and deep hemispheric magma oceans has a bearing on the survival of secondary atmospheres on ultrashort-period exoplanets \citep{2021ApJ...922L...4D,2023ApJ...954..202B} and the difference between day- and nightside mineralogy of tidally-locked rocky exoplanets \citep{2012ApJ...752....7H,Essack2020,Fortin2022}. The upcoming JWST cycles will enable direct comparison with such models, but for this undertaking detailed simulations of the thermal evolution of the most highly-observable planets are required, taking into accounts variations of melting behaviour, for example due to different redox states and accompanying differences in melting behaviour \citep{2021PNAS..11810427L}, core-mantle ratios \citep{2008ApJ...688..628E}, and chemical equilibration between core and mantle \citep{2021ApJ...914L...4L}. Constraining and understanding these processes will define the next years of exoplanet research, and we discuss an outlook with regards to mission development in Section \ref{sec:outlook}. Historically, however, theoretical progress has been made by expanding from the previously known regimes of the terrestrial planets, extending into novel regimes that may govern compositionally and thermally diverse exoplanetary regimes.

\subsubsection{Volatile cycling on temperate exoplanets}
\label{sec:interior_cycling}

Much of the previous discussion in this section has focused on specific physical or chemical processes that operate within an isolated layer of the planet, such as the mantle. However, it is thought that the surface habitability of the Earth is established through the coupling of mantle and atmosphere in the carbonate-silicate cycle. The basis of this conceptual framework is the geochemical evidence for approximately constant surface temperatures between 0--50$^\circ$C since the Archean eon and accompanying decrease in atmospheric carbon dioxide levels over billions of years \citep{2020SciA....6.1420C}. This phenomenon can be explained by the carbonate-silicate cycle, in which weathering of calcium and magnesium silicates in rocks and soils release ions that are transported to the seafloor via runoff and precipitation. Due to ongoing subduction on mobile lid planets, such as the Earth, this system can establish a negative temperature feedback that explains Earth's past surface temperature evolution. Substantial effort in the past years has gone into expansion of this framework to exoplanets, with a focus on carbon in- or outgassing \citep[e.g.,][]{2018A&A...614A..18D,2020ApJ...902L..10H,2020A&A...643A..44S,2021PEPI..32006788G,2021PSJ.....2..208K,2021A&A...651A.103N,2021PSJ.....2...49H,2023ApJ...942L..20H}, water cycles and ocean levels \citep[e.g.,][]{2014ApJ...781...27C,2015ApJ...801...40S,2020MNRAS.496.3786M,2022PSJ.....3...66G,2022AsBio..22..713M}, or tidal heating \citep{2017E&PSL.474...13M,2018A&A...613A..37B,2019A&A...624A...2D,2022A&A...663A..79A,2024A&A...684A..49F}, incorporating a range of physical and chemical effects. Importantly, however, the carbonate-silicate cycle is foremost dependent on atmospheric and surface fluxes \citep[e.g.,][]{2013ApJ...765..131K,2020ApJ...896..115G,2022JGRE..12707456G}. Typically, carbon storage in the mantle is assumed in such models, and only recently several works have expanded to couple surface volatile levels with geodynamic simulations \citep[e.g.,][]{2019A&A...627A..48H,2020A&A...643A..44S,2021A&A...649A..15O,2022ApJ...930L...6U}. Fully coupled models that take into account the deep carbon cycle with self-consistent mantle convection and atmospheric evolution remain to be established. From an atmospheric point-of-view, a variety of poorly constrained parameters can sensitively affect the surface balance \citep[e.g.,][]{2021NatGe..14..143G}, such that observational evidence from exoplanet science is required to better constrain coupled climate-interior models \citep{2017ApJ...841L..24B,2020NatCo..11.6153L,Triaud2023,2024SpScT...4...75G}.

\section{Towards the characterization of exo-Earths} \label{sec:outlook}

The major goals of exoplanet science -- from the authors' perspective -- are (i) to explore the evolutionary diversity of worlds in the universe, (ii) constrain the uniqueness of the Earth and Earth-like planets across planetary systems, and (iii) establish quantitative evidence for the abundance or absence of life beyond the Solar System. All three are connected and build on each other. Without a heuristic understanding of planetary diversity, the frequency of truly Earth-like worlds will remain a mystery. Without an interdisciplinary understanding of the main processes and evolutionary trajectories of Earth-like worlds there is no baseline scenario in which to interpret putative biosignatures in sparse remote signals. 

\subsection{Contrasting unhabitable, habitable, and prebiotic worlds} \label{sec:habitability}

The next stages in reaching these goals will be to understand the dividing processes between unhabitable, potentially habitable, and prebiotic (or urable) worlds \citep{2016AsBio..16...89C,2021plha.book.....K,2022AsBio..22..889D}. The first distinction may appear clear: finding lines between worlds that cannot sustain life and those that potentially can. However, the recent debate surrounding potential life in the Venusian atmosphere certainly showcased the inherent difficulties in remote interpretation of planetary environments \citep{2004AsBio...4...11S,2021NatAs...5..655G,2020A&A...644L...2S,2022PNAS..11921702B,2023AsBio..23.1189F}. From the perspective of planetary habitability, however, the potential abundance of liquid water has been used as key environmental sign for habitability, emphasizing the geodynamics discussion in Section \ref{sec:interior_cycling}. In the next few years with JWST, the ELTs, and transit survey missions, we will foremost learn more about short-period exoplanets, which will elevate our understanding of the interaction between the core, mantle, and atmosphere on inhospitable worlds, hence motivating the major emphasis of this review on the processes expected for high-entropy worlds. It is important to separate this debate from the discussion surrounding habitability, as deeper exploration of planetary physics and chemistry will lead the way to a more advanced understanding of planetary environments \citep{2022arXiv221014293M}. 

\paragraph{Unhabitable vs. habitable environments}

The trajectory of planetary astrophysics until this point has been often motivated by the classical definition of the habitable zone, defining a region around stars where liquid water is possible. Sophisticated climate models enable to map out the habitable zone \citep[e.g.,][]{2011ApJ...734L..13P,2017ApJ...837L...4R,2018ApJ...858...72R,2021MNRAS.504.1029B,2023NatAs...7.1070Y} with ever-increasing precision for roughly Earth-like conditions and composition. However, if exoplanet statistics so far is a guidance, then the underlying compositional and thermal variation of the exoplanet census is far wider than we have imagined. In particular, volatile and energy fluxes across the main planetary reservoirs of core, mantle, and atmosphere are poorly understood for partially molten planets. This will change in the upcoming years with ever-increasing characterization of ultrashort-period exoplanets. We see a number of major questions that are potentially addressable in the upcoming years. To start with, it needs to be constrained whether short-period super-Earths have observable atmosphere or not \citep{2009ApJ...703.1884S,2019ApJ...886..140K,2019ApJ...886..141M}. If there are no residual observable atmospheres in the $\gtrsim$ bar regime, then the question is why not. Are they lost by atmospheric escape and thus planetary surfaces are oxidized by this process \citep{2018AsBio..18..630M,2018SSRv..214...65T,2019AJ....158...26L}? Or is the planetary volatile budget hidden in the interior, for example in the form of high-pressure ices or partitioned into magmatic layers (Section \ref{sec:atmospheres1})? If residual atmospheres exist, are they oxidized (O-rich) \citep{2021ApJ...909L..22K,2021ApJ...922L...4D} or reduced (H-rich) \citep{2021ApJ...914L...4L,2022PSJ.....3..127S}? For the specific case of M-dwarf rocky exoplanets, the abundance of observable atmospheres can likely be addressed by a dedicated large-scale survey, combining the unique capabilities of with JWST and HST \citep{DDT_Report2024}, and therefore the next few years of JWST science operations will be crucial for distinguishing the prevalence and nature of atmospheric volatiles on rocky exoplanets.

Because of the prevalence of rocky exoplanets inside the runaway greenhouse threshold, observations of planets with super-runaway (but not yet desiccated/escaped) atmospheres will yield insights into the evolutionary trajectory of magma ocean worlds on longer timescales \citep{Miller-Ricci2009,2014ApJ...784...27L,2019A&A...621A.125B,2024PSJ.....5....3S}. An opportunity with JWST/NIRSPec will be to observe the 4 $\mu$m atmospheric window of water vapour in a steam atmosphere \citep{2019ApJ...875...31K,2021ApJ...919..130B} to constrain the energy loss from magma ocean planets. As is known for terrestrial-type planets, the oxidation state of the planetary interior has a first order effect on the magmatically generated gases at the surface, thus governing their long-term climate \citep[e.g.,][]{2021PEPI..32006788G,2022JGRE..12707123L,2023MNRAS.525.3703G}. Quite recently, with the dawn of JWST science operations, this concept has come into sharper focus for sub-Neptune exoplanets: if sub-Neptunes are dominated by magma oceans in their deep interior \citep{2018ApJ...869..163V,2021ApJ...914L...4L,2022PSJ.....3..127S}, then the internal redox state can potentially be probed with spectroscopy. In that case, a number of gases are potentially diagnostic of both the atmosphere-interior phase state (liquid or solid) and its composition (rock or volatile). \citet{2024ApJ...962L...8S} showed that the absence of \ce{NH3} and simultaneous presence of \ce{CH4} and \ce{CO2} in the atmosphere of K2-18 b can be self-consistently explained by an internal reduced (H- or Fe-rich) magma ocean, challenging the Hycean (water-ocean) interpretation of temperate sub-Neptunes \citep{2020ApJ...891L...7M,2023ApJ...956L..13M,2021ApJ...914...38Y,2021ApJ...921L...8H,2021ApJ...922L..27T}. With new observations, the miscibility of deep internal and outer atmospheric layers has come into sharper focus \citep{2024arXiv240303325B,2024A&A...683L...2H}. 

The next few years will be critical for finding atmospheric markers that can distinguish internal phase and redox state for sub-Neptunes, leading the way toward smaller super-Earths and terrestrial exoplanets. The \ce{CO}/\ce{CO2} ratio and the abundance of nitrogen species are examples \citep{2024ApJ...962L...8S,2024ApJ...963..157T}, but further diagnostics are needed, motivating cross-disciplinary work exploring more general partitioning coefficients \citep{2023FrEaS..1159412S,2023PSJ.....4...30H} at high pressure, temperature, and varying redox state \citep{2022ARA&A..60..159W}. On a population scale, testing the radius inflation effect of runaway greenhouse atmospheres can be a probe for the climatic effects of oxidized volatile endowments with ESA PLATO \citep{2024PSJ.....5....3S}. For tidally-locked and atmosphere-less planets the extent and depth of dayside magma oceans, and the mineralogy of residual surfaces \citep{2012ApJ...752....7H} and evaporating tails \citep{2024MNRAS.528.1249C} will be informative on interior geophysics and geochemistry, yielding new constraints on geodynamical models \citep{2021ApJ...908L..48M,2023A&A...678A..29M,2022ApJ...936..148B} and thus ultimately expanding coupled climate-interior frameworks of temperate rocky exoplanets. Beyond the lifetime of currently planned missions are a number of important avenues that will technically remain unfeasible until the arrival of large-scale direct imaging surveys \citep{2019AJ....158...83A,2020arXiv200106683G,2022A&A...664A..21Q,2022A&A...664A..22D}. To these belong the detection of surface oceans \citep{2009ApJ...700..915C,2017AJ....154..189F}, the mapping of the climate diversity of young, Hadean analog worlds \citep{2019A&A...621A.125B,2019JGRE..124.2015K,2021JGRE..12606711L}, rule out abiotic false positive biomarkers \citep{2022AJ....163..299D,2023AJ....166..254H,2023ApJ...951L..39K}, and statistical signatures of the carbonate-silicate cycle \citep{2017ApJ...841L..24B,2020NatCo..11.6153L,Triaud2023}. These and similar questions that we discussed throughout this review article will crucially inform our understanding of the abiotic processes that shape terrestrial and non-terrestrial worlds,  bringing us closer to an understanding of what defines Earth-like planets in terms of physics and chemistry. Achieving a better understanding of planetary processes is a uniting endeavour of planetary science and exoplanet astronomy \citep{2021JGRE..12606643K,2024SpScT...4...75G}.

\paragraph{Prebiotic environments}

The habitability of a planetary environment, however, is not the only pre-requisite for the chemical origins of life. Laboratory studies of prebiotic chemistry indicate that under present-day conditions the emergence of life as we know it could not develop. Rather, the chemical environment needs to be substantially more reduced to favour key precursor compounds, such as \ce{HCN}, \ce{CH2O}, \ce{CN2H2}, and \ce{C3HN} \citep{benner2020did,2020SciA....6.3419S}. This presents a major question for planetary science, as it relates to atmospheric, surface, and interior fluxes of energy and chemical compounds. \citet{2022AsBio..22..889D} therefore coined the term 'urability' to distinguish the planetary conditions that are required for the chemical origin of life, a more stringent set of criteria than surface habitability through the presence of liquid water. The question of differing chemical, potentially more reduced environments on early terrestrial-like planets interconnects the discussion we have provided in this review. How must physics and chemistry on a planetary scale operate, such that a terrestrial planet will undergo a sufficiently reducing surface environment for the origins of life, but eradicate all nascent biology immediately? Different suggestions addressing parts of this challenge have been proposed. For instance, planets may go through an initially reduced atmospheric phase as a result of atmospheric escape \citep{2001Sci...293..839C,2012Icar..219..267W,2021MNRAS.505.2941Y}, or reduced impacts may contribute to the surface environment \citep{2015Icar..257..290K,2017E&PSL.480...25G,2020PSJ.....1...11Z,2020GGG....2108734K,2022PSJ.....3..115I,2023PSJ.....4..169W}. Even though M star planets may differ fundamentally from Earth-like planets, they offer unique opportunities with regards to the impact flux on Hadean-like worlds \citep{2022ApJ...938L...3L,2023A&A...671A.114J}, as their extended pre-main sequence phases offer ample chances to study the desiccation of magma ocean atmospheres, and the possible transition from reduced to oxidized planets. Deciphering this key chemical and structural transition in the geophysical evolution of rocky planets will be a major stepping stone to interpret possible signs of extant life \citep{2015SciA....1E0047S,2017ARA&A..55..433K}. What are the main factors that oxidize or reduce planetary surfaces during their lifetime, and how variable is the distribution of surface geochemistires on rocky extrasolar planets?

\subsection{Biomarkers: finding extrasolar life} \label{sec:biomarkers}

The pathway toward convincing evidence for extrasolar life will rely on continued public support through novel space missions that have the technical capabilities to do so. No presently selected and planned exoplanet mission will be able to decipher atmospheric or surface biosignatures on temperate terrestrial exoplanets around Sun-like stars because the robust detection of key atmospheric and surface signals requires extraordinary resolution and star-light suppression techniques \citep{2018AsBio..18..739F}. Importantly, astronomical telescopes need to be able to distinguish key pairs of molecules in the atmospheres of exoplanets that may tell us about the likelihood of biological life on these worlds. A key concept is to search for chemical disequilibrium in exoplanet atmospheres \citep{1975RSPSB.189..167L,2018AsBio..18..663S,2018AsBio..18..709C,2020ApJ...892..127W,2022NatAs...6..189K,2024NatAs...8..101Y}, such as \ce{O2} \citep{2014AsBio..14...67M,2018AsBio..18..630M,2022A&A...665A.156K}, \ce{NH3} \citep{2022AsBio..22..171H}, \ce{PH3} \citep{2020AsBio..20..235S}, \ce{C5H8} \citep{2021AsBio..21..765Z}, the \ce{CO2}--\ce{CH4} pair \citep{2018SciA....4.5747K,2020ApJ...901..126W,2020PSJ.....1...58W,2022PNAS..11917933T,2023ApJ...944..209L}, or the \ce{N2}--\ce{O2} pair \citep{2019AsBio..19..927L}. Other suggestions include, for example, atmospheric seasonality \citep{2004NewA...10...67G,2018ApJ...858L..14O} or step-like spectral features of biology \citep['vegetation's red edge',][]{2005AsBio...5..372S,2018AsBio..18.1123O}.

Uncovering these signatures in remote data is an even greater challenge than deciphering the main abiotic processes shaping the climate and geodynamics of rocky exoplanets and will thus require strategic developments toward this goal, in particular coverage of key spectral windows in the mid-infrared \citep{2018ExA....46..543D,2018ExA....46..475D,2021AJ....161..180F,2015IJAsB..14..279Q,2022A&A...664A..21Q} and at shorter wavelengths \citep{2019arXiv191206219T,2020arXiv200106683G,2022ExA....54.1237S}. Ground-based direct imaging via the upcoming Extremely Large Telescopes (ELTs) may enable atmospheric characterization for the few very nearest exoplanetary systems \citep{2014arXiv1412.1048U,2015A&A...576A..59S,2019BAAS...51c.162L}, however, exploration beyond that would require extensive, likely unfeasible telescope time allocation \citep{2023AJ....165..267H,2023PSJ.....4...83C,2023ASPC..534..799C}. Therefore, a sufficient number of rocky exoplanets will only be accessible with space-based surveys. In the context of long-term planning strategies, ESA has announced strategic developments toward space-based mid-infrared capabilities for characterizing temperate terrestrial exoplanets \citep{ESAVoyage2050,2022A&A...664A..21Q,2022ExA....54.1197Q}, while NASA will develop a large-scale imaging mission with a focus on reflected light \citep{NASADecadalSurvey2021}. With nominal mission development timelines and continued community support, these large-scale direct imaging observatories will be able to start their surveys in the 2040s. The development path towards the exploration of potentially 'Earth-like' exoplanets will lead right through an informed understanding of the abiotic environment of rocky and probably inhospitable exoplanets.

\section{Summary} \label{sec:summary}

In this review we discussed the major processes involved in shaping the physics and chemistry of super-Earths and potentially Earth-like exoplanets with a focus on the observational and theoretical state of the field right at the start of  JWST science operations. Current major questions surrounding the nature of low-mass exoplanets are related to their composition, phase state, and thermal evolution. Statistical surveys have shown that the exoplanet distribution is not smooth, but fragmented into distinct, but blurry, regimes, with clustering of planets into denser, Earth-similar compositions, and lighter ones. These latter under-dense planets in the super-Earth size-regime are either dominated by hydrogen atmospheres or volatile-rich bulk compositions. Such compositions on short-period orbits, and thus under high irradiations, suggest that their interiors are in partially to fully molten phase states, which has significant consequences for the interaction of deep planetary interior and atmospheric layers. Characterising distinct thermodynamic and compositional regimes among the lower-density, potentially molten super-Earths will be important to understand key aspects of planet formation, including migration and the main accretion process, as well as planetary evolution. The coupled feedback processes between interior and atmosphere of rocky exoplanets govern the chemical segregation of the planetary structure, including core, mantle, potentially mixed volatile-refractory layers, climate, and ultimately surface conditions. 

For the Solar System there has been a long-standing debate about the dynamics of energy transfer during magma ocean stages and the concurrent redistribution of major redox-active elements. Exoplanet science in the next few years can help advancing this debate by uncovering the distribution of redox states among planetary mantles and their secondary atmospheres. Key observational challenges include the discovery of secondary atmospheres on rocky exoplanets, the amount of volatiles locked up in planetary interiors, the prevalence of atmosphere-stripped rocky exoplanets, the geometry and depth of dayside magma oceans, the phase-state of super-Earth and sub-Neptune interiors, and the geochemical properties of exoplanetary surfaces. While exoplanet science is observationally driven, key advances critically depend on Earth-based laboratory simulations of novel compositional and thermodynamic phase space. Of particular relevance are expansions of the equations of state for seemingly exotic rock compositions with elemental fractionations unknown from the Solar System. In addition, novel measurements of volatile distribution between different phase states at thermodynamic conditions most relevant for the atmosphere-interior interface of super-Earths and sub-Neptunes ($P \sim$ 10–1000 bar, $T \gtrsim$ 1500 K, strongly varying $f$O$_2$, $\sim$ IW-6 to IW+6, and fractionation in C-H-O-N-S and O-Si-Fe-Mg-Ca abundances) are required to enable evolutionary, multi-dimensional models of planetary structure, geodynamics, and climate that can be compared against high-resolution observations of rocky exoplanets. 

The next few years of rocky exoplanet science will be dominated by exploration of high-temperature regimes of planetary evolution. This gives us the prime opportunity to understand phases of planetary evolution that are long gone in the Solar System. In particular, the dynamics of planetary magma oceans and the accompanied formation of planetary atmospheres is a unifying interdisciplinary question that will help to constrain key processes governing the nature of early planetary surfaces, interior geophysics, and climate science that are degenerate for laboratory simulations and present-day observations of Solar System objects. While no single observation of individual exoplanets will reveal the full picture of planetary evolution, the possibility to characterize rocky exoplanets across the full range of thermodynamic regimes and ages yields a unique power to create a unified theory of planetary evolution. On the road to discover and characterize potential Earth analogs around Sun-like stars, the lessons learned from more extreme planets, such as those on ultrashort-period orbits, around low-mass M stars, and those at the end of their life cycles, will be important for the findings of future large-scale direct imaging surveys. This novel class of exoplanet surveys will be required to reach the necessary starlight suppression, resolution, and spectral coverage to identify traces of key molecules in the atmospheres of temperate terrestrial exoplanets, to distinguish clement and potentially habitable exoplanets from those that are not, probe the distribution of Earth-like worlds, and ultimately constrain the prevalence of life as we know it across the multitude of worlds in our galaxy.

\vspace{-0.3cm}
\acknowledgments
We thank Claire Zurkowski for contributing Fig. \ref{fig:geochemistry}, Mantas Zilinskas and Sebastian Zieba for contributing Fig. \ref{fig:rockvapour}, and Hannah Diamond-Lowe, Charles-Édouard Boukaré, Johanna Teske, Oliver Shorttle, Eliza Kempton, Mark Hammond, Lena Noack, Junjie JJ Dong, Brad Foley, Aline Vidotto, Conel Alexander, and Martin Schlecker for discussions and suggestions that significantly improved the manuscript. TL acknowledges support by the Branco Weiss Foundation, the Alfred P. Sloan Foundation (AEThER project, G202114194), and NASA's Nexus for Exoplanet System Science research coordination network (Alien Earths project, 80NSSC21K0593). YM acknowledges support from the Dutch Science Foundation (NWO) Planetary and Exoplanetary Science (PEPSci) grant. This review made use of NASA's Astrophysics Data System and Exoplanet Archive, and the NCCR PlanetS DACE platform.

\bibliography{references}{}
\bibliographystyle{aasjournal}

\end{document}